\documentclass[superscriptaddress,showpacs,amsmath,amssymb,aps,prc,floatfix,twocolumn,showkeys]{revtex4-1}

\usepackage{graphicx}% Include figure files
\usepackage{dcolumn}% Align table columns on decimal point
\usepackage{bm}% bold math
\usepackage[mathlines]{lineno}% Enable numbering of text and display math
\begin{document}

\title{Measurement of two-photon exchange effect by comparing elastic $e^\pm p$ cross sections}

%%%%%%%%%%%%%%% Latex Macros for institute addresses  %%%%%%%%%%%%%%%%%%%%%%%%% 
\newcommand*{\FIU}{Florida International University, Miami, Florida 33199}
\newcommand*{\FIUindex}{10}
\affiliation{\FIU}
\newcommand*{\ODU}{Old Dominion University, Norfolk, Virginia 23529}
\newcommand*{\ODUindex}{29}
\affiliation{\ODU}
 \newcommand*{\ANL}{Argonne National Laboratory, Argonne, Illinois 60439}
\newcommand*{\ANLindex}{1}
\affiliation{\ANL}
\newcommand*{\UTFSM}{Universidad T\'{e}cnica Federico Santa Mar\'{i}a, Casilla 110-V Valpara\'{i}so, Chile}
\newcommand*{\UTFSMindex}{36}
\affiliation{\UTFSM}
\newcommand*{\JLAB}{Thomas Jefferson National Accelerator Facility, Newport News, Virginia 23606}
\newcommand*{\JLABindex}{35}
\affiliation{\JLAB}
\newcommand*{\RPI}{Rensselaer Polytechnic Institute, Troy, New York 12180-3590}
\newcommand*{\RPIindex}{30}
\affiliation{\RPI}
\newcommand*{\CSUDH}{California State University, Dominguez Hills, Carson, CA 90747}
\newcommand*{\CSUDHindex}{2}
\affiliation{\CSUDH}
\newcommand*{\CANISIUS}{Canisius College, Buffalo, NY}
\newcommand*{\CANISIUSindex}{3}
\affiliation{\CANISIUS}
\newcommand*{\CMU}{Carnegie Mellon University, Pittsburgh, Pennsylvania 15213}
\newcommand*{\CMUindex}{4}
\affiliation{\CMU}
\newcommand*{\CUA}{Catholic University of America, Washington, D.C. 20064}
\newcommand*{\CUAindex}{5}
\affiliation{\CUA}
\newcommand*{\SACLAY}{CEA, Centre de Saclay, Irfu/Service de Physique Nucl\'eaire, 91191 Gif-sur-Yvette, France}
\newcommand*{\SACLAYindex}{6}
\affiliation{\SACLAY}
\newcommand*{\CNU}{Christopher Newport University, Newport News, Virginia 23606}
\newcommand*{\CNUindex}{7}
\affiliation{\CNU}
\newcommand*{\UCONN}{University of Connecticut, Storrs, Connecticut 06269}
\newcommand*{\UCONNindex}{8}
\affiliation{\UCONN}
\newcommand*{\FU}{Fairfield University, Fairfield CT 06824}
\newcommand*{\FUindex}{9}
\affiliation{\FU}
\newcommand*{\FSU}{Florida State University, Tallahassee, Florida 32306}
\newcommand*{\FSUindex}{11}
\affiliation{\FSU}
\newcommand*{\Genova}{Universit$\grave{a}$ di Genova, 16146 Genova, Italy}
\newcommand*{\Genovaindex}{12}
\affiliation{\Genova}
\newcommand*{\GWUI}{The George Washington University, Washington, DC 20052}
\newcommand*{\GWUIindex}{13}
\affiliation{\GWUI}
\newcommand*{\ISU}{Idaho State University, Pocatello, Idaho 83209}
\newcommand*{\ISUindex}{14}
\affiliation{\ISU}
\newcommand*{\INFNFE}{INFN, Sezione di Ferrara, 44100 Ferrara, Italy}
\newcommand*{\INFNFEindex}{15}
\affiliation{\INFNFE}
\newcommand*{\INFNFR}{INFN, Laboratori Nazionali di Frascati, 00044 Frascati, Italy}
\newcommand*{\INFNFRindex}{16}
\affiliation{\INFNFR}
\newcommand*{\INFNGE}{INFN, Sezione di Genova, 16146 Genova, Italy}
\newcommand*{\INFNGEindex}{17}
\affiliation{\INFNGE}
\newcommand*{\INFNRO}{INFN, Sezione di Roma Tor Vergata, 00133 Rome, Italy}
\newcommand*{\INFNROindex}{18}
\affiliation{\INFNRO}
\newcommand*{\INFNTUR}{INFN, Sezione di Torino, 10125 Torino, Italy}
\newcommand*{\INFNTURindex}{19}
\affiliation{\INFNTUR}
\newcommand*{\ORSAY}{Institut de Physique Nucl\'eaire, CNRS/IN2P3 and Universit\'e Paris Sud, Orsay, France}
\newcommand*{\ORSAYindex}{20}
\affiliation{\ORSAY}
\newcommand*{\ITEP}{Institute of Theoretical and Experimental Physics, Moscow, 117259, Russia}
\newcommand*{\ITEPindex}{21}
\affiliation{\ITEP}
\newcommand*{\JMU}{James Madison University, Harrisonburg, Virginia 22807}
\newcommand*{\JMUindex}{22}
\affiliation{\JMU}
\newcommand*{\KNU}{Kyungpook National University, Daegu 702-701, Republic of Korea}
\newcommand*{\KNUindex}{23}
\affiliation{\KNU}
\newcommand*{\LPSC}{LPSC, Universit\'e Grenoble-Alpes, CNRS/IN2P3, Grenoble, France}
\newcommand*{\LPSCindex}{24}
\affiliation{\LPSC}
\newcommand*{\MISS}{Mississippi State University, Mississippi State, MS 39762-5167}
\newcommand*{\MISSindex}{25}
\affiliation{\MISS}
\newcommand*{\UNH}{University of New Hampshire, Durham, New Hampshire 03824-3568}
\newcommand*{\UNHindex}{26}
\affiliation{\UNH}
\newcommand*{\NSU}{Norfolk State University, Norfolk, Virginia 23504}
\newcommand*{\NSUindex}{27}
\affiliation{\NSU}
\newcommand*{\OHIOU}{Ohio University, Athens, Ohio  45701}
\newcommand*{\OHIOUindex}{28}
\affiliation{\OHIOU}
\newcommand*{\Rich}{University of Richmond, Richmond, Virginia 23221}
\affiliation{\Rich}
\newcommand*{\ROMAII}{Universita' di Roma Tor Vergata, 00133 Rome Italy}
\newcommand*{\ROMAIIindex}{31}
\affiliation{\ROMAII}
\newcommand*{\MSU}{Skobeltsyn Institute of Nuclear Physics, Lomonosov Moscow State University, 119234 Moscow, Russia}
\newcommand*{\MSUindex}{32}
\affiliation{\MSU}
\newcommand*{\SCAROLINA}{University of South Carolina, Columbia, South Carolina 29208}
\newcommand*{\SCAROLINAindex}{33}
\affiliation{\SCAROLINA}
\newcommand*{\TEMPLE}{Temple University,  Philadelphia, PA 19122 }
\newcommand*{\TEMPLEindex}{34}
\affiliation{\TEMPLE}
\newcommand*{\EDINBURGH}{Edinburgh University, Edinburgh EH9 3JZ, United Kingdom}
\newcommand*{\EDINBURGHindex}{37}
\affiliation{\EDINBURGH}
\newcommand*{\GLASGOW}{University of Glasgow, Glasgow G12 8QQ, United Kingdom}
\newcommand*{\GLASGOWindex}{38}
\affiliation{\GLASGOW}
\newcommand*{\VT}{Virginia Tech, Blacksburg, Virginia   24061-0435}
\newcommand*{\VTindex}{39}
\affiliation{\VT}
\newcommand*{\VIRGINIA}{University of Virginia, Charlottesville, Virginia 22901}
\newcommand*{\VIRGINIAindex}{40}
\affiliation{\VIRGINIA}
\newcommand*{\WM}{College of William and Mary, Williamsburg, Virginia 23187-8795}
\newcommand*{\WMindex}{41}
\affiliation{\WM}
\newcommand*{\YEREVAN}{Yerevan Physics Institute, 375036 Yerevan, Armenia}
\newcommand*{\YEREVANindex}{42}
\affiliation{\YEREVAN}

\newcommand*{\NOWJLAB}{Thomas Jefferson National Accelerator Facility, Newport News, Virginia 23606}
\newcommand*{\NOWUK}{University of Kentucky, Lexington, Kentucky 40506}
\newcommand*{\NOWGLASGOW}{University of Glasgow, Glasgow G12 8QQ, United Kingdom}
\newcommand*{\NOWINFNGE}{INFN, Sezione di Genova, 16146 Genova, Italy}
\newcommand*{\NOWUF}{University of Florida, Gainesville, Florida 32611}

 %%%%%%%%%%%%%%% END OF Latex Macros for institute addresses  %%%%%%%%%%%%%%%%%%%%%

\author{D. Rimal}
\altaffiliation[Current address: ]{\NOWUF}
\affiliation{\FIU}
\author{D. Adikaram}
\affiliation{\ODU}
\author{B.A. Raue}
\altaffiliation[Corresponding author: baraue@fiu.edu]{}
\affiliation{\FIU}
\author{L.B. Weinstein}
\affiliation{\ODU}
\author{J. Arrington}
\affiliation{\ANL} 
\author{W.K. Brooks}
\affiliation{\UTFSM}
\author {M.~Ungaro} 
\affiliation{\JLAB}
\affiliation{\RPI}
\author {K.P. ~Adhikari} 
\affiliation{\MISS}
\affiliation{\ODU}
\author {A.V.~Afanasev}
\affiliation{\GWUI}
\author {Z.~Akbar} 
\affiliation{\FSU}
% ...

\collaboration{The CLAS Collaboration}
\noaffiliation

\date{\today}

\begin{abstract}

\begin{description}
\item[Background] The electromagnetic form factors of the proton measured by unpolarized and
polarized electron scattering experiments show a significant disagreement that grows with the squared
four momentum transfer ($Q^{2}$). Calculations have shown that the two measurements can be largely
reconciled by accounting for the contributions of two-photon exchange (TPE).  TPE effects are not 
typically included in the standard set of radiative corrections since theoretical calculations of the TPE effects are 
highly model dependent, and, until recently, no direct evidence of significant TPE effects has
 been observed.

\item[Purpose] We measured the ratio of positron-proton to electron-proton elastic-scattering cross sections in order to 
determine  the TPE contribution to elastic electron-proton scattering and thereby resolve the proton electric form factor discrepancy.

\item[Methods]
We produced a mixed simultaneous electron-positron beam in Jefferson Lab's Hall B by passing the 5.6 GeV primary electron beam 
through a radiator to produce a bremsstrahlung photon beam and then passing the photon beam through a convertor to produce 
electron/positron pairs.  The mixed electron-positron (lepton) beam with useful energies from approximately 0.85 to 3.5 GeV then struck a 30-cm 
long liquid hydrogen (LH$_2$) target located within the CEBAF Large Acceptance Spectrometer (CLAS). By detecting both the scattered leptons 
and the recoiling protons we identified and reconstructed elastic scattering events and determined the incident lepton energy.  A detailed 
description of the experiment is presented.

\item[Results] We present previously unpublished results for the quantity $R_{2\gamma}$, the TPE
correction to the elastic-scattering cross section, at $Q^2\approx 0.85$ and 1.45 GeV$^2$ over a large range of virtual photon polarization 
$\varepsilon$.   

\item[Conclusions]  Our results, along with recently published results from VEPP-3, demonstrate a non-zero contribution from TPE effects 
and are in excellent agreement with the calculations that include TPE effects and largely reconcile the form-factor discrepancy up to 
$Q^2\approx 2$ GeV$^2$. These data are consistent with an increase in $R_{2\gamma}$ with decreasing $\varepsilon$ at 
$Q^2\approx 0.85$ and 1.45 GeV$^2$.  There are indications of a slight increase in $R_{2\gamma}$  with $Q^2$.

\end{description}
\end{abstract}


\maketitle

\section{Introduction}
\label{sec-intro}
The electromagnetic form factors are the fundamental observables that 
contain information about the spatial distribution of the charge and magnetization 
inside the proton. The electric ($G_{E} (Q^2)$) and magnetic ($G_{M} (Q^2)$)
form factors have been extracted by analyzing data from both unpolarized and polarized 
electron scattering experiments assuming an exchange of a virtual photon between 
the electron and the proton while accounting for soft radiative effects and external hard photons.

The unpolarized electron scattering experiments use the 
Rosenbluth separation method~\cite{walker94, andivahis94, berger71, litt70, christy04, qattan05}, 
where the $e^{-}p$ elastic cross section is measured at fixed four-momentum transfer, $Q^{2}$ ($Q^2=-q^2=4EE'\sin^2(\theta/2)$, 
where $E$ is the 
incident electron beam energy, $E'$ is the scattered electron energy, and $\theta$ is the angle of the scattered electron), while varying 
the electron scattering angle and the incident energy of the electron.  The form factors are then extracted from the reduced cross section, given by
\begin{equation}
\label{eq:red_sigma}
\sigma_R=\frac{d\sigma}{d\Omega} \frac{\left(1+\tau\right)\varepsilon}{\sigma_{\text{Mott}}\tau}=\frac{\varepsilon}{\tau}G^2_E\left(Q^2\right)+G^2_M\left(Q^2\right),
\end{equation}
where $\sigma_{\text{Mott}}$ is the cross section for elastic scattering from a point-like proton, 
$\varepsilon=\left[1+2\left(1+\tau\right)\tan^2\left(\theta/2\right)\right]^{-1}$ is the virtual photon polarization, 
 $\tau=Q^2/4M_p^2$,  and $M_p$ is the proton mass.  
 $G^2_E(Q^2)$ is then proportional to the $\varepsilon$-dependence of $\sigma_R$ and $G^2_M(Q^2)$ is proportional to the cross section 
 extrapolated to $\varepsilon=0$. 

\begin{figure}[tb]  
\includegraphics[width=0.45\textwidth]{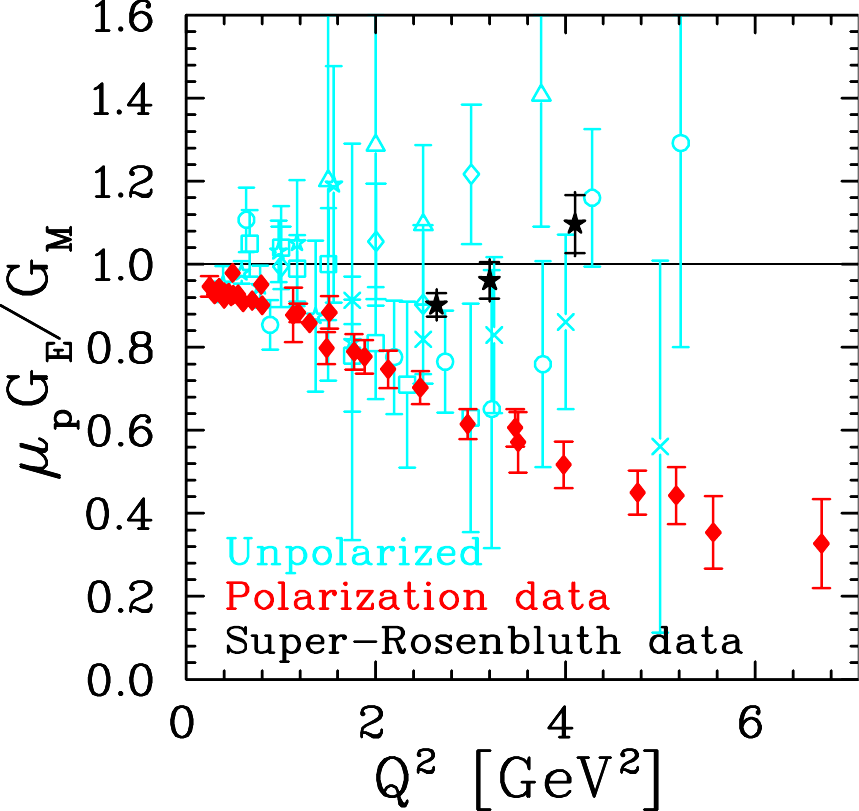} 
\begin{center} 
\caption{\label{fig-GE-GM} 
(Color Online) Ratio of $\frac{\mu_{p}G_{E} (Q^2) }{G_{M} (Q^2)}$ from Rosenbluth
\cite{arrington03a} (open cyan symbols) and ``Super Rosenbluth'' \cite{qattan05} (black stars) measurements and from polarization 
measurements \cite{punjabi05, puckett10, puckett11,zhan11,Ron11} 
(filled red diamonds) measurements.}
\end{center} 
\end{figure}

Recoil polarization experiments~\cite{punjabi05, puckett10, puckett11, zhan11, Ron11} measure the polarization of the recoiling 
proton after scattering a polarized electron off an unpolarized proton target.  The ratio of the electric and magnetic form factors 
$G_E(Q^2)/G_M(Q^2)$ is proportional to the ratio of the transverse and longitudinal polarization of the recoil proton. The form-factor ratio
can also be extracted from spin-dependent elastic scattering of polarized electrons from polarized protons~\cite{crawford07}.
The ratio of the electric to magnetic form factors, $\frac{\mu_{p}G_{E} (Q^2) }{G_{M} (Q^2)}$, where $\mu_p$ is the proton magnetic moment, 
extracted from polarized and unpolarized electron scattering shows a significant discrepancy that grows with $Q^{2}$, as seen in Fig.~\ref{fig-GE-GM}.

A popular explanation is that the observed discrepancy results from neglecting hard two-photon exchange (TPE)
corrections~\cite{guichon03, blunden03, chen04, arrington04a}, a higher-order contribution to the radiative
corrections~\cite{arrington07c, carlson07, arrington11b}.  In TPE, the first exchanged virtual photon can excite the proton to a higher state 
and the second virtual photon de-excites the proton back to its ground state.  TPE will affect the cross section through its interference with the 
single photon exchange (First Born Approximation) amplitude.   This should be smaller than the Born cross section by a factor of 
$\alpha\approx 1/137$.  However, the size of the TPE contribution to the cross section is expected to have a significant $\varepsilon$
dependence~\cite{blunden05, afanasev05} that grows with $Q^2$, while the $\varepsilon$-dependent part of the unpolarized
cross section in the Born Approximation becomes very small at large $Q^2$.

%A popular explanation for the observed discrepancy is that it is the result of neglecting higher order 
%radiative corrections such as contributions from the two-photon exchange (TPE) effect.  TPE is expected to affect the   
%$\varepsilon$ dependence 
%of the elastic cross section measurements~\cite{guichon03, blunden03, chen04, arrington04a, afanasev05}. 
%If that is the case, the first exchanged virtual photon can excite the proton to a higher state and 
%the second exchange of a virtual photon de-excites the proton back to its ground state. Theoretical 
%investigations suggest that contributions due to the TPE effects may be enough to reconcile the 
%puzzling form-factor discrepancy~\cite{arrington07c, carlson07, arrington11b}. Calculations have shown that the 
%TPE contribution can have a significant dependence on the lepton scattering angle, or more precisely, the transverse polarization of the 
%virtual photon ($\varepsilon$). The magnitude of the TPE effect is expected to be the largest at small $\varepsilon$
%\cite{guichon03, blunden03, chen04,arrington04a, afanasev05, arrington07b, blunden09, sibirtsev10}. 

\begin{figure}[htb]  
\includegraphics[width=0.5\textwidth]{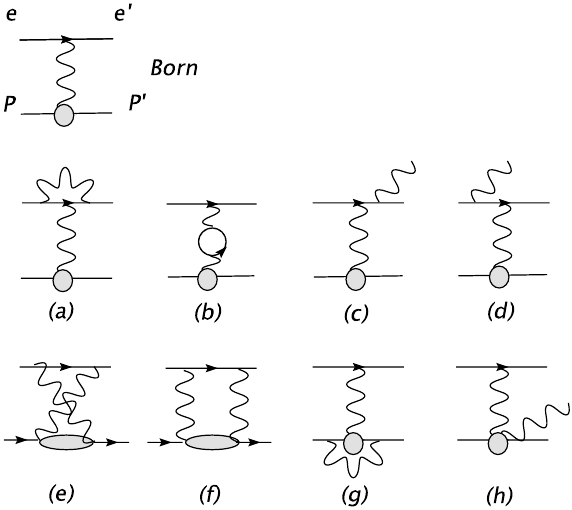} 
\begin{center} 
\caption{\label{fig-diag} 
Feynman diagrams for elastic lepton-proton scattering, including the first-order QED radiative corrections.  Diagrams (a) and (g) show the electron and
proton vertex renormalization terms, diagram (b) shows the photon propagator renormalization term, diagrams (c) and (d) show the electron 
bremsstrahlung term, diagram (h) shows the proton bremsstrahlung term, and diagrams (e) and (f) show the two-photon exchange terms, 
where the intermediate state can be an unexcited
proton, a baryon resonance or a continuum of hadrons.} 
\end{center} 
\end{figure}

Calculations of the box and crossed TPE diagrams (Figs.~\ref{fig-diag}(f) and~\ref{fig-diag}(e)) in elastic $e^{-}p$ scattering are complicated 
since such calculations require complete knowledge of intermediate hadronic states~\cite{blunden05, afanasev05b, kondratyuk06,
kondratyuk07, belushkin07, borisyuk08,borisyuk12, borisyuk14, tomalak14, Zhou2014}. As a result, these calculations have significant
model dependence. 

A model-independent way of measuring the size of the TPE effect is by comparing  
$e^{-}p$ and $e^{+}p$ elastic scattering cross sections~\cite{arrington04b, arrington09b}. 
The interference between one- and two-photon exchange diagrams has the opposite sign
for electrons and positrons while most of the other radiative corrections are identical for electrons and positrons 
and cancel to first order in the ratio. Apart from TPE, the only other charge-dependent 
contribution comes from the interference between the lepton and proton bremsstrahlung radiation 
terms, which is of comparable size to the TPE effect.  Note that the TPE contributions are typically neglected in the correction of electron 
scattering data except for the infrared-divergent contribution, which is needed to cancel the IR-divergent terms associated with low-energy 
bremsstrahlung. There are different conventions for how to include the IR-divergent TPE contributions \cite{mo69,maximon00}, and these yield 
slight differences in the meaning of the remaining finite TPE contributions \cite{arrington11b}, referred to here as $\delta_{2\gamma}$. In this work, 
we apply radiative corrections from Ref.~\cite{ent01}, which follows the Mo and Tsai convention \cite{mo69}, as do most published extractions of 
the elastic cross section (with the notable exception of Ref.~\cite{bernauer10}).

The ratio of the $e^\pm p$ elastic scattering cross sections can be written as 
\begin{eqnarray}
	R = \frac{\sigma(e^+p)}{\sigma(e^-p)} \approx
	\frac{1+\delta_{even}-\delta_{2\gamma}-\delta_{e.p.brem}}
     	{1+\delta_{even}+\delta_{2\gamma}+\delta_{e.p.brem}} \label{eq:R1} \\
	      \approx  1 - 2 ( \delta_{2\gamma} + \delta_{e.p.brem})/(1+\delta_{even}) ~,
\label{eq:R2}
\end{eqnarray} 
where $\delta_{even}$ is the total charge-even radiative correction
factor and $\delta_{2\gamma}$ and $\delta_{e.p.brem}$ are the
TPE and lepton--proton bremsstrahlung interference contributions.  See Ref.~\cite{Moteabbed} for more details.
The signs of $\delta_{2\gamma}$ and $\delta_{e.p.brem}$ are chosen
by convention such that they appear as additive corrections for
electron scattering. 
Typically, the experimental ratio $R$ is corrected for the calculated $\delta_{e.p.brem}$ and $\delta_{even}$ to isolate the TPE contribution:
\begin{equation}
\label{eq:R2g}
	R_{2\gamma} \approx 1 - 2 \delta_{2\gamma} .
\end{equation}
The measured TPE correction ($\delta_{2\gamma}$) can be directly used to correct the measured reduced unpolarized elastic scattering cross section, 
$\sigma_R$ (Eq.~\ref{eq:red_sigma}), as 
\begin{equation}\label{eq:corrected_sigmaR}
	\sigma_R^{corr}=\sigma_R \left(1- \delta_{2\gamma}\right)
\end{equation}
and then used to extract the TPE-corrected $G_E$ and $G_M$.

%Examinations of the cross section data and comparisons of the cross section and polarization results allow us
%to set some limited constraints on TPE contributions.  Deviations from a linear $\varepsilon$ dependence in
%$\sigma_R$ would indicate contributions beyond the single-photon exchange contribution. Examination of these
%data, in particular the precise data of Ref.~\cite{qattan05}, indicates that the TPE contributions are roughtly
%linear in $\varepsilon$~\cite{tvaskis06}. Assuming a linear TPE contribution, the polarization-Rosenbluth
%discrepancy can be used to estimate the size of the TPE contributions.
%For $Q^2$ above 3-4~GeV$^2$, an $\varepsilon$-dependent correction of approximately F would be enough to explain the
%observed discrepancy~\cite{guichon03, arrington03a, arrington04a, qattan11b}, while at lower $Q^2$ values the discrepancy is
%smaller and provides only weak limits on TPE contributions~\cite{qattan15}.  

An analysis of Rosenbluth separation data~\cite{tvaskis06} found no non-linear effects in the relationship between $\sigma_R$ and $\varepsilon$ in elastic 
\cite{qattan05}, inelastic, or deep inelastic scattering. Assuming a TPE contribution linearly dependent on $\varepsilon$, the polarization-Rosenbluth 
discrepancy can be used to estimate the size of the TPE contributions needed to reconcile them.
For $Q^2$ above 3-4~GeV$^2$, an $\varepsilon$-dependent correction of approximately 5\% could explain the
observed discrepancy~\cite{guichon03, arrington03a, arrington04a, qattan11b}. At $Q^2<2$ GeV$^2$ the discrepancy is
smaller and provides a less sensitive constraint on TPE contributions~\cite{qattan15}, though it is consistent with a few \% correction.

%%%%%%%%%%%%%%%%%%%%%%%%%%%%%%%%%%%%%%%%%%%%%%%%%%%%%%%%%%%

In the 1960s and 1970s there were several attempts to determine the TPE corrections to electron-proton elastic scattering.  Early measurements
comparing electron and positron elastic-scattering cross sections~\cite{yount62, browman65, anderson66, bartel67, cassiday67, 
anderson68, bouquet68, mar68, hartwig75} were largely limited to low $Q^2$ and/or high $\varepsilon$, where 
calculations~\cite{drell57, drell59, greenhut69} suggest that TPE contributions are small. Given the limited experimental 
sensitivity of these early measurements, none of the experiments observed a significant deviation from $R_{2\gamma}=1$. 
A global analysis~\cite{arrington04b} of these measurements showed only limited evidence for non-zero TPE contributions.
Improved measurements of these contributions, in particular for large $Q^2$ and small $\varepsilon$ values,
are required to reconcile the form factor discrepancy.
%As a result, most electron-scattering results neglected the TPE 
%corrections but applied an uncertainty in the radiative correction procedure of roughly
%1--1.5\% that was dominated by the uncertainty in TPE corrections.  

There have been several recent attempts to make improved TPE measurements by comparing 
$e^\pm p$ scattering. The VEPP-3~\cite{VEPP-3, Rachek2015}  and OLYMPUS~\cite{olympus, Henderson:2016dea} experiments used alternating 
electron and positron beams in storage rings incident on internal gas targets.   In these experiments, data for $e^\pm p$ scattering are 
taken at a fixed beam energy leading to known event kinematics. These experiments measure $R_{2\gamma}$ as a function of
lepton scattering angle, which varies both  $Q^2$ and $\varepsilon$ simultaneously, and do not measure the $\varepsilon$
dependence at fixed $Q^2$.
%
% JRA: we bring this up as a potential 'issue' with these experiments, but if we want to do so we should make sure
% that we sufficiently describe how the address this.  But I don't think it's a critical point.
%
Because the target thickness \cite{OLtgt} and hence the luminosity was not well known, both experiments planned to normalize their data to 
$R_{2\gamma}=1$ at low $Q^2$ and high $\varepsilon$.
The VEPP-3 experiment utilizes a non-magnetic spectrometer while the OLYMPUS experiment utilizes the upgraded BLAST detector that
was previously located at MIT-BATES.

The MUSE Collaboration~\cite{museproposal} will compare $e^\pm p$ and $\mu^\pm p$ 
scattering at very low $Q^2$. This is motivated by the ``proton radius puzzle'', the 
difference between proton radius extractions involving muonic hydrogen~\cite{pohl10,pohl13} and those
involving electron-proton interactions~\cite{mohr08, bernauer10, zhan11}. The MUSE experiment 
will compare electron and muon scattering to look for indications of lepton non-universality, but will 
also examine TPE corrections, which are important in the radius extraction from electron scattering 
data~\cite{rosenfelder00, blunden05, arrington11c, bernauer11,arrington13,arrington-sick,Higinbotham15,Griffioen15}.  

We applied a very different approach to compare $e^+ p$ and $e^-p$ scattering. Rather than alternating 
between mono-energetic $e^+$ and $e^-$ beams, we generated a mixed beam of positrons 
and electrons covering a wide range of energies and used the large-acceptance CLAS spectrometer in 
experimental Hall B at Jefferson Lab to detect both the scattered lepton and the struck proton.
The over-constrained elastic-scattering kinematics allowed us to reject inelastic events and to determine the energy of the 
incident lepton in each event. This allows a simultaneous 
measurement of electron and positron scattering, while also covering a wide range in $\varepsilon$ and $Q^2$.  This paper is a follow up to 
our previously published results~\cite{Adikaram15} and includes corrections for $\delta_{even}$ along with previously unpublished results.

\section{Experimental Details}
\label{Sec:ExptDetails}

\begin{figure}[t]
\begin{center}
\includegraphics[width=0.47\textwidth]{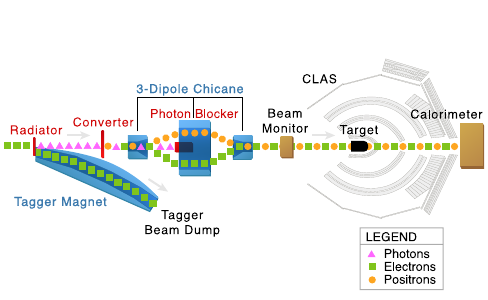}
\caption{(Color online) Beamline sketch for the CLAS TPE
  experiment. The chicane bends the electron and positron trajectories in the horizontal plane, rather than the vertical plane as
  shown in the figure.  The electron and positron directions are selected by the chicane polarity. The TPE Calorimeter was removable and only 
  placed in the beam for special calibration runs. Not shown in the figure is the DFM that is attached to the front of the calorimeter. Drawing is not to scale.}
\label{fig:TPEsketch}
\end{center}
\end{figure}

This experiment was conducted at the Thomas Jefferson National Accelerator Facility (Jefferson Lab). 
A simultaneous mixed beam of electrons and positrons was produced using the 5.6 GeV primary electron 
beam from the accelerator (see Fig.~\ref{fig:TPEsketch}). Bremsstrahlung photons were produced by bombarding a $9\times 10^{-3}$ radiation
length (RL) gold radiator with a 110-140-nA electron beam. The resulting photon beam traversed a 12.7-mm inner-diameter 
nickel collimator, while the electrons were diverted into the tagger beam dump by the Hall B tagger 
magnet \cite{clastagger}. The photon beam then struck a 0.09 RL gold 
converter to produce electron-positron pairs. The mixed lepton-photon beam then passed through a three-dipole magnet chicane.
The chicane bent electrons and positrons in the opposite directions, spatially separating 
them in the horizontal plane (shown as a vertical separation in Fig.~\ref{fig:TPEsketch}). The photon beam was stopped by a 
4-cm-wide and 35-cm-long tungsten block placed at the upstream face of the second dipole. The electron 
and positron beams were then recombined into a single beam by the third dipole. The mixed lepton beam then passed through a
pair of collimators en route to a 6 cm-diameter, 30-cm long liquid hydrogen (LH$_{2}$) target. The scattered leptons and the 
protons were detected in the CEBAF Large Acceptance Spectrometer (CLAS)~\cite{clasNIM}. 

The first and third dipoles of the TPE chicane were
operated with a magnetic field of $B\approx \pm 0.4$ T and were
about 0.5 m long. They were powered in series by a single power supply. The second dipole had
a field of $B\approx \mp 0.38$ T and was about 1 m long. The momentum acceptance of the chicane is fixed 
by the width of the photon blocker and the apertures of the second dipole. The width of the photon blocker ($\pm 2$ cm) 
fixed the maximum lepton momentum and the aperture of approximately $\pm 20$ cm
fixed the minimum lepton momentum. In the ideal case the three dipoles are left-right symmetric and
the two lepton beams should be identical. 
The final useful lepton beam energy ranged from approximately 0.5 to 3.5 GeV.

This experiment ran with a much higher primary electron beam current and much thicker radiator than is normally used in CLAS photoproduction 
experiments and the process of producing a tertiary mixed beam produced a large rate of background radiation in the hall.  To 
protect CLAS from this radiation a number of shielding structures (not shown in Fig.~\ref{fig:TPEsketch}) were installed in the hall.   
Two large shielding structures were constructed between the first and second dipoles of the chicane and between the second and
third dipoles of the chicane.  A 1-m by 1-m by 0.1-m thick lead wall was placed immediately downstream of the chicane.  The lepton beams 
passed through a 1.75-cm diameter tungsten collimator in this wall.  Further downstream just before CLAS was a 4-m by 
4-m by 2.5-cm thick steel wall.   A second lepton beam clean-up collimator made of lead with a 4-cm diameter aperture was located at the entrance to CLAS.
The shielding around the CLAS tagger beam dump was increased during a 2004 test run \cite{Moteabbed} and remained in place for this experiment.
This shielding was designed to remove backgrounds from the beamline and beam dump that would otherwise overwhelm the CLAS detector 
systems. % with the higher-than-normal beam currents used in this experiment.  

\begin{table}[htb]
 \begin{center}
 \begin{tabular}{|c|c|}\hline
Primary Beam & $110\leq I\leq 140$ nA \\
             & $E=5.6$ GeV   \\ \hline
Radiator (gold) & $9\times 10^{-3}$ RL \\
Dist. from target & 21.76 m \\ \hline
Photon Collimator  & 12.7  mm ID \\
Dist. from target & 15.88 m \\ \hline
 %                             &  length = 30 cm \\ \hline
 Converter (gold) & $9\times 10^{-2}$ RL \\ 
 Dist. from target & 15.51 m \\ \hline
1st and 3rd Dipoles & $B\approx 0.4$ T \\
 & $L\approx 0.5$ m \\ \hline
2nd Dipole & $B\approx 0.38$ T \\
 & $L\approx 1$ m \\ \hline
Lepton Collimator 1 (tungsten) & 1.75 cm ID \\ 
Dist. from target & 9.64 m \\ \hline
Beam Monitor & 3.12 m \\ 
Dist. from target & \\ \hline
Lepton Collimator 2 (lead) & 4 cm ID \\ 
Dist. from target & 3.02 m \\ \hline
{LH2 target}    & diameter=6 cm \\
                      &length=30 cm \\ \hline
CLAS Torus Current & $\pm 1500$ A \\ \hline
Mini-Torus Current & 4000 A \\ \hline
\end{tabular}
 \end{center}
 \caption{Running conditions. ID = Inner Diameter, RL=radiation lengths.}
\label{tab:beam}
\end{table}

\begin{figure}[t] 
\begin{center}
\includegraphics[width=0.47\textwidth]{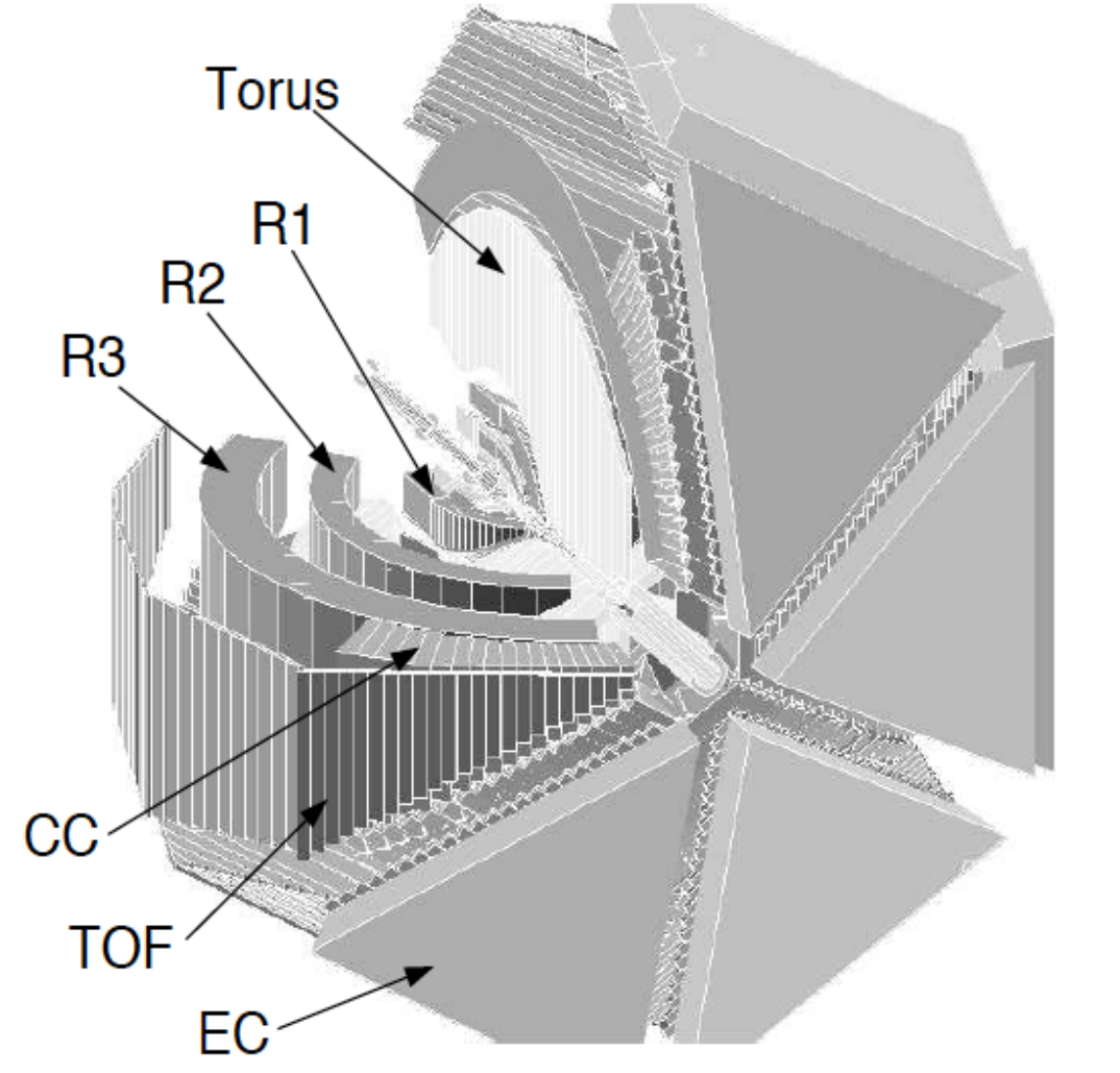}
\caption{\label{fig:clas} 
Three dimensional view of CLAS showing the beamline, drift chambers (R1, R2,
and R3), the Cherenkov Counter (CC), the Time of Flight system (TOF) and the
Electromagnetic Calorimeter (EC).  In this view, the beam enters the picture
from the upper left corner.}
\end{center}
\end{figure}

CLAS (see Fig.~\ref{fig:clas}) is a nearly $4\pi$ acceptance detector divided into six segments known as sectors. 
Six superconducting coils produce a toroidal magnetic field in the azimuthal direction. The magnetic 
field bends the charged particles towards (in-benders) or away (out-benders) from the beamline.  
Each CLAS sector contains three regions (R1, R2, and R3) of drift 
chambers to determine charged particle trajectories~\cite{clasDCNIM}, a Cherenkov counter (CC) for 
electron identification~\cite{clasCCNIM}, time-of-flight (TOF) scintillator counters for timing 
measurements~\cite{clasSCNIM}, and an electromagnetic calorimeter (EC) for energy measurements of 
charged and neutral particles~\cite{clasECNIM}. The CC and EC cover only the forward region of CLAS ($8^\circ<\theta < 45^\circ$). 
The CLAS event trigger required at least some minimum ionizing energy deposited in the EC in any sector and a hit in the 
opposite sector TOF. The CC was not used because it is optimized for in-bending 
particles only and would therefore create a systematic charge bias in lepton detection. Data from the EC was not necessary for particle 
identification and due to limited angular coverage and the possibility that it would bias the electron-positron comparison, the EC was not 
used in the analysis.   A compact mini-torus magnet (not shown) was placed close to the target to shield 
the drift chambers from M{\o}ller electrons. 

\begin{figure}
	\centering
	\includegraphics[width=0.47\textwidth]{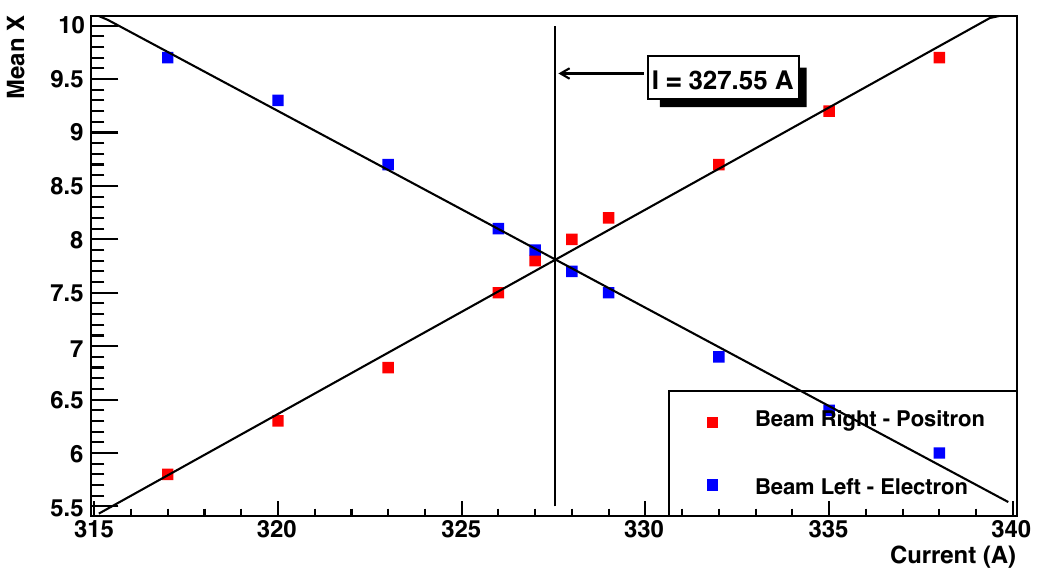}
	\caption {Positron and electron beam positions at the Beam Monitor as a function of the current in the first and third dipoles of the chicane.  
	The positron beam position was measured while the electron beam was blocked and vice versa. The 
	fits to the data points are shown by the diagonal black lines and their crossing is labeled by the vertical line at a current 
	of 327.55 A.}
	\label{fig:idscan}
\end{figure}

A sparse fiber beam monitor (labeled as Beam Monitor in Fig.~\ref{fig:TPEsketch}) was installed just upstream of CLAS to 
measure the position and spatial distribution of the two lepton beams and to 
monitor their stability during the experiment. The sparse fiber beam monitor contains two sets of 16, $1 \times 1$ mm$^2$ 
scintillating fibers forming vertical and horizontal grids with a fiber spacing of 5 mm.  
During commissioning and following each chicane magnetic field reversal, we blocked one of the lepton beams by inserting a remotely-controlled lead 
block at the entrance of the second chicane dipole.  By alternately blocking each one of the two lepton beams, we measured the centroid and shape 
of the other beam in two dimensions.
In order to center both lepton beams at the same position, we determined the position of each individual beam 
as a function of the current in the first and third chicane dipoles.  Figure~\ref{fig:idscan} shows the location of the positron
and electron beams as a function of the dipole current. We set the final current at the crossing of the fits to the individual 
beam positions for both chicane polarities.

We periodically reversed the polarity of the CLAS torus magnets and the beamline chicane magnets to control
systematic uncertainties. Periodic torus field reversal provides control on the systematics due to potential 
detector acceptance related bias for the oppositely charged leptons. Similarly, reversing the chicane current swaps spatial 
positions of the oppositely charged lepton beams.  Data from three such complete polarity cycles and one partial cycle were used 
in the final analysis. This is discussed in more detail in Sec.~\ref{sec:acceptance}.

We determined the energy-dependent lepton fluxes by measuring the energy distributions of the electron and positron beams with 
the ``TPE calorimeter'' installed downstream of CLAS.  
To measure the energy distribution of one lepton beam, we inserted the calorimeter into the beamline, emptied the target, 
blocked the other beam and reduced the beam intensity by a factor of about $10^{-4}$ by reducing the primary beam current to 1 nA 
and reducing the radiator thickness to $10^{-4}$ RL.   

\begin{figure}[b]
	\centering
	\includegraphics[width=0.47\textwidth]{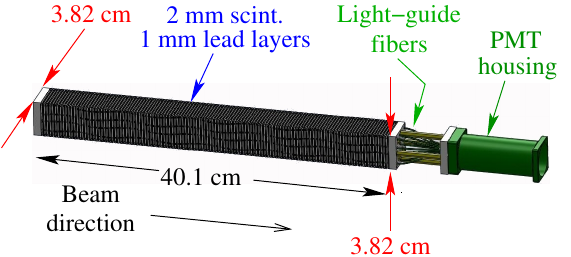}
	\caption {(Color online) Drawing of a single shashlik module.  The downstream TPE Calorimeter consists of 30 of these modules arranged in a stack 5
	modules high and six modules across contained within a light-tight box. }
	\label{fig:shashlik}
\end{figure}

The TPE Calorimeter consisted of 30 shashlik modules \cite {Badier94} arranged in five rows  of six modules each.  The individual shashlik modules (Fig.~\ref{fig:shashlik}) 
are $3.82\times 3.82\times 45$ cm$^3$ and consist of alternating $3.82\times 3.82$ cm$^2$ layers of 1-mm thick lead and 2-mm thick plastic scintillator.  
Each module has 16 wavelength shifting light-guide fibers, each 1.5 mm in diameter and spaced 7.7 mm apart.  The wavelength shifting fibers transmit the light from 
the individual scintillator layers to photomultiplier tubes.  In front of the shashlik modules was a dense fiber monitor (DFM) consisting of a  
closely-packed array of $1\times 1$ cm$^2$ scintillating fibers arranged both horizontally and vertically, with an area that covered
the face of the calorimeter. We used the DFM to make sure that both lepton beams had the same centroid at the upstream Beam Monitor and at 
the DFM and were therefore parallel.

\begin{figure}[htb]
  \begin{center}    
\includegraphics[width=0.47\textwidth]{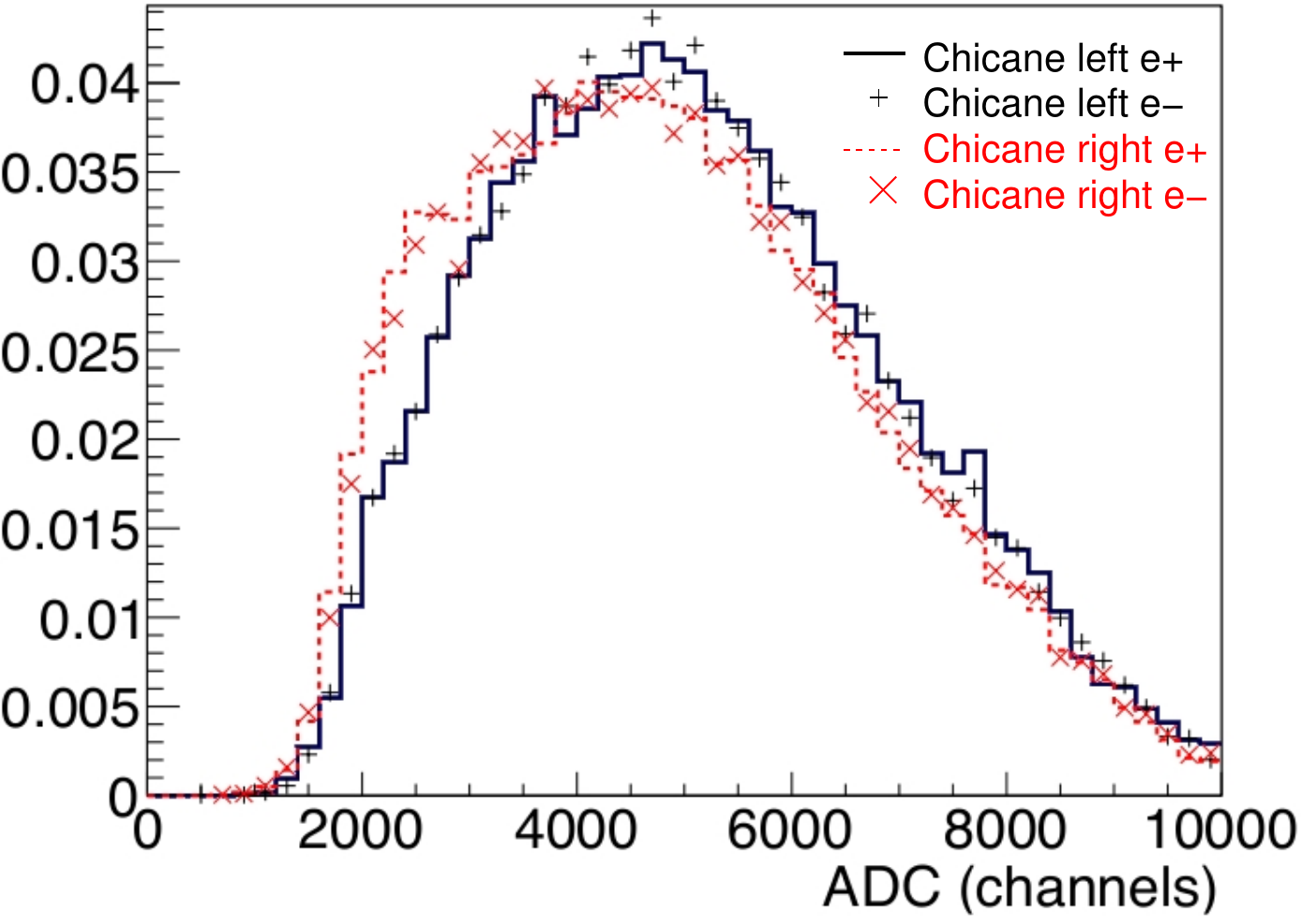}
  \caption{(Color online) The beam energy distribution for electrons and positrons as they pass on the left or the right side of the chicane as indicated by the key. 
  The horizontal axis is in ADC channel number, where channel 1000
  corresponds approximately to 370 MeV.  The distributions are normalized to unity.  Note the energy distributions for electrons and positrons passing on one 
  side of the chicane are very similar to each other but the energy distributions for the two sides of the chicane differ from each other, indicating that the chicane 
  was not symmetric.}
  \label{fig:EdistTpeCal}
  \end{center}    
\end{figure}

We measured the beam-energy distribution for each lepton beam before and after each chicane magnet polarity reversal
(see Fig.~\ref{fig:EdistTpeCal}).  The energy distributions for electrons and positrons 
passing through the left side of the chicane are very similar to each other as are the distributions for when the electrons and positrons pass through the right side of 
the chicane.  However, the distributions for leptons passing through the left side of the chicane differ from the distributions of leptons passing through the right
side of the chicane, indicating that the chicane was not perfectly left/right symmetric.

\begin{figure}[htb]
  \begin{center}    
\includegraphics[width=0.49\textwidth]{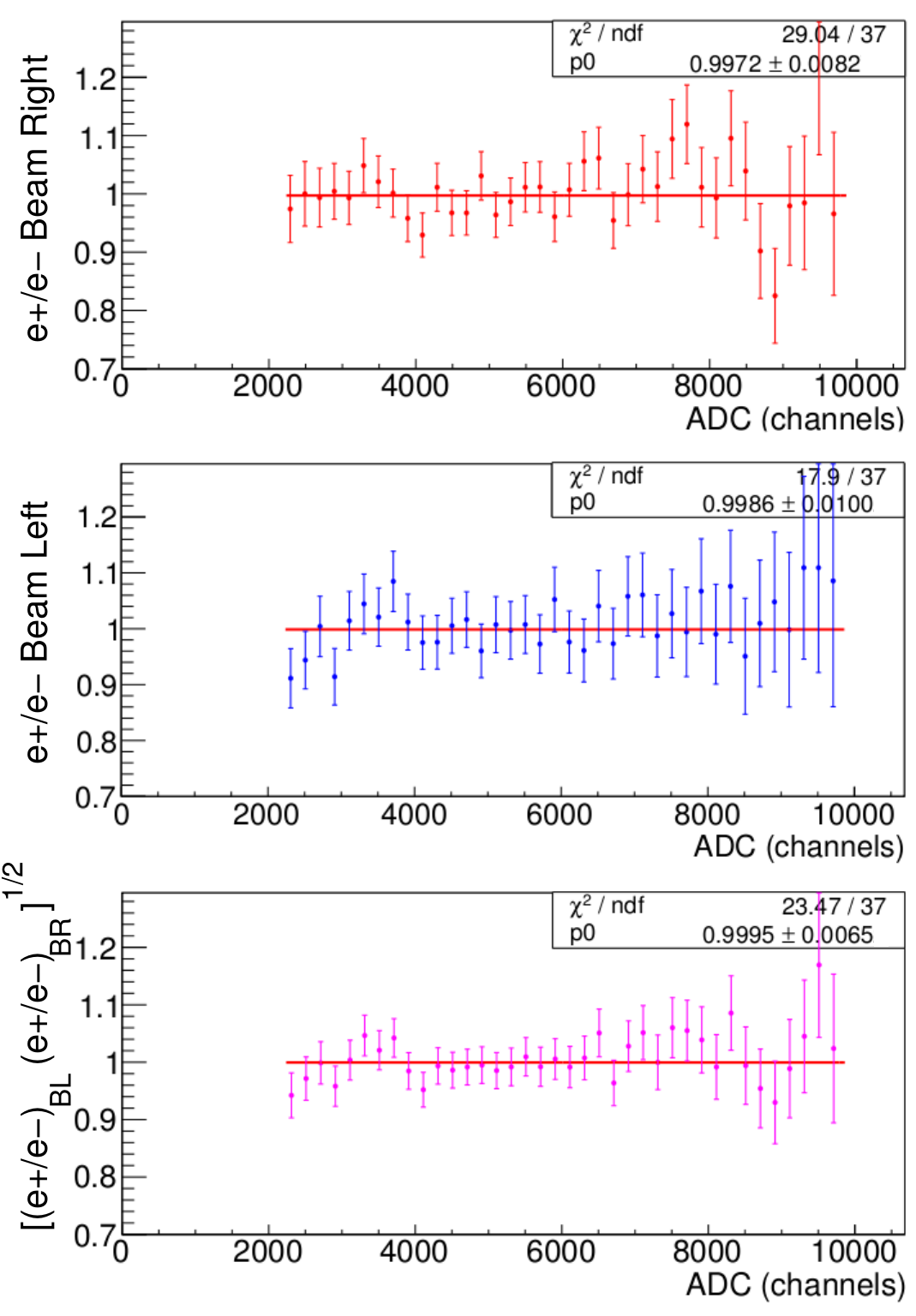}
  \caption{(Color online) The ratio of the incident positron energy distribution to the incident electron energy distribution versus incident energy
  (measured in channels where channel 1000 corresponds approximately to 370 MeV) for leptons passing on the right side of 
  the chicane (top panel) and for leptons passing on the left side of the chicane (middle panel), and the square root of the product of the two ratios
   (bottom panel).  The distributions are normalized to unity.   The statistics boxes show the results of one-parameter (constant) fits 
   to the ratios.}
  \label{fig:EdistRatioB}
  \end{center}    
\end{figure}

In order to know our relative electron and positron luminosities, we rely on several pieces of information:
\begin{itemize}
	\item At GeV energies, electron-positron pair production on the nucleus is the dominant cross section by a factor of $10^3$~\cite{PDG} 
	and is charge-symmetric.
	\item At energies over 500 MeV, electron and positron interactions with matter are identical (i.e., the annihilation cross 
	section is negligible and  the difference between M{\o}ller and Bhabha cross sections is negligible)~\cite{Messel}. 
	\item The magnet current of the beamline chicane where the two lepton beams had the same average location was reproducible to 0.1 A for each magnet cycle.
	\item The ratios of the positron to  electron energy distributions for particles passing on one side of the chicane (either left or right)   
	as measured by the TPE Calorimeter are energy independent. This is shown in Fig.~\ref{fig:EdistRatioB} where we have plotted
	the ratio of the incident positron energy distribution to that of the incident electron versus energy for beams through the left (top) 
	and right (middle) sides of the chicane.  Monte Carlo simulations of the beamline reproduce this behavior.
	\item The product of the ratios of the positron to  electron energy distributions for positive and negative chicane settings  as measured by 
	the TPE Calorimeter is also energy independent as seen in the bottom panel of Fig.~\ref{fig:EdistRatioB}. These electron-positron energy ratios 
	were measured for each chicane flip and were all consistent. Note that the distributions in Fig.~\ref{fig:EdistRatioB} are normalized to unity because the 
	separate measurements of $e^+$ and $e^-$ distributions making up the ratios could not be absolutely normalized since we did not have a measurement 
	of the incident primary electron beam charge precise to 1\% at the low primary beam currents used to measure the energy distributions.
\end{itemize}

\begin{figure}[t] 
\begin{center}
\includegraphics[width=0.50\textwidth]{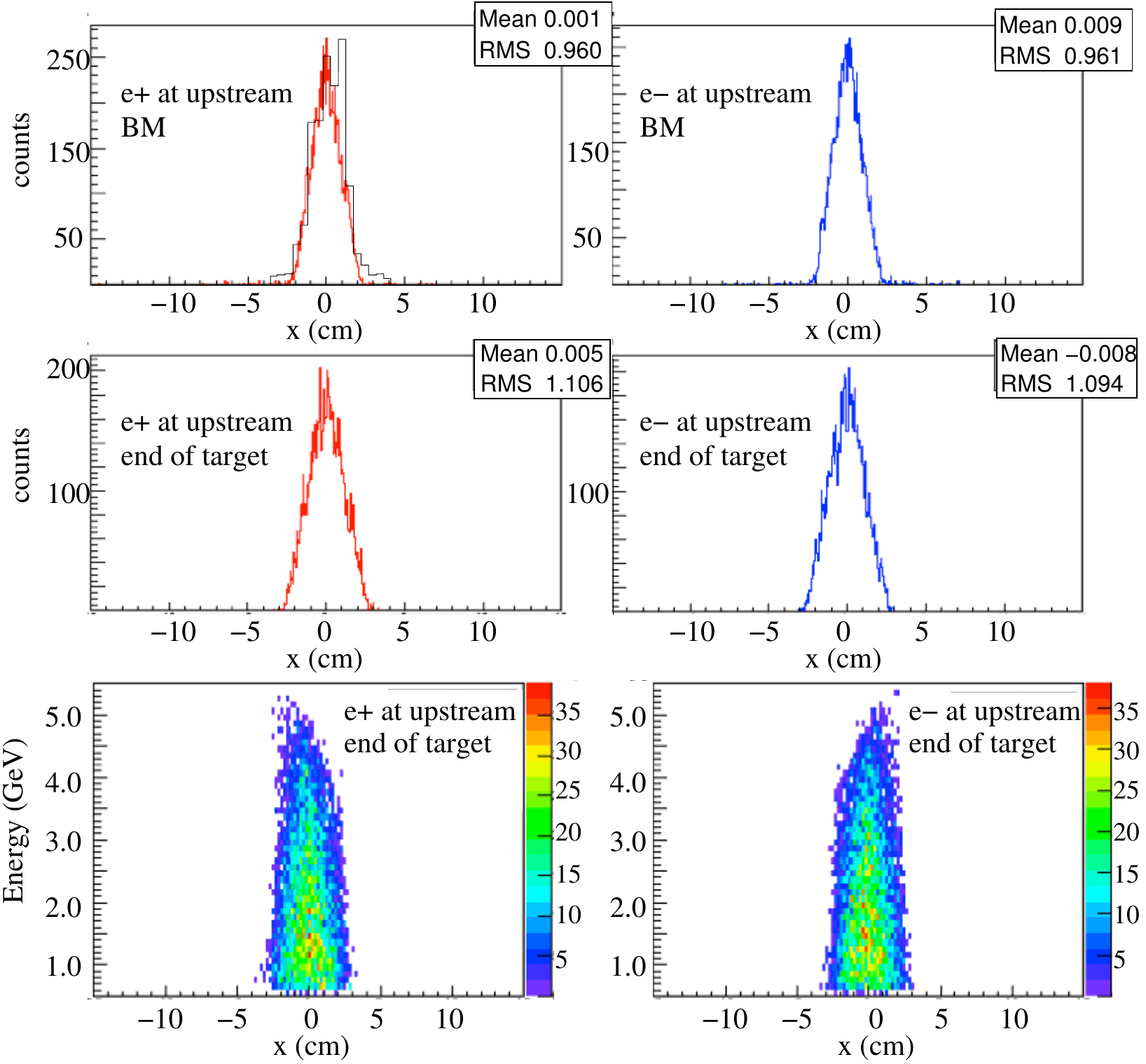}
\caption{\label{fig:BeamSim} 
(Color online) Results of Monte Carlo simulations of the horizontal ($x$) beam distribution at the sparse-fiber beam monitor (top) and the at the upstream face of the
target (middle) and the beam-energy versus $x$ distributions at the upstream face of the target for both positrons (left) and electrons (right) for a single chicane polarity. 
The upper left panel also has a measured spatial distribution for a {\it combined positron/electron} beam taken during the run overlaid on the simulation results.  The 
spike in the histogram to the right of the peak is due to improperly gain matched fiber. }
\end{center}
\end{figure}

Detailed GEANT Monte Carlo simulations of the lepton beam transport that included all of the beamline components and materials were conducted prior to the experiment 
to determine the optimal beamline configuration and to ensure symmetry of the flux and energy of positrons and 
electrons. The simulations included all electron and positron interactions with matter, including the aforementioned 
M{\o}ller and Bhabha scattering. Various combinations of radiator, converter, and collimation were tested in the simulation to achieve the highest possible lepton flux 
while also minimizing background. Fig.~\ref{fig:BeamSim} shows the horizontal ($x$) spatial distributions for electrons 
and positrons at the upstream sparse-fiber beam monitor (BM) and at the target for a single chicane polarity.  The RMS of the simulated distributions for both leptons 
at the beam monitor is 0.96 cm and agreed with online measurements using the beam monitor. An example of a BM measurement for a {\it combined positron/electron}
beam has been overlaid on the simulated positron histogram (upper left panel).  The spike in the histogram to the right of the peak is due to an improperly gain matched 
fiber. The $x$-distribution RMS increases to 1.1 cm at the upstream face of 
the target. Fig.~\ref{fig:BeamSim} also shows that the energy versus $x$ distributions are very similar up to about 4.0 GeV but show an asymmetric tilt above 4.0 GeV.
However, as stated above, the useful energy range of the lepton beam was limited to about 3.5 GeV.  Furthermore, since we measured 
the electron-proton and positron-proton yields for both positive chicane and negative chicane, any asymmetries 
in the chicane cancel (see Eq.~\ref{eq-Rmeas} in Sect.~\ref{sec:acceptance}) and the resulting lepton luminosities are equal. 

\begin{figure}[t] 
\begin{center}
\includegraphics[width=0.48\textwidth]{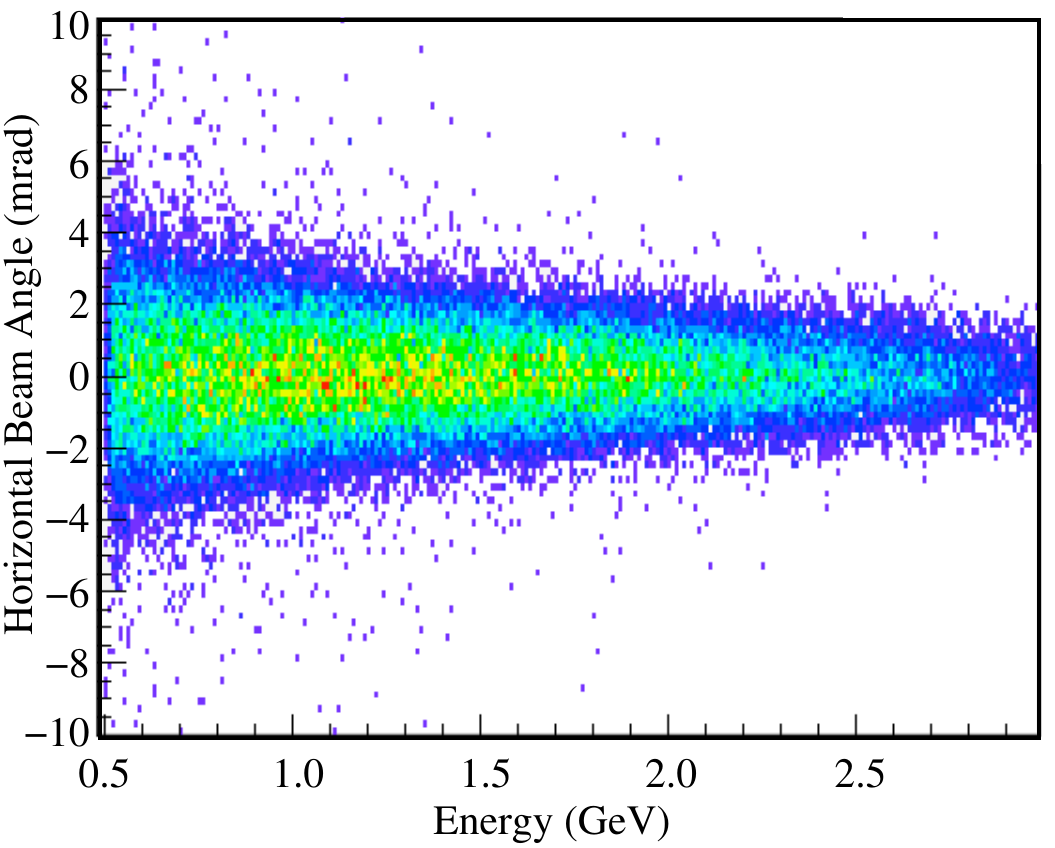}
\caption{\label{fig:BeamAngle} 
(Color online) Results of Monte Carlo simulations of the horizontal angular dispersion of the beam distribution at the upstream face of the
target  as a function of beam energy. }
\end{center}
\end{figure}

 Figure~\ref{fig:BeamAngle} shows the simulated horizontal angular dispersion of the beam at the upstream face of the target as a function of beam energy for a 
 single chicane setting.  The mean angle is less than 1$\mu$rad  while the width of the distributions varied from $\sigma=1.7$ mrad at $E=0.8$ GeV down to 
 $\sigma\approx 0.7$ mrad for $E>2.8$ GeV.

\section{Data Analysis}
\label{sec-analysis}
The identification of elastic $e^\pm p$ events with no charge bias required us to make a series of cuts and corrections and to 
test the charge independence of our analysis procedures.  This section will discuss the steps taken in the analysis process. These include
applying momentum and energy loss corrections, applying data
selection cuts, determining dead detector corrections, subtracting backgrounds, and applying radiative corrections.

\subsection{Energy loss and momentum corrections}
\label{sec:EMCorr}

%\begin{figure}[tbh]
%  \centering
%  \includegraphics[width=0.47\textwidth]{./images/MomCorr.pdf}
%  \caption {(Color online) Typical invariant mass distribution, $W$, for elastic electron-proton scattering before (black histogram) 
%  	and after (red histogram) momentum corrections for a negative-polarity torus run. 
%  	The heavy curve is a Gaussian fit of the momentum-corrected distribution and the fit parameter p1 (in the inset box) corresponds to 
%	the mean of 0.9386 MeV, which is good agreement with the proton mass, indicated by the vertical line.} 
%  \label{fig:MomCorr}
%\end{figure}

%Eloss
As a charged particle traverses CLAS, it loses energy through interactions with the target and detector materials. The CLAS reconstruction 
software returns an effective momentum without accounting for this energy loss. For the low momentum protons, this loss could have 
a significant impact on event reconstruction kinematics. The standard CLAS ELOSS package \cite{ELOSS} corrects for this lost energy using the 
Bethe-Bloch equation to relate the material characteristics and path length to the energy loss. Energy-loss corrections ranged from $\approx 4-5$ MeV for 
protons with momenta above 0.5 GeV up to $\approx 25$ MeV for momenta down to 0.2 GeV.  No energy loss corrections were done for leptons.

%Momentum
Because of incomplete knowledge of the magnetic field and drift chamber positions in CLAS, the reconstructed momenta show some 
systematic deviations. 
%This is evidenced by shifted and broadened missing and invariant mass distributions. For example 
%(see Fig.~\ref{fig:MomCorr} black histogram) the centroid of the invariant mass, $W=\sqrt{M_{p}^2+2M_p\nu-Q^2}$, distribution of the 
%$p(e,e')$ elastic peak is shifted from its expected value $W = M_{p} = 0.9383$ GeV.
To determine the momentum corrections, a set of runs was taken with a 2.258-GeV primary electron beam incident directly 
on the CLAS target.  Data were taken with both torus polarities. We then used exclusive events where all the final-state particles were detected 
and employed four-momentum conservation to determine the correct scattering angles and magnitudes of the momenta.  The events used were
$p(e,e'p)$  and $p(e,e'p\pi^+\pi^-)$ events. This combination of particles provided the same scattering-angle and 
momenta ranges as seen in the final data as well as providing events with both positive and negative charge.  The momentum
corrections were less than 1\% of the momentum and ultimately lead to an invariant mass distribution for electron-proton elastic scattering that is consistent 
with the proton mass to within less than 1 MeV. Imprecision in the momentum corrections was unimportant because we used the measured lepton and proton 
momenta to select elastic scattering events (see below) but not to calculate any of the kinematic quantities of the elastic events.
%The same distribution for positron-proton elastic scattering showed no significant difference from that of
%electron-proton elastic scattering.

\subsection{Data selection cuts}
\label{sec:Data}

We applied a series of cuts to the data to select elastic $e^\pm p$ events. In addition to the kinematic cuts described below, a 
28-cm target vertex cut was applied to both lepton and proton candidates to remove events from the target walls.  We explored using cuts on the transverse target
vertex and the distance of closest approach between the lepton and proton but saw no effect on the final data set. A set of momentum-dependent 
fiducial cuts on the angles (both $\theta$ and $\phi$) were applied to select the region of CLAS with uniform acceptance.  The $\phi$ cuts remove 
the sector edges were the detection efficiency varies rapidly. The $\theta$ cuts are necessary because the $\theta$ acceptance of CLAS is different 
for the two lepton charges and were selected such that the angular acceptance of both positrons and electrons were identical for both torus polarities.  
The $\theta$ cut was chosen to be the minimum angle for the out-bending particle and varied from about 15$^\circ$ for leptons of 
1.5 GeV to about 20$^\circ$ for leptons of 0.8 GeV (the minimum energy used in this analysis). 

This analysis did not use the usual EC- and CC-based CLAS lepton identification scheme.  These detector components cover only a limited 
range of scattering angles. We instead employed elastic scattering kinematics, which are over-constrained by the simultaneous detection of 
both the lepton and the proton.

\begin{figure*}[htb]
  \centering
  \includegraphics*[width=\textwidth]{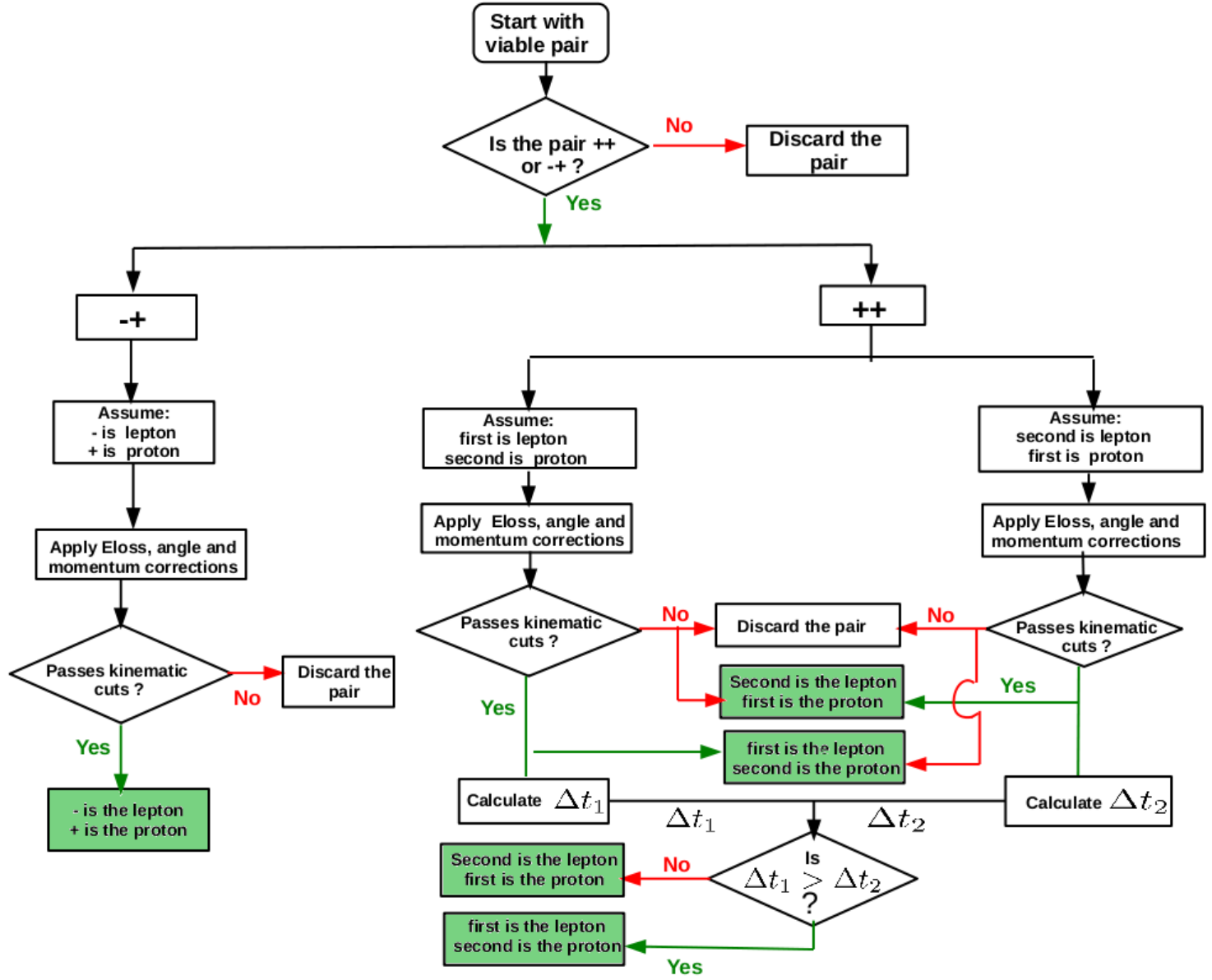}
  \caption {(Color online) Flow chart showing the decision process in selecting elastic events.  The green filled boxes correspond to identified elastic events.}
    \label{fig:flowchart}
\end{figure*}

The elastic event identification algorithm is shown in Fig.~\ref{fig:flowchart} and started with the selection of the events with at least two good 
tracks in opposite sectors of CLAS. Ideally, events with only two tracks would be selected. However, events triggered by accidental hits in
conjunction with a valid elastic event could have more than two tracks.  In that case, pairs of viable tracks were formed by looping over all 
possible good track pairs in the event that had either a negative/positive or positive/positive charge combination. For a pair with a 
negative/positive charge combination: the negative track was considered as a $e^{-}$ candidate and the positive track 
as a $p$ candidate. If the pair passed all elastic kinematic cuts discussed in the next section, the 
pair was identified as the elastic $e^-p$ pair. If not, the next track pair of the event was considered.  For positive/positive pairs, we 
first considered one of the tracks to be the $e^+$ candidate and the other to be $p$ candidate. We then checked 
to see whether the pair passed elastic kinematic cuts as $e^+p$ or as $pe^+$. If the pair passed 
kinematic cuts both as $e^{+}p$ and as $pe^{+}$, an additional minimum-timing cross check was done.  This cross check used the difference between
the TOF of the particle pairs ($\Delta t_{meas}=$ proton TOF $-$ lepton TOF) and compared it to TOF difference ($\Delta t_{calc}$) calculated 
assuming the pair was $e^+p$ (pair 1) or $pe^+$ (pair 2). Whichever pair
assumption that led to the smallest difference $\Delta t_n=\Delta t_{meas} - \Delta t_{calc}$ ($n=1$ or 2) was assigned to the event. Overall,
a negligible fraction of events ($10^{-5}$) had more than one pair passing all cuts.  We note that no TOF cuts were applied and that all cuts for $e^-p$ and 
$e^+p$ events were identical in order to avoid introduction of a charge bias. 

\subsubsection{Elastic Kinematic Cuts}
\label{sec:KinCuts}

\begin{figure}[htb]
		\begin{center}
		\includegraphics[width= 0.48\textwidth]{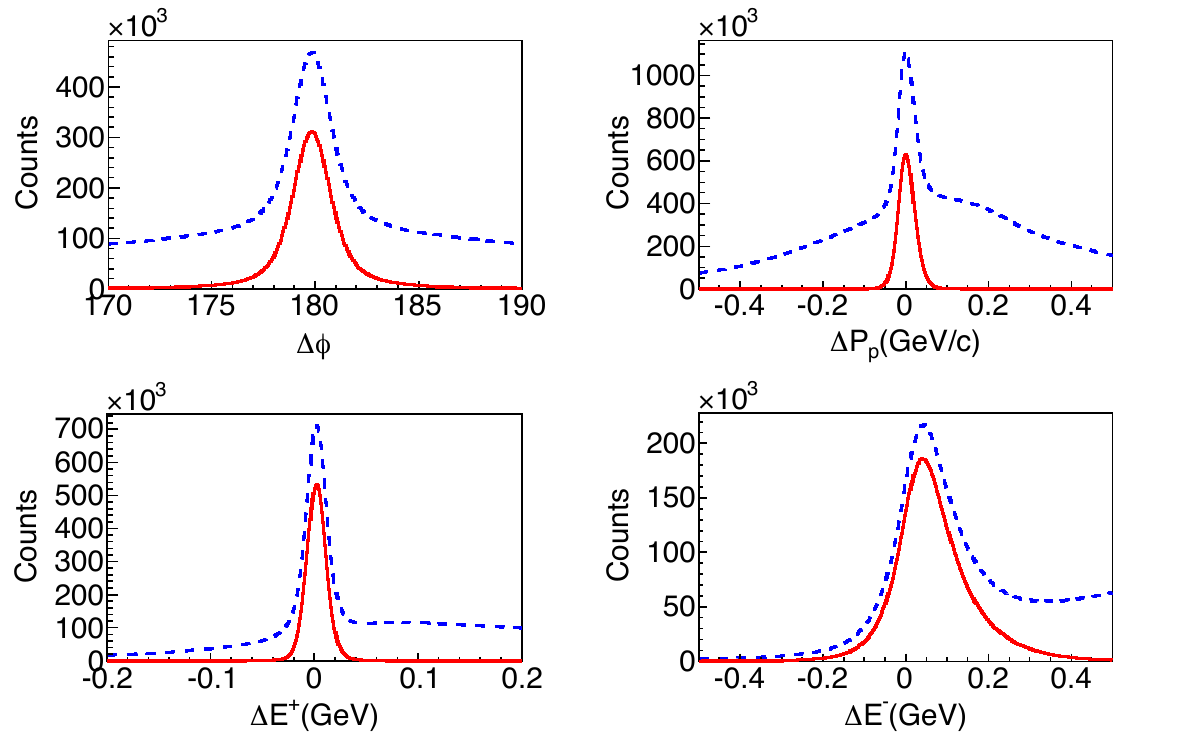}
		\caption{(Color online) The four kinematic variables,  $\Delta\phi$, $\Delta p_p$ and $\Delta E^\pm$ before (blue) 
		and after (red) applying the other three kinematic cuts.  Distributions are summed over the entire kinematic range of the data for 
		both $e^+$ and $e^-$ events and both torus polarities. No significant differences in the distributions were observed between $e^+$ and $e^-$ 
		events or between different torus polarities.}
		\label{fig:kin}
		\end{center}
\end{figure}

Because elastic scattering kinematics are overdetermined by measuring momenta and angles for both leptons and protons, we can identify elastic events
and determine the incident lepton energy by a series of four kinematic cuts.
\begin{enumerate}
	\item Co-planarity cut: The elastically scattered lepton and proton are co-planar. As a result, the azimuthal angle difference between the 
	lepton and the proton ($\Delta \phi = \phi_l -\phi_p$) was sharply peaked at 180$^\circ$  (Fig.~\ref{fig:kin}, upper left).

	\item Lepton Energy Cuts: The unknown energy of the incident lepton can be 
	reconstructed using the scattering angles of the lepton ($\theta_l$) and the proton ($\theta_{p}$) as,
		\begin{equation}
		\label{eq:E1}
  			E^{\text{angles}}_l = M_p\left( \cot \left(\frac{\theta_l}{2}\right)\cot \theta_p - 1\right).
		\end{equation}
		The incident lepton energy can also be calculated using the
		momenta of the lepton ($p_l$) and the proton ($p_p$) and their scattering angles as,
		\begin{equation}
		\label{eq:E2}
  			E^{\text{mom}}_l = p_l \cos \theta_l + p_p\cos\theta_p.
		\end{equation}
		
		$E^{\text{angles}}_l$ has better precision and accuracy than $E^{\text{mom}}_l$ because the scattering angles 
		are better determined by CLAS than the momentum. Kinematic variables 
		such as $Q^{2}$ and $\varepsilon$, which require knowledge of the beam and scattered lepton energies were calculated using 
		$E^{\text{angles}}_l$ and $E'_{\text{calc}}$.
		Figure~\ref{fig:IncidentBeamSum}  shows the beam energy for $e^{+}$ and $e^{-}$ reconstructed using Eq.~\ref{eq:E1}.  A beam-energy 
		cut of $E^{\text{angles}}_l>0.85$ GeV was applied to avoid the lower energies where the energy distribution is changing rapidly.
		
		\begin{figure}[htb]
			\begin{center}
			\includegraphics[width=0.92\linewidth]{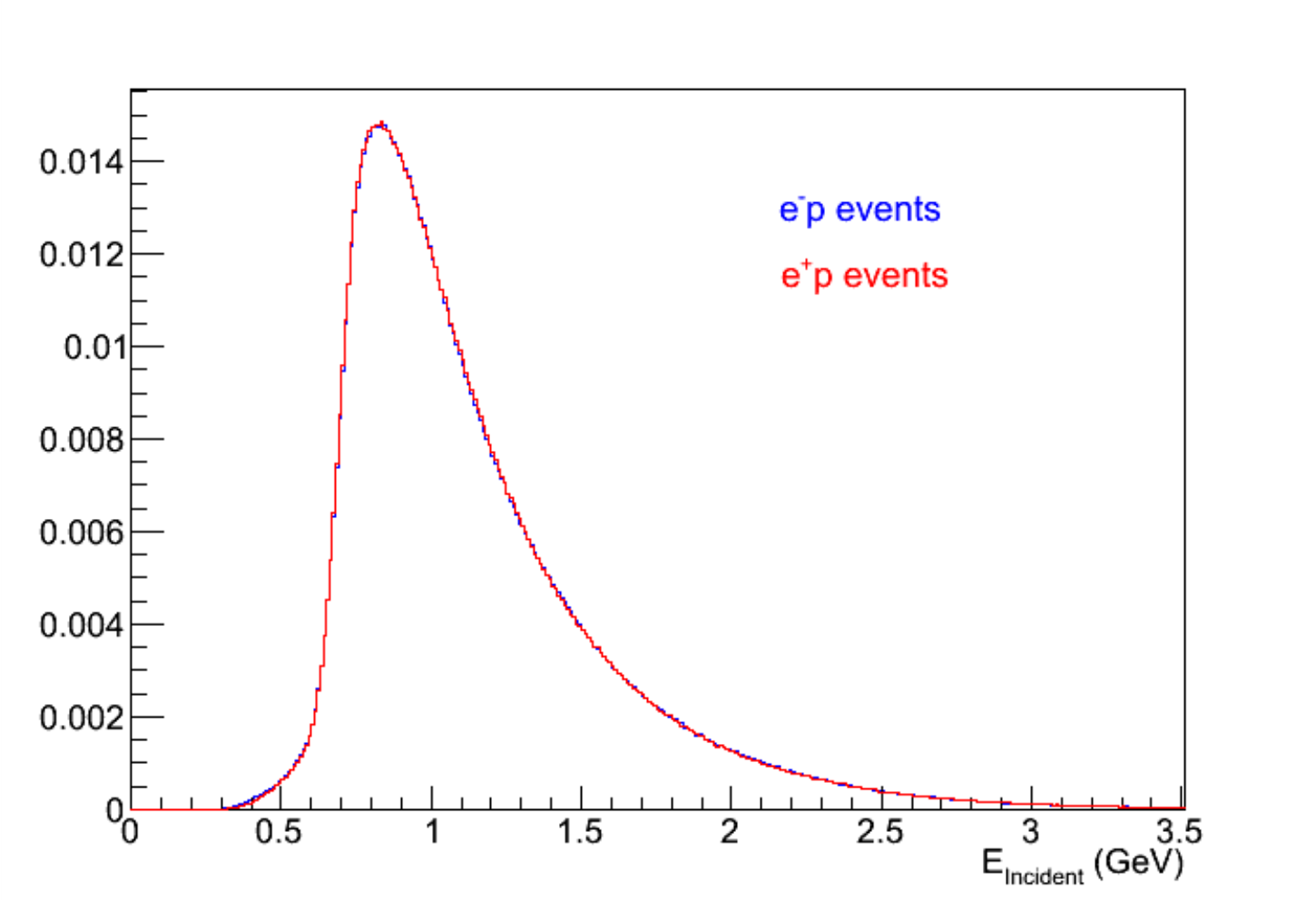}
			\caption{(Color online) Reconstructed incident beam energy distributions of all elastic scattering events using scattering angles.
			The positron (red) and electron (blue) distributions have been scaled by the total number of counts in the distributions and show
			almost imperceptible differences.  This figure differs from Fig.~\ref{fig:EdistTpeCal} in that it shows the incident energy distribution for
			elastic scattering events rather than the overall beam energy distribution.}
			\label{fig:IncidentBeamSum}
			\end{center}
		\end{figure}
		
		For perfect momentum and angle reconstruction, Eqs.~\ref{eq:E1} and~\ref{eq:E2} yield the same result,
		\begin{equation}
			\Delta E_l = E^{\text{angles}}_l - E^{\text{mom}}_l =0.
		\end{equation}
		The energy of the elastically scattered lepton can be calculated using the incident energy and the scattering angle as,
		\begin{equation}
		\label{eq:Eprime}
  			E'_{\text{calc}} = \frac{E^{\text{angles}}_l M_p}{M_p + E^{\text{angles}}_l (1 -\cos\theta_l)}.
		\end{equation}
		For perfect reconstruction, the difference between the CLAS-measured scattered lepton energy  ($E'_{\text{meas}}$)
		and the energy calculated by Eq.~\ref{eq:Eprime} should be zero:
		\begin{equation}
		\label{eq:deprime} 
			\Delta E'= E'_{\text{meas}} - E'_{\text{calc}}=0.
		\end{equation}
		Fig.~\ref{fig:EbeamEeCorr} shows that $\Delta E_l$ and $\Delta E'$ are linearly correlated. Rather than applying cuts to these variables,
		the optimal, uncorrelated cuts are on their sums ($\Delta E^+=\Delta E_l + \Delta E'$)  and  their
		differences ($\Delta E^-=\Delta E_l - \Delta E'$).  Distributions for $\Delta E^+$ and $\Delta E^-$ are shown in the bottom
		panels of Fig.~\ref{fig:kin}.
		\begin{figure}[htb]
		\begin{center}
			\includegraphics[width= 0.42\textwidth]{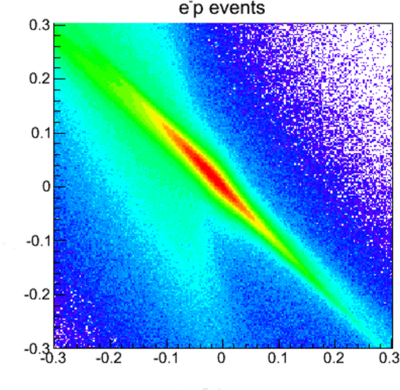}
			\put(-100, -2) {$\Delta E'$ (GeV)}
			\put(-210, 70) {\rotatebox{90}{$\Delta E_{l}$ (GeV)}}
			\caption{(Color online) $\Delta E_{\text{l}}$ and $\Delta E'$ distributions for {\it candidate} $e^{-}p$ events prior to application of kinematic cuts
			showing the linear correlation between $\Delta E_{\text{l}}$ vs. $\Delta E'$.  An identical correlation is seen for $e^{+}p$ events.}
		\label{fig:EbeamEeCorr}
		\end{center}
		\end{figure}		
				 
	\item Proton Momentum Difference Cut: The momentum of the recoil proton was calculated using the lepton
	 and proton scattering angles along with the angle-determined recoil lepton energy as
	 \begin{eqnarray}
	 	p_{p}^{\text{calc}} = \frac{E'_{calc}\sin\theta_{l}}{\sin\theta_{p}}. 
	\label{eq:pprot}
	 \end{eqnarray}
	A cut was placed on the difference between the measured and calculated proton momenta 
	($\Delta p_{p} =  p_{p}^{\text{meas}} - p_{p}^{\text{calc}}$). The difference $\Delta p_p$ is shown in the upper right panel of Fig.~\ref{fig:kin}.
\end{enumerate}

In each case, the widths of the distributions vary with  $Q^2$ and $\varepsilon$.  Based on the means and widths of Gaussian fits to the 
peaks of the distributions, $Q^2$- and $\varepsilon$-dependent, parameterized cuts were set to $\pm 3\sigma$. Fig.~\ref{fig:kin} 
shows distributions of the four cut variables before and after applying cuts on other three variables.  The effect of the other three cuts on any
one variable leads to distributions that are remarkably free of background for all but kinematic regions corresponding to large electron angles
(see Sec.~\ref{sec:Background}). The non-Gaussian shape of the $\Delta E^-$ distribution in Fig 11 is due to summing over the entire kinematic range, 
where the width and background distributions are changing.    The positive offset in $\Delta E^-$ is due to the fact that $\Delta E_l$ (Eq.~\ref{eq:Eprime}) is 
offset in the negative direction because of imperfections in the momentum corrections leading to $E'_{meas}$ being less than $E'_{calc}$.  For each 
kinematic bin (see, e.g., Fig.~\ref{fig:BGsamp}) the signal peak is Gaussian, but the background is not.

\subsection{Kinematic coverage and binning}
\begin{figure}[htbp]
  \begin{center}    
  \includegraphics[width=0.49\textwidth]{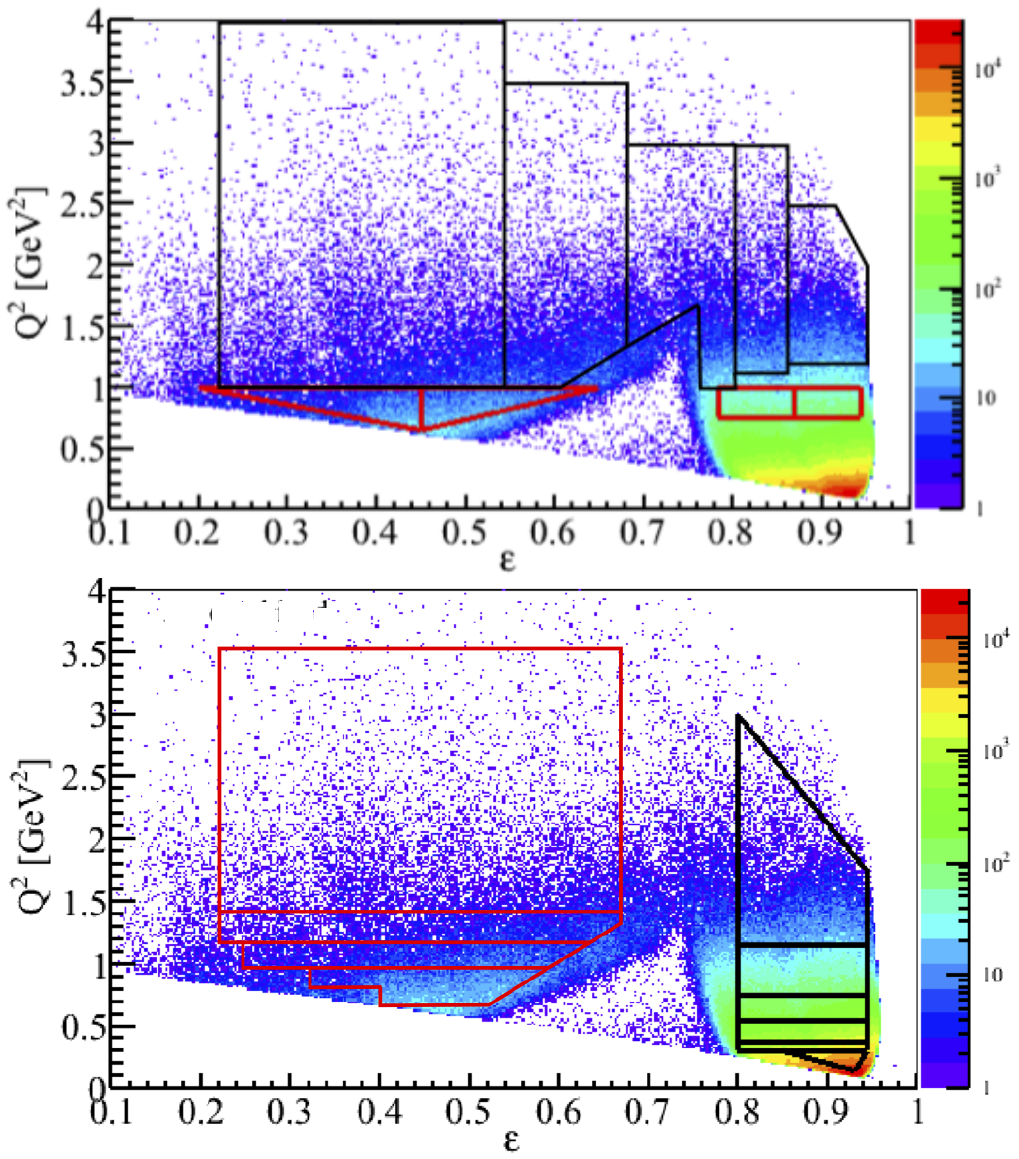}
  \caption{(Color online) Data binning in $Q^2$ and $\varepsilon$ overlaid on positive torus $e^{+}p$ events.
  	The upper plot shows the two sets of bins for the $\varepsilon$ dependence (red and black boxes for $\langle Q^2\rangle=0.85$ and
	1.45 GeV$^2$, respectively), while the lower plot shows the two binning choices for the $Q^2$ dependence (red and black boxes for 
	$\langle\varepsilon\rangle=0.45$ and 0.85, respectively.)}
  \label{fig:bins}
  \end{center}    
\end{figure}

Figure~\ref{fig:bins} shows the $Q^2$ and $\varepsilon$ distribution of $e^+p$ elastic scattering events for positive torus polarity. 
The wide coverage of $Q^2$ and $\varepsilon$ is apparent.  
There is a hole in the distribution at $\varepsilon\approx 0.7$ and lower values of $Q^2$.  This hole is due to the trigger used in the 
experiment, which required one particle track hitting the forward TOF and the EC.  Events where neither particle had a lab-frame scattering 
angle of less than about 45$^\circ$ did not trigger the CLAS readout.  The trigger hole is largest for $e^+p$, positive torus events, which 
ultimately limits our kinematic coverage.

The data bins (Fig.~\ref{fig:bins}) were selected to measure the $Q^2$ dependence of $R_{2\gamma}$ at two values of $\varepsilon$ and the 
$\varepsilon$ dependence of $R_{2\gamma}$ at two values of $Q^2$ with roughly equal statistical uncertainties in each range. 
We avoided the edges of the 
distributions, where the acceptance for in-bending and out-bending particles vary rapidly. The binning choice leads to some overlap in the 
data bins. The average values, $\langle Q^2\rangle$ and $\langle\varepsilon\rangle$, are given in Tables~\ref{tab:sys} and \ref{tab:FinalResults}.

\subsection{Dead detector removal and acceptance matching}
\label{sec:acceptance}

In addition to the fiducial cuts mentioned above, we also removed dead, 
broken, and/or inefficient detector elements of CLAS as these components could lead to charge-dependent biases in the 
lepton detection efficiency.  Events that hit inefficient TOF paddles were removed.  The forward region of one of the six sectors of 
CLAS (sector 3) had a large number of holes due to dead drift chamber and EC channels.  All data with either particle entering this
region of sector 3 were removed from the analysis as such events would have insufficient information for event reconstruction.

As mentioned above, the polarities of the CLAS torus magnets and the beamline chicane magnets were periodically reversed during the course of the
experiment. 
For a given torus polarity, $t=\pm$, and chicane polarity, $c=\pm$, 
we measured the ratio of detected elastically-scattered positrons, $N^+_{tc}$, and electrons, $N^-_{tc}$:
\begin{equation}
\label{eq-R1a}
	R_{tc}=\frac{N^+_{tc}}{N^-_{tc}}.
\end{equation}
Any proton acceptance and detector efficiency factors were the same for both lepton charges and cancel in 
this ratio.
The yield $N^\pm_{tc}$ is proportional to the elastic-scattering cross section, $\sigma^\pm$ (here $\pm$ refers to the lepton charge),
the lepton-charge-related detector efficiency and acceptance function, $f^\pm_t$, as well as chicane-related luminosity factors, 
$L^\pm_c$,  so that 
\begin{equation}
\label{eq-R1b}
	R_{tc}=\frac{\sigma^+ f^+_t L^+_c}{\sigma^-f^-_t L^-_c}.
\end{equation}
Taking the square root of the product of measurements done with both torus polarities but a fixed chicane polarity gives
\begin{eqnarray}
	R_c&=& \sqrt{R_{+c}R_{-c}}= \sqrt{\frac{N^+_{+c}}{N^-_{+c}} \frac{N^+_{-c}}{N^-_{-c}}}  \nonumber \\
	&=& \sqrt{\frac{\sigma^+ f^+_+ L^+_c}{\sigma^- f^-_+ L^-_c} \frac{\sigma^+ f^+_- L^+_c}{\sigma^- f^-_- L^-_c}} \nonumber \\
	&=&\frac{\sigma^+}{\sigma^-}  \frac{L^+_c}{L^-_c},  \label{eq-R2}
\end{eqnarray}
where we assume that  $f^+_+=f^-_-$ and $f^+_-=f^-_+$.  That is, the unknown detector efficiency and acceptance functions for positrons
cancel those for electrons when the torus polarity is switched and are expected to cancel out in this double ratio.  The validity of this cancellation is 
discussed in more detail below. %in Sec.~\ref{sec:acceptance}.

Reversing the chicane current swaps the spatial positions of the oppositely charged lepton beams so that $L^+_+=L^-_-$ and $L^+_-=L^-_+$.
Then taking the square root of the product of the double ratios defined in Eq.~\ref{eq-R2} leads to 
\begin{eqnarray}
	R&=& \sqrt{R_{++}R_{-+}R_{+-}R_{--}}
		=\sqrt{\frac{N^+_{++}}{N^-_{++}} \frac{N^+_{-+}}{N^-_{-+}} \frac{N^+_{+-}}{N^-_{+-}} \frac{N^+_{--}}{N^-_{--}}} \nonumber \\
	&=&	\sqrt{\frac{\sigma^+ L^+_+}{\sigma^- L^-_+} \frac{\sigma^+ L^+_-}{\sigma^- L^-_-}} 
		=\frac{\sigma^+}{\sigma^-}. \label{eq-Rmeas}
\end{eqnarray}
By taking data with both chicane polarities, any flux-dependent 
differences between the two lepton beams is eliminated within the uncertainty. 
Each complete cycle of chicane and torus polarity reversal contained all four configurations
($tc=++$, $+-$, $-+$, $--$).

This experiment relies on the fact that the electron and positron acceptance factors ($f_\pm^\pm$) cancel out in Eq.~\ref{eq-R2}.   
However, inefficient detectors can bias the lepton detection efficiencies.
This effect was taken into account by implementing a ``swimming'' algorithm to ensure the same detection efficiencies in each TOF paddle. 
For each event, this algorithm traced the particle trajectories through the CLAS geometry and 
the magnetic field (including the mini-torus field) and predicted the hit positions on the detectors. The algorithm was then
rerun with the conjugate lepton charge, keeping the momentum and scattering angle unchanged.  The
event was accepted only if both the actual lepton and its conjugate are within
the fiducial acceptance region and hit a good TOF paddle. Otherwise, the 
event was rejected.  The typical change to the final results from applying the swimming algorithm was about $\pm0.2$\%.

%The results of swimming is shown pictorially in Fig.~\ref{fig:swim}.
%
%\begin{figure}[htpb]
%	\centering
%	\includegraphics[width=10cm]{./figuresOriginalFormat/swimming.png}
%	\caption{Swimming a detected $e^-p$  event and its conjugate
%         $e^+$ through CLAS.}
%	\label{fig:swim}
%\end{figure}

The angles and the momenta of the lepton and proton in each event are not independent of each other.  These correlations can potentially 
interfere with the acceptance canceling as described in Eqs.~\ref{eq-R1b} and \ref{eq-R2}. In addition, the minitorus magnetic field, used to 
deflect Moller electrons, was never reversed.  We simulated events using a Monte Carlo program in order to determine the 
magnitude of these effects on our quadruple ratios.

The energy distributions of the incident lepton beams were taken from a detailed GEANT-4 simulation of the beamline, including the 
radiator, convertor, tagger and chicane magnets, collimators, and shielding.  Lepton-proton elastic scattering events were thrown uniformly in 
phase space and then weighted by the cross section.  This allowed us to get a realistic distribution of events with high statistics for all bins in a 
reasonable time period.  Once generated, the Monte Carlo data were analyzed with the same analysis routine as the experimental data.

For each bin, we calculated the acceptances for positive and negative torus fields and for electron-proton and positron-proton events separately 
as the ratio of weighted reconstructed events (selected with the same analysis procedure as the data) to weighted generated events: 
%\[
\begin{equation}
\label{eq-acc1}
	f^\pm_\pm = N'_{rec}/N'_{gen}=\frac{\sum_{i=1}^{N_{rec}} w_i^{rec}}{\sum_{i=1}^{N_{gen}} w_i^{gen}},
\end{equation}
%\]
where the subscript on $f^\pm_\pm$ refers to the torus polarity and the superscript refers to the lepton charge.
We calculated the uncertainty for each acceptance using weighted binomial uncertainties and then combined the acceptances to get the acceptance correction factor 
as
\begin{equation}
	Acc = \sqrt{\frac{f_-^+}{f_-^-}\frac{f_+^+}{f_+^-}}.
\end{equation}
We then divided the quadruple ratios (Eq.~\ref{eq-Rmeas}) with this acceptance correction factor.

\begin{figure}[htpb]
	\centering
	\includegraphics[width=0.45\textwidth]{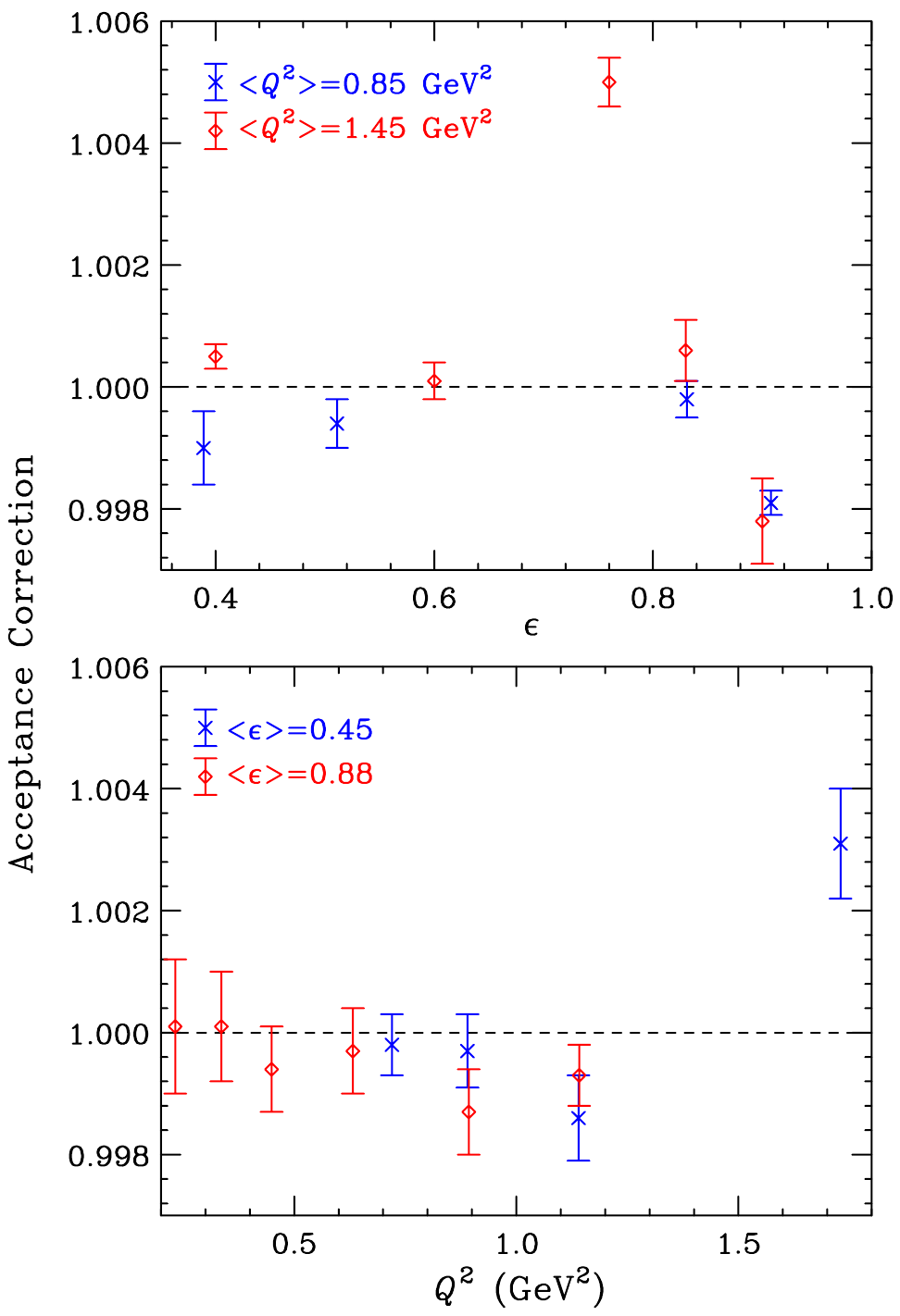}
	\caption{(Color online) Quadruple ratio of acceptance correction factors for the two $Q^2$ ranges as indicated in the upper plot and the 
		two $\varepsilon$ ranges as indicated in the lower plot.   Measured $e^+p/e^-p$
		cross section ratios are divided by these correction factors.}
	\label{fig:Acc}
\end{figure}

The acceptance correction factors for the final kinematic points are shown in Fig.~\ref{fig:Acc}.  The acceptance 
correction factors are all within 0.5\% of unity and almost all are compatible with unity.  The statistical uncertainties are all less than or 
equal to 0.1\%.  Therefore, the effects of the minitorus and of lepton-proton kinematic correlations are very small.

\subsection{Background Subtraction}
\label{sec:Background}

\begin{figure}[tb]
 	\centering
 	\includegraphics[width= 0.45\textwidth]{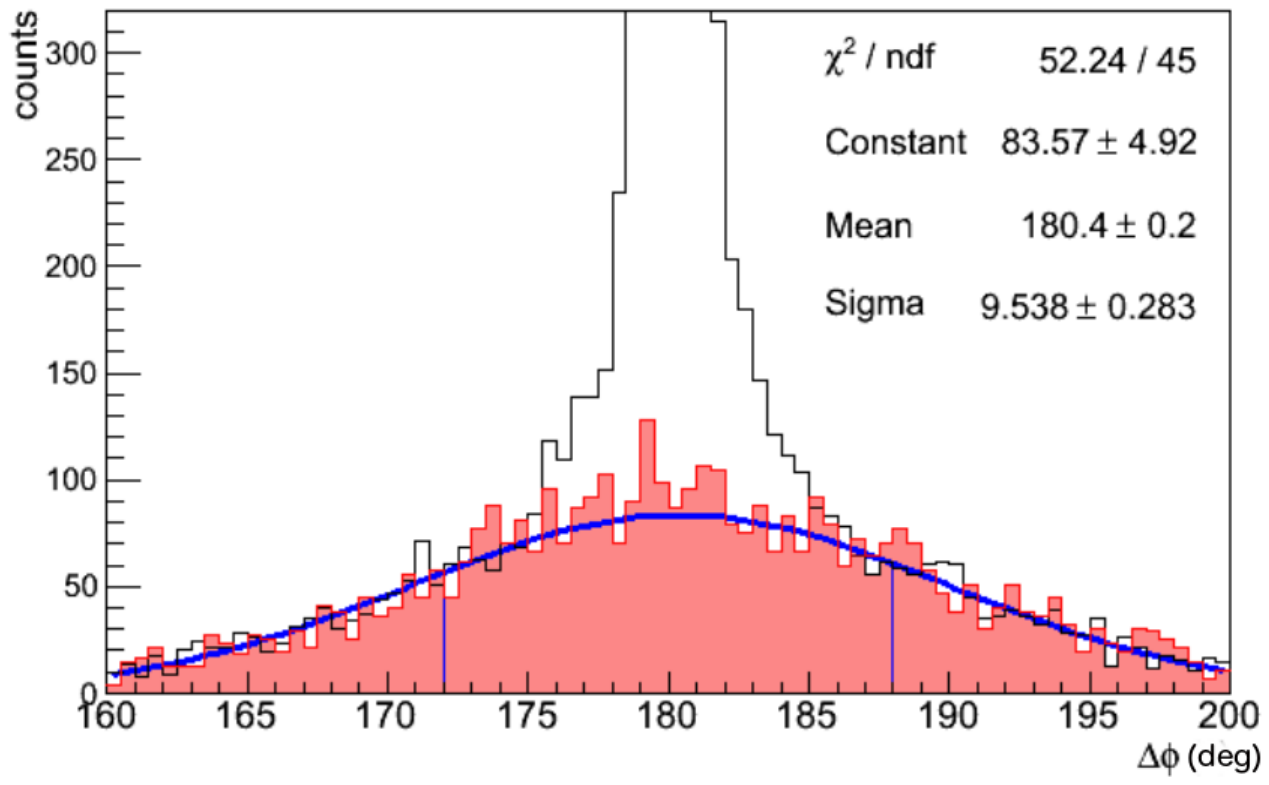}
 	\caption {(Color online) Black histogram is the $\Delta \phi$ distribution for $e^-p$ elastic events at $\langle\varepsilon\rangle=0.40$ and 
	$\langle Q^2\rangle=1.44$ GeV$^2$, the bin with the largest background. The other three kinematic	cuts have been applied. 
	Tails of the distribution to the left of 172$^{\circ}$ and to the right of 188$^{\circ}$ (shown by vertical lines) 
	were fit with a Gaussian function shown in blue.  The filled red histogram is a scaled background sample from Fig.~\ref{fig:BGsamp}.}
 	\label{fig:BGfit}
 \end{figure}

After applying all event selection cuts some background remains, particularly at low $\varepsilon$ and high $Q^2$.  The background was 
found to be symmetric about $\Delta\phi=0$ but not symmetric in $\Delta P_p$ or $\Delta E^{\pm}$. Therefore, we used the $\Delta\phi$ 
distributions to determine the background.  $\Delta\phi$ distributions were made for each bin and for $e^+p$ and $e^-p$ 
events separately. The tails of the $\Delta\phi$ distributions (over the regions $160^\circ-172^\circ$ and $188^\circ-200^\circ$) were fit with a 
Gaussian.  Fig.~\ref{fig:BGfit} shows the Gaussian background fit for the bin with the most background. 

\begin{figure}[htbp]
	\begin{center}
	\includegraphics[width=0.45\textwidth]{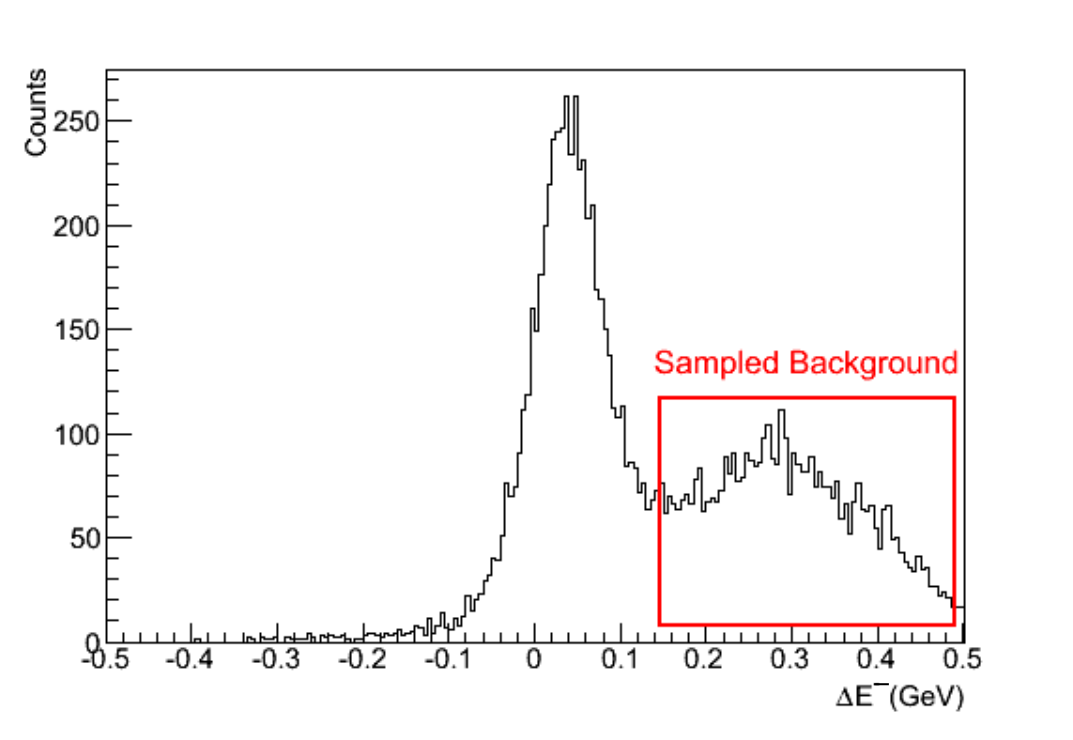}
	\caption{$\Delta E^-$ distribution for $e^-p$ events at $\langle\varepsilon\rangle=0.40$ and $\langle Q^2\rangle=1.44$ GeV$^2$, 
	the bin with the largest background. The other three kinematic	cuts have been applied. The box shows the region of the distribution 
	that was sampled for background.}
	\label{fig:BGsamp}
	\end{center}
\end{figure}

To verify the Gaussian shape of the background, we used a sampling method to determine the shape of the background at low $\varepsilon$. 
Figure~\ref{fig:BGsamp} shows the $\Delta E^-$ distribution for $e^-p$. The sample was selected from the right-hand tail of the distribution and 
scaled to match the tails of the $\Delta\phi$ distributions.  The sampled background shown by the red histogram in Fig.~\ref{fig:BGfit} shows excellent 
agreement with the tails of the $\Delta\phi$ distribution and also with the Gaussian background fit.  The $\Delta E^-$ distribution for $e^+p$ events (not shown) 
at the same kinematics is similar in shape but with background that is 5-10\% smaller than for the $e^-p$ events.  However, the sampled background for $e^+p$ 
events also matches Gaussian background fit. At higher $\varepsilon$ the $\Delta E^-$ peak broadened significantly and the background was much smaller so 
it was not possible to use the sampling method. In bins where it was possible to use both methods we found that the final result for $R_{2\gamma}$ was the same 
to within statistical uncertainties, therefore, the Gaussian fit was employed for all bins.

%\vfill\eject

\subsection{Radiative Corrections}
%%%%%%%%%%%%
%From Analysis note %
%%%%%%%%%%%%
Higher order QED diagrams beyond the Born approximation have a significant,
but generally well-calculable, impact on the elastic charged lepton--proton
scattering cross sections. The largest contributions are the charge-even
terms, which are the same for electrons and positrons. The charge-odd terms cause the difference between the positron and electron 
scattering cross sections while the charge even terms dilute this difference. 

There are two leading order corrections that are odd in the product of the
beam and target charges. The first is the TPE contribution (or more
correctly, the interference between one- and two-photon exchange amplitudes), which is highly
model-dependent, and which we aim to extract.  The second is the interference
between real photon emission from the proton and from the incident or
scattered electron.  The latter is considered a background for this
measurement and needs to be computed to isolate the TPE contribution.

The bremsstrahlung interference term is somewhat model dependent, as the
proton bremsstrahlung contribution has some sensitivity to the proton internal
structure. However, this sensitivity is relatively small and the amplitude
for photon emission from the proton is also small at low $Q^2$, where
the proton is not highly relativistic.

While the key contribution is the charge-odd bremsstrahlung term, the
charge-even terms also need to be applied, as they dilute the charge-odd term as shown in 
Eq.~\ref{eq:R1}.  For 
both contributions, the  bremsstrahlung contributions are typically calculated 
assuming a fixed energy loss or $W^2$ cut used to determine which events 
are included as elastic and which are in the excluded radiative tail. In our case, 
we apply our elastic event identification kinematic cuts, rather than a $W^2$
cut. The primary difference between the two approaches is that our cuts do not 
remove events where the incoming lepton radiates a photon; this radiation just 
changes the incident lepton energy.

We simulated radiative effects following the prescription of Ref.~\cite{ent01}, taking the 
``extended peaking approximation" approach. In this approach, radiated photons are 
generated only in the directions of the charged particles, but both the incoming and 
outgoing leptons and the struck proton are all allowed to radiate.  The sum of the radiated 
photon energy thus has a fairly realistic angular distribution, as shown in Ref.~\cite{ent01,weissbach06}.

%The radiative corrections are applied using the full formalism of Ref.~\cite{ent01} for 
%electron--proton scattering, and then again with the sign of the bremsstrahlung interference 
%term changed. The average of these two yield the bremsstrahlung contribution to $\delta_{even}$, 
%while the ratio of these yields the charge asymmetry, corresponding to the no-TPE  limit 
%$R = 1 - 2(\delta_{e.p.brem})/(1+\delta_{even})$. 

The Monte Carlo simulation was run twice for electrons with the radiative effects turned on and off, then twice more for positrons with the
radiative effects turned off and on, resulting in ratios of yields given by
\begin{equation}
\label{eq:MCY}
	R_{e^\pm}=\frac{Y^\pm_{rad}}{Y^\pm_{Born}}.
\end{equation}
In each of the simulations we assumed no TPE effects. 
We then define a charge-odd correction factor
\begin{eqnarray}
	C_{odd}&=&\frac{R_{e^+}}{R_{e^-}} \label{eq:Codd1} \\
	&=&\frac{1+\delta_{even}-\delta_{e.p.brem}}{1+\delta_{even}+\delta_{e.p.brem}}. \label{eq:Codd2}
\end{eqnarray}	
To within any detector acceptance effects, the terms of $Y^\pm_{Born}$ cancel in this ratio. One sees that $C_{odd}$ still has a contribution 
from $\delta_{even}$. 

\begin{figure}[htb]
\begin{center}
\includegraphics[width=0.49\textwidth]{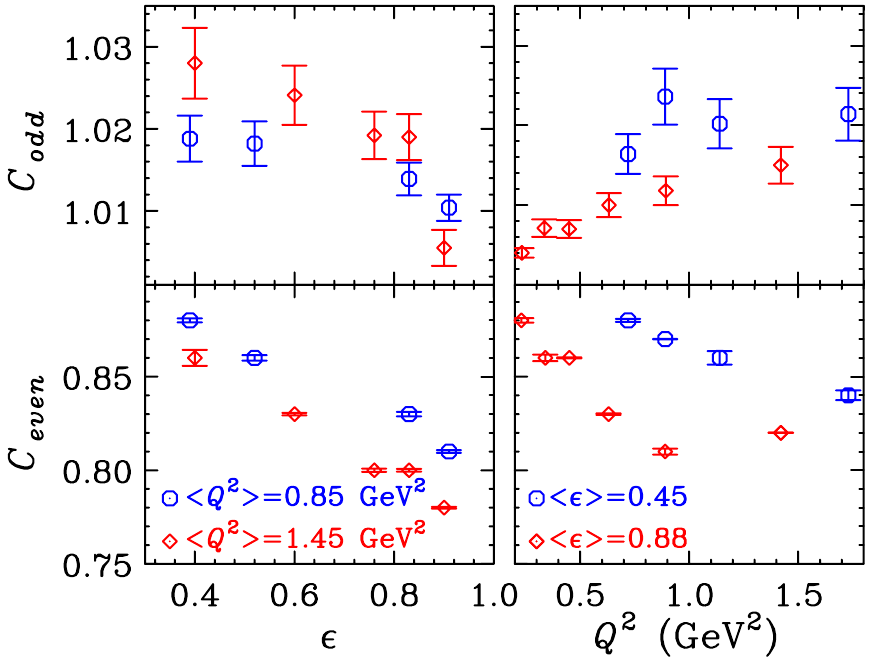}
\caption{(Color online) Bin-averaged radiative correction factors.  The top panels show $C_{odd}$, the ratio of simulated radiated $e^+p$ to $e^-p$ 
cross-section ratio to that of the unradiated (Born) $e^+p$ to $e^-p$ cross-section ratio.  The bottom panels show $C_{even}=1+\delta_{even}$.  
The error bars in both cases are the uncertainty contributions to 
the final result for $R_{2\gamma}$ rather than error bar on the value itself.  This was done because in the case of the even correction we have assumed a 15\% 
uncertainty, which would overwhelm the plot but nonetheless leads to a small contribution to the uncertainty on $R_{2\gamma}$.}
\label{fig:RadCor}
\end{center}
\end{figure}

We obtained the charge-even radiative correction by averaging the results of the simulation leading to 
\begin{eqnarray}
	C_{even}&=&\frac{R_{e^+}+R_{e^-}}{2}, \nonumber \\
	&=&1+\delta_{even}. \label{eq:Ceven1} 
\end{eqnarray}	
This can be used to extract the charge-odd term, $\delta_{e.p.brem}$, from Eq.~\ref{eq:Codd2}.
Figure~\ref{fig:RadCor} shows the charge-odd (top panels) and charge-even (bottom panels) bin-averaged radiative corrections. 
We can then extract $\delta_{2\gamma}$ from the measured $e^+p$ to $e^-p$ cross section ratio of Eq.~\ref{eq:R1} using $\delta_{e.p.brem}$ and
$C_{even}$ and use that to determine $R_{2\gamma}$ as defined in Eq.~\ref{eq:R2g}. 

%
%\begin{equation}
%\label{eq:radcor}
%	R_{2\gamma}=1+(R'-1)(1+\delta_{even}).
%\end{equation}

%The difference between the uncorrected and radiatively-corrected results, $R_{meas}-R_{2\gamma}$, where $R_{meas}$ is the experimentally measured 
%equivalent to $R$ in Eq.~\ref{eq:R}, and $R_{2\gamma}$ is shown in Eq.~\ref{eq:R2g}, varies from 0.003 at high $\varepsilon$ and low $Q^2$ to 0.034 at
%low $\varepsilon$ and high $Q^2$.
%Our radiative correction factor at $Q^2=1.45$ GeV$^2$ and $\varepsilon=0.4$ is consistent 
%with the radiative corrections applied to the Novosibirsk results at similar $Q^2$ and $\varepsilon$ \cite{Rachek2015}.

%We divided our measured $e^+p/e^-p$ cross section ratios by the calculated ratio of the radiated 
%$e^+p$ cross section to the radiated $e^-p$ cross section assuming no TPE. 

Any error due to the radiative corrections prescription 
is likely to have a correlated effect between different kinematics. Because the correlation 
is unknown, we approximate this by applying an overall scale uncertainty of 0.3\% (roughly 15\% of the correction at the high $Q^2$ kinematics), 
with an additional point-to-point uncertainty at each setting equal to 15\% of the correction for that point.  

%\newpage

\begin{table*}[htbp]
\centering
\small
\begin{center}
\begin{tabular}{c c c c c c c c c c c c}
\hline\hline
 Bin No. & $\langle Q^2\rangle$ & $\langle\varepsilon\rangle$ & $\delta R_{sector}$ & $\delta R_{cycle}$ & $\delta R_{track}$ & $\delta R_{kin}$ & $\delta R_{BG}$ & $\delta R_{vz}$ & $\delta R_{fid}$ & $\delta R_{acc}$ & $\delta R_{sys}$  \\
\hline
1 & 0.84 & 0.39 & 0.0100 & 0.0030 & 0.0013 & 0.0159  & 0.0054 & 0.0075 & 0.0001 & 0.001 & 0.0212\\
2 & 0.86 & 0.51 & 0.0034 & 0.0030 & 0.0013 & 0.0074  & 0.0010 & 0.0112 & 0.0001 & 0.001 & 0.0143\\
3 & 0.85 & 0.83 & 0.0034 & 0.0030 & 0.0013 & 0.0021  & 0.0030 & 0.0027 & 0.0014 & 0.001 & 0.0068\\
4 & 0.85 & 0.91 & 0.0034 & 0.0030 & 0.0013 & 0.0015  & 0.0024 & 0.0005 & 0.0014 & 0.001 & 0.0058\\ \hline
5 & 1.44 & 0.40 & 0.0034 & 0.0030 & 0.0013 & 0.0070 & 0.0023 & 0.0031 & 0.0003 & 0.001 & 0.0093\\
6 & 1.45 & 0.60 & 0.0034 & 0.0030 & 0.0013 & 0.0069 & 0.0021 & 0.0004 & 0.0005 & 0.001 & 0.0087 \\
7 & 1.46 & 0.76 & 0.0034 & 0.0030 & 0.0013 & 0.0075 & 0.0024 & 0.0021 & 0.0005 & 0.001 & 0.0095 \\
8 & 1.47 & 0.83 & 0.0034 & 0.0030 & 0.0013 & 0.0012 & 0.0014 & 0.0015 & 0.0046 & 0.001 & 0.0071\\
9 & 1.47 & 0.90 & 0.0034 & 0.0030 & 0.0013 & 0.0043 & 0.0021 & 0.0024 & 0.0057 & 0.001 & 0.0092\\ \hline \hline
10 & 0.72 & 0.45 & 0.0034 & 0.0030 & 0.0013 & 0.0033 & 0.0033 & 0.0003 & 0.0001 & 0.001 & 0.0067\\
11 & 0.89 & 0.45 & 0.0034 & 0.0030 & 0.0013 & 0.0132 & 0.0034 & 0.0057 & 0.0001 & 0.001 & 0.0155\\
12 & 1.14 & 0.45 & 0.0034 & 0.0030 & 0.0013 & 0.0037 & 0.0071 & 0.0015 & 0.0004 & 0.001 & 0.0095\\
13 & 1.73 & 0.45 & 0.0034 & 0.0030 & 0.0013 & 0.0063 & 0.0115 & 0.0012 & 0.0007 & 0.001 & 0.0140\\ \hline
14 & 0.23 & 0.92 & 0.0034 & 0.0030 & 0.0013 & 0.0012 & 0.0028 & 0.0003 & 0.0013 & 0.001 & 0.0059\\
15 & 0.34 & 0.89 & 0.0034 & 0.0030 & 0.0013 & 0.0005 & 0.0005 & 0.0002 & 0.0006 & 0.001 & 0.0049\\
16 & 0.45 & 0.89 & 0.0034 & 0.0030 & 0.0013 & 0.0007 & 0.0010 & 0.0002 & 0.0002 & 0.001 & 0.0050\\
17 & 0.63 & 0.88 & 0.0034 & 0.0030 & 0.0013 & 0.0011 & 0.0052 & 0.0006 & 0.0005 & 0.001 & 0.0072\\
18 & 0.89 & 0.88 & 0.0034 & 0.0030 & 0.0013 & 0.0017 & 0.0032 & 0.0008 & 0.0011 & 0.001 & 0.0062\\
19 & 1.42 & 0.87 & 0.0034 & 0.0030 & 0.0013 & 0.0016 & 0.0022 & 0.0016 & 0.0041 & 0.001 & 0.0071\\
 \hline\hline
\end{tabular}
\caption{Uncorrelated systematic uncertainties on the measured $e^+p/e^-p$ ratio for all data bins due to various sources as described in the text. 
``sector'' refers to CLAS detector imperfections, ``cycle'' refers to the differences in the electron and positron luminosities, ``track'' refers to charge 
independence of track reconstruction, ``kin'' refers to elastic event selection, ``BG'' refers to background fitting, ``vz'' refers to target vertex cuts, ``fid'' 
refers to fiducial cuts, ``acc'' refers to acceptance corrections, and ``sys'' is the quadrature sum of all listed uncertainties.
Bins 1-9 are selected to study the $\varepsilon$ dependence of $R_{2\gamma}$ at two values of $Q^2$ and bins 10-19 are selected to study the 
$Q^2$ dependence of $R_{2\gamma}$ at two values of $\varepsilon$.}
\label{tab:sys}
\end{center}
\end{table*}

\subsection{Systematic Uncertainties}
As discussed earlier, our experimental design helped to cancel or minimize most of the 
systematic uncertainties in the measurement of $R_{2\gamma}$. Any remnant systematic uncertainties 
are discussed below.  Table~\ref{tab:sys} lists the various sources of systematic uncertainty on the measured ratio before doing radiative corrections. 
The effect of these corrections is to reduce the measured ratio by a factor of $1-\delta_{even}$, so it similarly will reduce the total systematic uncertainty in 
$R_{2\gamma}$.

\begin{enumerate}
\item CLAS imperfections: We compared our final cross section ratio measured in 
different sectors of CLAS. The variations in these ratios quantify the systematic effects 
due to detector imperfections. Since we removed the forward going lepton or proton  
events from sector 3, we had five independent cross-section ratios for each bin. 
We calculated the weighted average and the chi-squared based on the scatter 
of the five independent ratios. We then added the same systematic uncertainty to each 
of the sector-based quadruple ratios and recalculated the chi-squared and the confidence 
level.  We chose a 0.75\% systematic uncertainty for each sector measurement to give an 
average confidence level of $\sim50$\% for all of the bins. This gives a sector-to-sector overall 
systematic uncertainty of $0.75\%/\sqrt{5} = 0.34\%$ for each bin except bin 1 as it showed a
larger sector dependence than the other bins. This uncertainty is listed in Table~\ref{tab:sys} under $\delta R_{sector}$.

\item Differences in the $e^{+}$ and $e^{-}$ luminosities: With electron-positron pair production 
being inherently charge symmetric, the $e^{+}$ and $e^{-}$ beam fluxes should be identical. In
the experiment, the only differences in the two beams could come from differences in beam transport 
from the converter to the target. The chicane magnet setting was periodically reversed several times 
during the run period to minimize the differences and we measured the energy distributions of the electron and positrons  
with TPE Calorimeter after each reversal. Fig.~\ref{fig:IncidentBeamSum} shows that the reconstructed energy distributions
of the incident $e^{+}$ and $e^{-}$ are identical. Any difference in the incident lepton flux primarily 
appears as the variation in the cross section ratios for the different chicane cycles. The systematic uncertainty was calculated similarly to 
that for the CLAS imperfections. For each of the independent chicane cycles we 
determined the double ratios (Eq.~\ref{eq-R2}).  We added the same systematic uncertainty to each double ratio to give an average
confidence level of 50\% for all bins. The overall systematic uncertainty due to lepton luminosity differences was estimated to be 0.3\% 
for each bin. It is listed in Table~\ref{tab:sys} under $\delta R_{cycle}$.

\item Charge independence of track reconstruction:  A series of special runs were conducted with the CLAS minitorus turned off in 
order to make sure that our track reconstruction and analysis code was independent of the charge of the particles.  We determined 
the number of $e^+p$ elastic events for positive and negative torus settings and a 
fixed chicane setting. We then replayed the same runs assuming the opposite torus polarity, thus 
reversing the roles of negatively and positively charged tracks, and determined the number of elastic events where both particles had a ``negative'' charge.
The analysis found equal numbers of events for the two analyses to within 0.13\%, which we have assumed as a systematic uncertainty
associated with the charge dependence of track reconstruction.  It is listed in Table~\ref{tab:sys} under $\delta R_{track}$.
 
%\begin{figure}[htb]
%  \begin{center}    
%\includegraphics[width=0.50\textwidth]{images/ChargeBlind.pdf}
%  \caption{(Color online) The double ratio of $e^+p$ to $e^-p$ elastic events for positive and negative torus polarities when the replay code is
%	given the correct torus polarity (red circles) compared to when the replay code is told the incorrect torus polarity (black squares).  
%	The ratios are plotted as a function of $\epsilon$ for a limited range in $Q^2$.  The correct-  and incorrect-torus ratios are 
%	almost identical.  The actual ratios are unimportant as the goal of this study was to verify a negligible difference in the ratios, 
%	indicating the analysis code is charge blind.}
%  \label{fig:ChargeBlind}
%  \end{center}    
%\end{figure}

\item Elastic event selection and background subtraction: For each bin, the systematic uncertainty due to elastic event selection 
cuts was estimated by increasing the width of the kinematic cuts from the nominal $\pm 3\sigma$ cuts to $\pm 3.5\sigma$ cuts.  Relaxing
these cuts doubled the background present in the data.  Thus the kinematic cut uncertainty includes the background subtraction uncertainty.
The deviation of the final ratio with the varied cuts from the ratio with the nominal cuts was assigned 
as the systematic uncertainty due to our event selection. It is listed in Table~\ref{tab:sys} under $\delta R_{kin}$.

\item Background fitting: We determined the systematic uncertainty due to background fitting 
by varying the fitting regions from the nominal fitting range. For each bin, we varied the fitting 
range by $-2^\circ$ (160$^\circ$ to 170$^\circ$ and 190$^\circ$ to 200$^\circ$) and 
$+2^\circ$ (160$^\circ$ to 174$^\circ$ and 186$^\circ$ to 200$^\circ$) and recalculated 
the final ratios. The systematic uncertainty due to the background subtraction was estimated to be the average 
deviation of the varied ratios ($R_{\pm 2^\circ}$) from that with the nominal fitting ranges ($R_{\text{Nom.}}$):
\begin{equation}
\label{eq:dRBG}
 \delta R_{BG} =\frac{\left(R_{\text{Nom.}}-R_{-2^\circ}\right)+\left(R_{\text{Nom.}}-R_{+2^\circ}\right)}{2}.
%($\Delta R_{\text{Nominal}-R_1}$,  $\Delta R_{\text{Nominal}-R_2}$).
\end{equation}

\item Target vertex cut: For each bin, the systematic uncertainty due to the target vertex cut
was estimated by varying the width of the nominal vertex cut 
of $-44 < v_{z} < -16$ cm to $-43 < v_{z} < -17$ cm. The deviation of the final ratio with the varied cuts from the 
ratio with the nominal cut was assigned as the systematic uncertainty due to the vertex cut. It is listed in Table~\ref{tab:sys} under $\delta R_{vz}$.

\item Fiducial cuts: The systematics effect due to the applied fiducial cuts were estimated
by increasing the lower limit of the $\phi$ cut by one degree and decreasing the upper limit of $\phi$ cut by one degree thereby reducing the fiducial volume.
The deviation of the final ratio with the tightened fiducial volume from that with the nominal fiducial volume was
assigned as the systematic uncertainty due to our fiducial cuts.
It is listed in Table~\ref{tab:sys} under $\delta R_{fid}$.

\item Acceptance correction: As seen above, the acceptance correction factors determined from the Monte Carlo simulation were close to unity with a high level
of uniformity.  We conservatively estimate an uncertainty of 0.1\% for all bins, which is 20\% of the largest deviation of the acceptance correction from unity. 
It is listed in Table~\ref{tab:sys} under $\delta R_{acc}$.

\end{enumerate}

For each bin, the contribution from all the sources were added in quadrature to obtain our total systematic 
uncertainties $\delta R_{sys}$. The total uncertainties are presented along with the final results in Table~\ref{tab:FinalResults}.

\section{RESULTS}
The final results are given in Table~\ref{tab:FinalResults} along with all associated uncertainties, and shown in Figs.~\ref{fig:RatioEpsDep} 
and \ref{fig:RatioQ2Dep}.  Table~\ref{tab:FinalResults} includes both $R_{meas}$, which is the experimentally measured equivalent to $R$ of Eq.~\ref{eq:R1}, and $R_{2\gamma}$ which is the radiatively-corrected result as shown in Eq.~\ref{eq:R2g}. Estimated systematic uncertainties 
associated with the $\delta_{e.p.brem}$ and $\delta_{even}$ corrections are also given in the table.
The numbers in the column labeled ``overlap''  indicate that a given bin contains part or all of the bins listed in that column of the table.  For example, bin 1
has an overlap with part of bin 10, while bin 10 overlaps both bins 1 and 2.  The reason for showing data from overlapping
kinematic bins is to separately study the $Q^2$ and $\varepsilon$ dependencies, though future use of our results in modeling TPE corrections should
take into account the fact that we are displaying non-independent results.  Quantitative model comparisons will be discussed in Sec.~\ref{sec-global}.

\begin{table*}[ht]
\centering
\begin{center}
\begin{tabular}{c c c c c c c c c c c}
\hline\hline
Bin No. &  $\langle Q^2 \rangle$  & $\langle \varepsilon \rangle$ & $R_{meas}$ & $R_{2\gamma}$ & $\delta R_{stat}$& $\delta R_{rad}$ 
& $\delta R_{sys}$ & $\delta R_{total}$  & $\delta_{RCnorm}$ & overlap \\
\hline
1   & 0.84 & 0.39 & 1.0268 & 1.0070 & 0.0122 & 0.0043 & 0.0182 & 0.0223 & 0.003 & 10 \\
2   & 0.86 & 0.52 & 1.0057 & 0.9896 & 0.0109 & 0.0024 & 0.0122 & 0.0166 & 0.003 & 10 \\
3   & 0.85 & 0.83 & 1.0226 & 1.0074 & 0.0066 & 0.0032 & 0.0055 & 0.0092 & 0.003 & 18 \\
4   & 0.85 & 0.91 & 1.0074 & 0.9976 & 0.0054 & 0.0015 & 0.0047 & 0.0073 & 0.003 & 18 \\ \hline
5   & 1.44 & 0.40 & 1.0623 & 1.0282 & 0.0102 & 0.0086 & 0.0075 & 0.0153 & 0.003 & 11,12,13 \\
6   & 1.45 & 0.60 & 1.0299 & 1.0047 & 0.0131 & 0.0047 & 0.0070 & 0.0155 & 0.003 & 11,12,13 \\
7   & 1.46 & 0.76 & 1.0120 & 0.9943 & 0.0109 & 0.0027 & 0.0075 & 0.0135 & 0.003 & \\
8   & 1.47 & 0.83 & 1.0134 & 0.9956 & 0.0122 & 0.0028 & 0.0056 & 0.0137 & 0.003 & 19 \\
9   & 1.47 & 0.90 & 1.0010 & 0.9965 & 0.0111 & 0.0007 & 0.0072 & 0.0132 & 0.003 & 19 \\ \hline \hline
10 & 0.72	& 0.45 & 1.0224 & 1.0052 & 0.0113 & 0.0036 & 0.0058 & 0.0132 & 0.003 & 1,2 \\	 			  
11 & 0.89	& 0.45 & 1.0246 & 1.0009 & 0.0110 & 0.0044 & 0.0132 & 0.0178 & 0.003 & 5,6 \\
12 & 1.14	& 0.45 & 1.0490 & 1.0239 & 0.0112 & 0.0067 & 0.0078 & 0.0152 & 0.003 & 5,6 \\
13 & 1.73	& 0.45 & 1.0427 & 1.0176 & 0.0118 & 0.0059 & 0.0113 & 0.0173 & 0.003 & 5,6 \\ \hline
14 & 0.23 	& 0.92 & 0.9950 & 0.9920 & 0.0020 & 0.0008 & 0.0052 & 0.0056 & 0.003 & \\
15 & 0.34 	& 0.89 & 0.9940 & 0.9888 & 0.0022 & 0.0012 & 0.0043 & 0.0050 & 0.003 & \\
16 & 0.45 	& 0.89 & 1.0040 & 0.9974 & 0.0022 & 0.0010 & 0.0043 & 0.0049 & 0.003 & \\
17 & 0.63 	& 0.89 & 1.0130 & 1.0025 & 0.0029 & 0.0020 & 0.0059 & 0.0069 & 0.003 & \\
18 & 0.89 	& 0.88 & 1.0240 & 1.0097 & 0.0036 & 0.0032 & 0.0049 & 0.0069 & 0.003 & 3,4 \\
19 & 1.42 	& 0.87 & 1.0150 & 1.0000 & 0.0067 & 0.0026 & 0.0057 & 0.0092 & 0.003 & 8,9 \\ \hline
\end{tabular}
\caption{Final measured ($R_{meas}$) and radiatively-corrected ($R_{2\gamma}$) cross section ratios and the associated statistical ($\delta R_{stat}$), 
	systematic ($\delta R_{sys}$), radiative correction ($\delta R_{rad}$),  and total uncorrelated uncertainties ($\delta R_{total}$).  
	The $\delta R_{RCnorm}$ column is a scale-type uncertainty common to the entire data set. The ``overlap'' column indicates overlapping bins.}
  \label{tab:FinalResults}
\end{center}
\end{table*}

\subsection{$\varepsilon$-dependence}

\begin{figure}[htb]
\begin{center}    
	\includegraphics[width=0.45\textwidth]{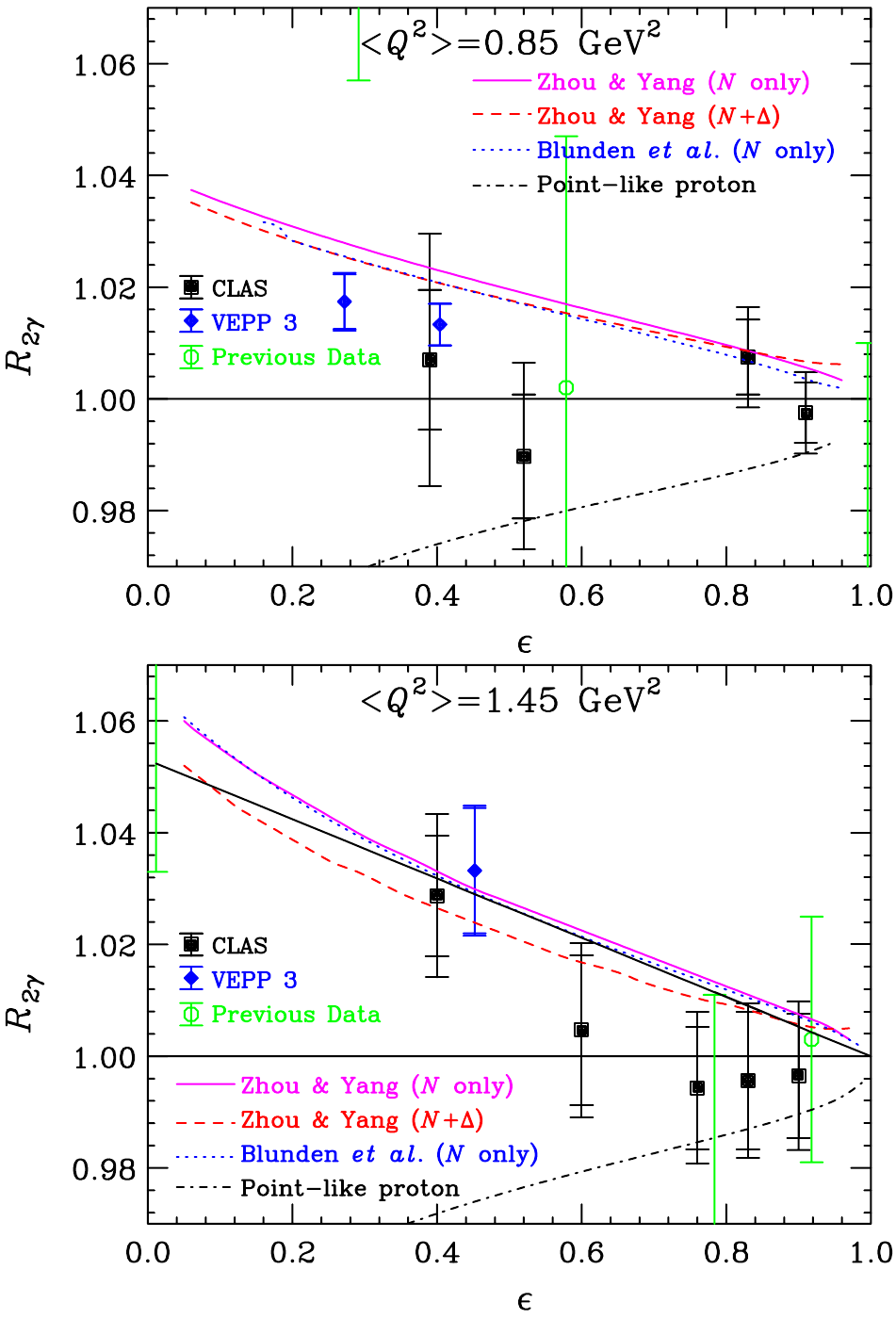}
	\caption{(Color online) $R_{2\gamma}$ as a function of $\varepsilon$ at $Q^2 \approx 0.85$  GeV$^2$ (top) and 1.45 GeV$^2$ (bottom) extracted from
	the measured ratio of $e^+p$/$e^-p$ cross sections corrected for both $\delta_{brem}$ and $\delta_{even}$. The filled black squares show 
	the results of this measurement. The inner error bars are the statistical uncertainties and the outer error bars are the statistical, systematic 
	and radiative-correction uncertainties added in quadrature. The line at $R_{2\gamma}=1$ is the limit of no TPE.  The magenta 
	solid and red dashed curves show the calculation by Zhou and Yang~\cite{Zhou2014} including $N$ only and $N+\Delta$ intermediate states,
	respectively.  The blue dotted curve shows the calculation by Blunden \textit{et al.}~\cite{blunden05}. The black dot-dashed line shows the 
	calculation of TPE effects on a structureless point proton \cite{arrington11b}.  The open green circles show the previous world data at 
	$0.7\leq Q^2 \leq 1.0$ GeV$^2$ and $1.2\leq Q^2\leq 1.53$ GeV$^2$ in the top and bottom plots, respectively~\cite{arrington04b}.  The filled blue 
	diamonds are from VEPP-3~\cite{Rachek2015} showing the combined statistical and systematic uncertainty.  The solid black line in the lower figure is 
	a linear fit to the all of the data shown and was constrained 
	to go to $R_{2\gamma}=1$ at $\varepsilon=1$.}
\label{fig:RatioEpsDep}
\end{center}    
\end{figure}

Fig.~\ref{fig:RatioEpsDep} shows the $\varepsilon$-dependence of $R_{2\gamma}$ at $Q^2\approx 0.85$ and 1.45~GeV$^2$, along with previous 
world data and the calculations of Refs.~\cite{blunden05,Zhou2014,arrington11b}.  Our results at $Q^2$ = 0.85 GeV$^2$  are consistent with no 
epsilon dependence, though inclusion of the VEPP-3 results at $Q^2=0.83$ and 0.976 GeV$^2$ may suggest a slight increase of $R_{2\gamma}$ with 
decreasing $\varepsilon$.
Our data at $Q^2$ = 1.45 GeV$^2$ when combined with the VEPP-3 $Q^2$ = 1.51 GeV$^2$ result show a moderate epsilon dependence.
Together with the VEPP-3 data, the results are inconsistent with the no-TPE ($R_{2\gamma}=1$) limit.

The data are compared to calculations of TPE in a hadronic framework~\cite{blunden05, Zhou2014}, and the analytic results for scattering from a 
structureless  (point-like) proton~\cite{arrington11b}. The data are significantly higher than the point-proton calculation and show the opposite $\varepsilon$ 
dependence. The data are consistent with the hadronic calculations which, for the $Q^2$ values presented here, are dominated by the elastic 
intermediate state.  
The hadronic calculations bring the form factor ratio extracted from Rosenbluth separation measurements into good agreement with the 
polarization transfer measurements up to $Q^2 \approx 2$ GeV$^2$~\cite{arrington11b}, so the data support the explanation of the discrepancy
in terms of TPE contributions. As discussed in Ref.~\cite{arrington07c}, confirmation that TPE contributions explain the discrepancy
is sufficient to allow extraction of the form factors without a significant uncertainty associated with the TPE corrections.

\subsection{$Q^{2}$-dependence}
\begin{figure}[htb]
\begin{center}    
	\includegraphics[width=0.45\textwidth]{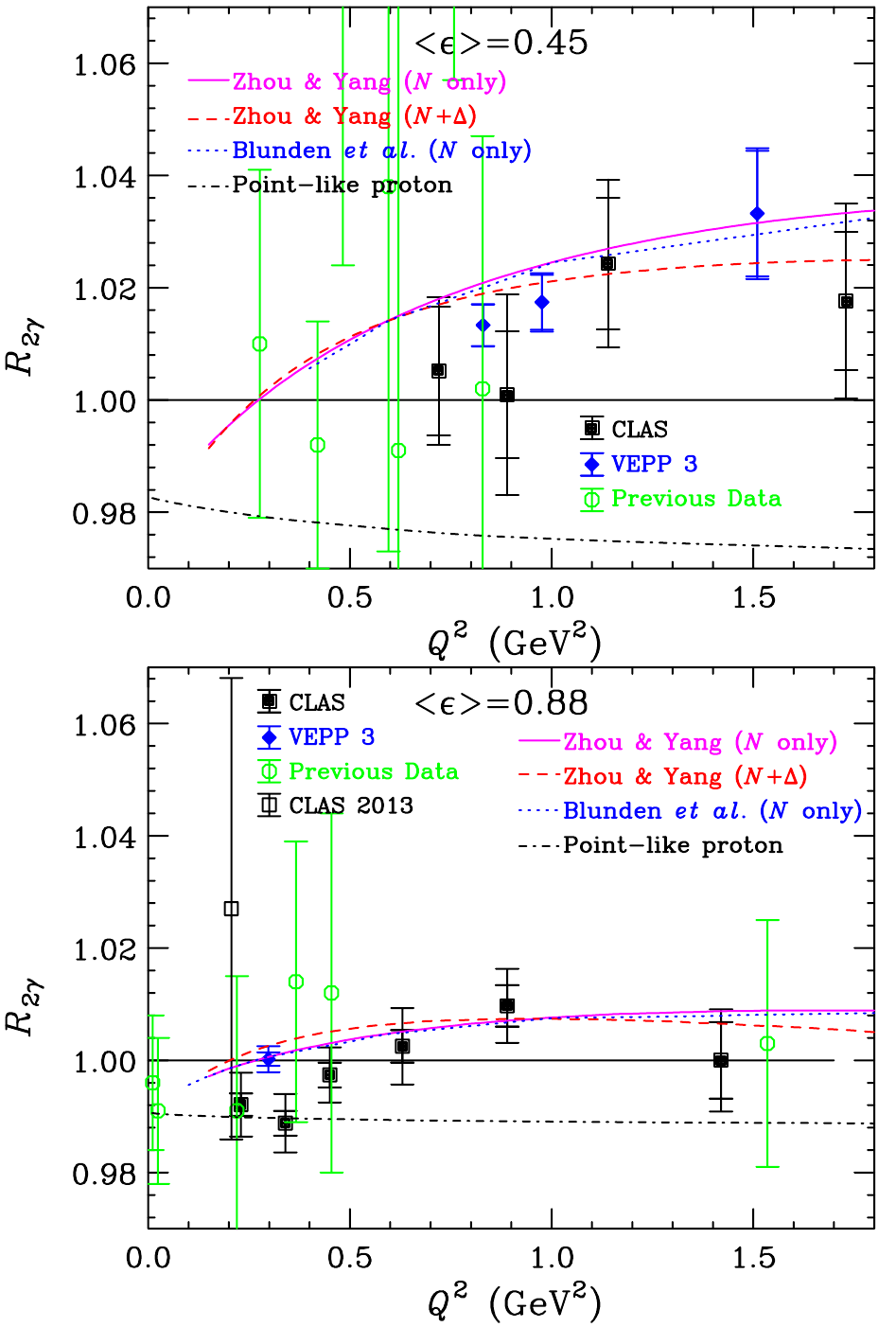}
	\caption{(Color online) Same as Fig.~\ref{fig:RatioEpsDep} except as a function of $Q^2$ at $\varepsilon\approx 0.45$  (top) and 0.88 (bottom).
	Also included is the CLAS 2013~\cite{Moteabbed} result (black open square), which has been averaged to a single point at $\varepsilon=0.893$. The open 
	green circles show the previous world data  at $0.2\leq\varepsilon\leq 0.7$ and $0.7\leq\varepsilon\leq 0.95$ in the top and bottom plots,
	respectively~\cite{arrington04b}.}
\label{fig:RatioQ2Dep}
\end{center}    
\end{figure}

Fig.~\ref{fig:RatioQ2Dep} shows the $Q^2$-dependence of the ratio at $\varepsilon\approx 0.45$ and 0.88 along with previous world data and 
the calculations of Refs.~\cite{blunden05,Zhou2014,arrington11b}.  In both cases our results are consistent with little or no $Q^2$ dependence, 
while the inclusion of the VEPP-3 data at $\varepsilon\approx 0.45$ indicates a gradual increase in $R_{2\gamma}$ with $Q^2$.  As before, the results 
are largely consistent with the calculations of Blunden {\it et al.}~and Zhou and Yang but not that for a point-like proton.

\subsection{TPE Corrected Rosenbluth Extraction at $Q^2=1.75$ GeV$^2$}
\label{sec:rosen}
\begin{figure}[htb] 
\begin{center}    
\includegraphics[width=0.45\textwidth,clip=true]{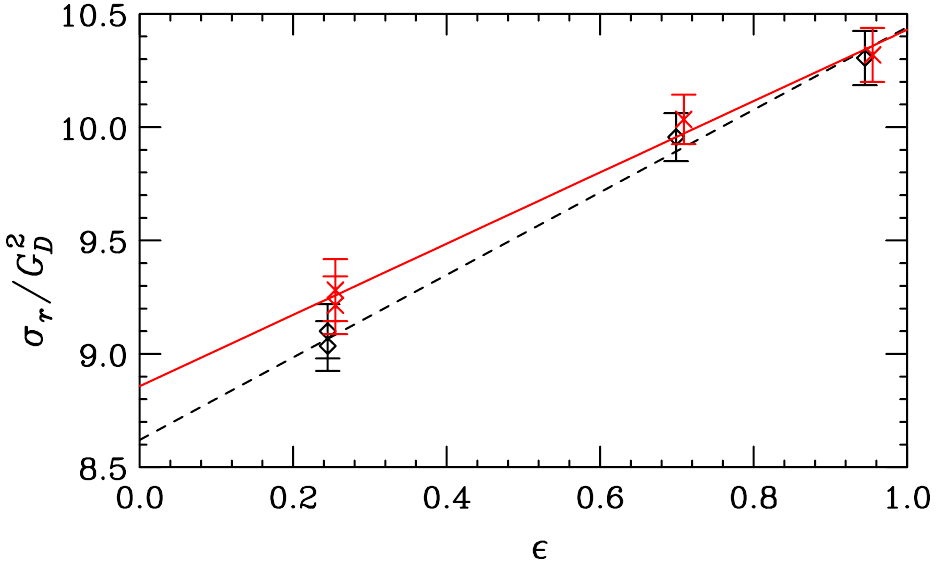}
\caption{(Color online) Reduced cross sections divided by the square of the dipole form factor, $G_D^2= \left(1 + \frac{ Q^2}{0.71}\right)$, plotted as a function of $
\varepsilon$. The black triangles show the original measurements from Andivahis \textit{et al.}~\cite{andivahis94} and the red circles show the TPE corrected 
measurements with uncertainties that include the uncertainties in the correction. The dashed black and solid red lines show the corresponding linear fits where the 
slope is proportional to $G_E^2$ and the intercept is proportional to $G_M^2$.}
\label{fig:tpeCorrRosenbluth}
\end{center}    
\end{figure}

From our results of $R_{2\gamma}$ at $Q^2\approx1.45$ GeV$^2$ we determined the correction factor $\delta_{2\gamma}\left(\varepsilon\right)$.  
We did a linear fit of all of the $R_{2\gamma}$ data shown in Fig.~\ref{fig:RatioEpsDep}  that was constrained to go to $R_{2\gamma}=1$ at $\varepsilon=1$.
We then applied the resulting correction factor (see Eq.~\ref{eq:corrected_sigmaR}), including fit uncertainties, to the unpolarized reduced cross section of 
Andivahis {\it et al.}~\cite{andivahis94} and did a Rosenbluth separation to extract $\mu_pG_E/G_M$ at $Q^2 = 1.75$ GeV$^2$. 
Figure~\ref{fig:tpeCorrRosenbluth} shows the original reduced cross 
section measurements from Andivahis {\it et al.}~and the CLAS TPE corrected values as a function of $\varepsilon$. The TPE corrections change the proton form 
factor ratio obtained from the unpolarized data from $\mu_pG_E/G_M = 0.910\pm0.060$ to $0.829\pm0.044$, bringing it into $1\sigma$ agreement with the 
polarization transfer result of  $0.789\pm0.042$ at $Q^2 = 1.77$ GeV$^2$ by Punjabi {\it et al.}~\cite{punjabi05}.   
% If there is an increase in $R_{2\gamma}$ with $Q^2$, as indicated by the calculations and possibly suggested by the data in Fig.~\ref{fig:RatioQ2Dep}, then one expects a larger TPE correction at $Q^2=1.75$ GeV$^2$ than we have extracted at $Q^2=1.45$ GeV$^2$ and improved agreement between the cross-section and polarization-transfer results.

\subsection{Global Analysis}
\label{sec-global}
In Ref.~\cite{Adikaram15}, we examined the sensitivity of the high-$Q^2$ and high-$\varepsilon$ data (without the VEPP-3 points), and found that they favored 
the hadronic TPE calculations~\cite{blunden05,Zhou2014} over the no-TPE hypothesis by 2.5$\sigma$.  The analysis here includes the full CLAS kinematic coverage,
which includes  additional data at lower $Q^2$ values. These additional data have large uncertainties and are in the kinematic region where the TPE calculations have
minimal disagreement, and so 
have a limited impact in testing different TPE hypotheses.  However, combining the VEPP-3 results, along with the full CLAS data set, yields a more significant test 
of the TPE calculations under the assumption that any missing charge-even corrections to the VEPP-3 results are minimal. Though other calculations of TPE effects 
are available (e.g. GPD-based calculations of ref.~\cite{chen04}), 
the hadronic calculations are expected to be more 
reliable at this low-to-moderate $Q^2$ range. To make a more quantitative comparison of the TPE calculations, we perform a global comparison of the
data to the hadronic calculations of Refs.\cite{blunden05,Zhou2014}, the no-TPE assumption, and the calculation based on a structureless proton~\cite{arrington11b}.

Our data points and the VEPP-3 measurements have uncertainties that are at the 0.5-1.8\% level. Previous measurements typically have uncertainties
greater than 3\%, and the measurements with better uncertainties are generally at $Q^2<0.5$~GeV$^2$ or $\varepsilon > 0.7$, where the calculations
all suggest minimal TPE contributions. Because of the large experimental uncertainties leading to low sensitivity, as well as incomplete knowledge of how radiative 
corrections were applied to extract $R_{2\gamma}$, we do not include these points in our analysis.

\begin{table}[b]
\centering
\begin{center}
\begin{tabular}{l c c}
TPE calculation							& $\chi^2_\nu$	& Conf. Level \\
\hline
Blunden ($N$)~\cite{blunden05}			& 1.23		& 23.5\%  \\
Zhou \& Yang ($N$)~\cite{Zhou2014} 		& 1.27		& 20.8\%  \\
Zhou \& Yang ($N+\Delta$)~\cite{Zhou2014} 	& 1.19		& 27.0\%  \\
$\delta_{2\gamma}=0$	(No TPE)			& 2.32		& 0.20\%  \\
Point-proton calculation					& 7.38		& $2.6\times10^{-15}$\% \\
\end{tabular}
\caption{Comparison of the 16 CLAS and VEPP-3 data points to various TPE calculations showing the reduced $\chi^2$ value and the confidence level.}
  \label{tab:Global}
\end{center}
\end{table}

For this analysis, we have to select a subset of our data, to avoid double counting of data included in more than one binning scheme. We take the high-$Q^2$ data
(bins 5--9) and the high-$\varepsilon$ data (bins 14--18, excluding bin 19 which overlaps bins 8 and 9). We also include the two low-$Q^2$, low-$\varepsilon$ data
points (bins 1 and 2), which do not overlap with the bins at high-$Q^2$ or high-$\varepsilon$. This yields a total of 12 data points from our measurement.  For 
the Novosibirsk data, we use the four non-normalization data points, including a 0.3\% systematic uncertainty applied to account for the model-dependence of the 
high-$\varepsilon$ normalization procedure. The comparison of the CLAS plus VEPP-3 data (16 data points total) to the various models is summarized in 
Table~\ref{tab:Global}. We find that the addition of the CLAS data points that were not presented in our previous publication \cite{Adikaram15} do not significantly impact 
the comparison to the models but the addition of the VEPP-3 data yields a significant improvement. The data are in good agreement with the hadronic calculations of 
Ref.~\cite{blunden05,Zhou2014} but of insufficient precision 
to make any definitive distinction between
them. However, the data exclude the no-TPE hypothesis at the $5.3\sigma$ level, and rule out the point-proton result at the $\sim 25\sigma$ level.   The point-proton
model is essentially equivalent to the $Q^2=0$ limit, which is insensitive to proton structure, used to approximate TPE corrections at low $Q^2$ values~\cite{bernauer10}.
The fit includes a variation of the normalization uncertainty associated 
with the model dependence of the radiative corrections, which increases all of the CLAS ratios by roughly 0.3\% for the fit to the hadronic
calculation and decreases it by a similar amount for the point-like comparison.

% Reducting blunden(N) TPE by 10% to approximate the N+Delta version, the chisquared gets better (0.91), and normalization comes down just a bit.

%\input{conclusions}
\section{Conclusions}
Our results, along with recently published results from VEPP-3, rule out the zero TPE effect hypothesis at the 99.8\% confidence level
and are in excellent agreement ($\chi^2_\nu=1.19$ to 1.27) with the calculations~\cite{blunden05,Zhou2014} that include TPE effects and largely reconcile the form-factor 
discrepancy. The combined CLAS and VEPP-3 data are consistent with an increase in $R_{2\gamma}$ with decreasing $\varepsilon$ at $Q^2\approx 0.85$ and 
1.45 GeV$^2$. A slight, non-statistically significant,  increase in $R_{2\gamma}$  with $Q^2$ is seen. 
Extracting the $\varepsilon$-dependent TPE correction factor, $\delta_{2\gamma}\left(\varepsilon\right)$, from our results for $R_{2\gamma}$ at 
$Q^2\approx 1.45$ GeV$^2$ and applying it to the extraction of $\mu_pG_E/G_M$ at $Q^2 = 1.75$ GeV$^2$ from the Ref.~\cite{andivahis94} 
reduced cross-section data brings it into good agreement with the polarization transfer measurement at $Q^2 = 1.77$ GeV$^2$ by 
Punjabi {\it et al.}~\cite{punjabi05}.

Our data, together with those of VEPP-3, show that TPE effects are present and are large enough to explain the proton electric form factor discrepancy 
up to $Q^2 \approx 2$ GeV$^2$. Since this paper was submitted, the OLYMPUS results have been published~\cite{Henderson:2016dea}. A recent review 
article~\cite{TPERev} in which all three of the modern data sets were included in a global analysis came to a similar conclusion.  However, the form factor 
discrepancy is small at the low momentum transfers of the new data.   Though there are currently no experiments 
planned to extend the measurements to $Q^2\geq 3$ GeV$^2$, where the form-factor discrepancy is the largest, such experiments are needed before one 
can definitively state that TPE effects are the reason for the discrepancy.

\begin{acknowledgments}
We thank Bernard Mecking, the former Jefferson Lab Hall B leader, for suggesting this innovative experimental technique. 
We  acknowledge the outstanding efforts of the Jefferson Lab staff  (especially  Dave Kashy and the CLAS technical staff)  that made this 
experiment possible. This work was supported in part by the U.S. Department of
Energy and National Science Foundation, the Italian Istituto Nazionale di Fisica
Nucleare, the Chilean Comisi\'on Nacional de Investigaci\'on Cient\'ifica y Tecnol\'ogica 
(CONICYT), the French Centre National de la Recherche Scientifique and
Commissariat \`{a} l'Energie Atomique, the Scottish Universities Physics Alliance (SUPA), the UK Science and Technology Facilities Council 
(STFC), and the National Research Foundation of Korea.  Jefferson Science Associates, LLC, operates the Thomas Jefferson National 
Accelerator Facility for the United States Department of Energy under contract
DE-AC05-060R23177.

\end{acknowledgments}

%\bibliography{TPEPRC}

\begin{thebibliography}{81}%
\makeatletter
\providecommand \@ifxundefined [1]{%
 \@ifx{#1\undefined}
}%
\providecommand \@ifnum [1]{%
 \ifnum #1\expandafter \@firstoftwo
 \else \expandafter \@secondoftwo
 \fi
}%
\providecommand \@ifx [1]{%
 \ifx #1\expandafter \@firstoftwo
 \else \expandafter \@secondoftwo
 \fi
}%
\providecommand \natexlab [1]{#1}%
\providecommand \enquote  [1]{``#1''}%
\providecommand \bibnamefont  [1]{#1}%
\providecommand \bibfnamefont [1]{#1}%
\providecommand \citenamefont [1]{#1}%
\providecommand \href@noop [0]{\@secondoftwo}%
\providecommand \href [0]{\begingroup \@sanitize@url \@href}%
\providecommand \@href[1]{\@@startlink{#1}\@@href}%
\providecommand \@@href[1]{\endgroup#1\@@endlink}%
\providecommand \@sanitize@url [0]{\catcode `\\12\catcode `\$12\catcode
  `\&12\catcode `\#12\catcode `\^12\catcode `\_12\catcode `\%12\relax}%
\providecommand \@@startlink[1]{}%
\providecommand \@@endlink[0]{}%
\providecommand \url  [0]{\begingroup\@sanitize@url \@url }%
\providecommand \@url [1]{\endgroup\@href {#1}{\urlprefix }}%
\providecommand \urlprefix  [0]{URL }%
\providecommand \Eprint [0]{\href }%
\providecommand \doibase [0]{http://dx.doi.org/}%
\providecommand \selectlanguage [0]{\@gobble}%
\providecommand \bibinfo  [0]{\@secondoftwo}%
\providecommand \bibfield  [0]{\@secondoftwo}%
\providecommand \translation [1]{[#1]}%
\providecommand \BibitemOpen [0]{}%
\providecommand \bibitemStop [0]{}%
\providecommand \bibitemNoStop [0]{.\EOS\space}%
\providecommand \EOS [0]{\spacefactor3000\relax}%
\providecommand \BibitemShut  [1]{\csname bibitem#1\endcsname}%
\let\auto@bib@innerbib\@empty
%</preamble>
\bibitem [{\citenamefont {Walker}\ \emph {et~al.}(1994)\citenamefont {Walker}
  \emph {et~al.}}]{walker94}%
  \BibitemOpen
  \bibfield  {author} {\bibinfo {author} {\bibfnamefont {R.~C.}\ \bibnamefont
  {Walker}} \emph {et~al.},\ }\href@noop {} {\bibfield  {journal} {\bibinfo
  {journal} {Phys. Rev. D}\ }\textbf {\bibinfo {volume} {49}},\ \bibinfo
  {pages} {5671} (\bibinfo {year} {1994})}\BibitemShut {NoStop}%
\bibitem [{\citenamefont {Andivahis}\ \emph {et~al.}(1994)\citenamefont
  {Andivahis} \emph {et~al.}}]{andivahis94}%
  \BibitemOpen
  \bibfield  {author} {\bibinfo {author} {\bibfnamefont {L.}~\bibnamefont
  {Andivahis}} \emph {et~al.},\ }\href@noop {} {\bibfield  {journal} {\bibinfo
  {journal} {Phys. Rev. D}\ }\textbf {\bibinfo {volume} {50}},\ \bibinfo
  {pages} {5491} (\bibinfo {year} {1994})}\BibitemShut {NoStop}%
\bibitem [{\citenamefont {Berger}\ \emph {et~al.}(1971)\citenamefont {Berger},
  \citenamefont {Burkert}, \citenamefont {Knop}, \citenamefont {Langenbeck},\
  and\ \citenamefont {Rith}}]{berger71}%
  \BibitemOpen
  \bibfield  {author} {\bibinfo {author} {\bibfnamefont {C.}~\bibnamefont
  {Berger}}, \bibinfo {author} {\bibfnamefont {V.}~\bibnamefont {Burkert}},
  \bibinfo {author} {\bibfnamefont {G.}~\bibnamefont {Knop}}, \bibinfo {author}
  {\bibfnamefont {B.}~\bibnamefont {Langenbeck}}, \ and\ \bibinfo {author}
  {\bibfnamefont {K.}~\bibnamefont {Rith}},\ }\href@noop {} {\bibfield
  {journal} {\bibinfo  {journal} {Phys. Lett. B}\ }\textbf {\bibinfo {volume}
  {35}},\ \bibinfo {pages} {87} (\bibinfo {year} {1971})}\BibitemShut {NoStop}%
\bibitem [{\citenamefont {Litt}\ \emph {et~al.}(1970)\citenamefont {Litt} \emph
  {et~al.}}]{litt70}%
  \BibitemOpen
  \bibfield  {author} {\bibinfo {author} {\bibfnamefont {J.}~\bibnamefont
  {Litt}} \emph {et~al.},\ }\href@noop {} {\bibfield  {journal} {\bibinfo
  {journal} {Phys. Lett. B}\ }\textbf {\bibinfo {volume} {31}},\ \bibinfo
  {pages} {40} (\bibinfo {year} {1970})}\BibitemShut {NoStop}%
\bibitem [{\citenamefont {Christy}\ \emph {et~al.}(2004)\citenamefont {Christy}
  \emph {et~al.}}]{christy04}%
  \BibitemOpen
  \bibfield  {author} {\bibinfo {author} {\bibfnamefont {M.~E.}\ \bibnamefont
  {Christy}} \emph {et~al.},\ }\href@noop {} {\bibfield  {journal} {\bibinfo
  {journal} {Phys. Rev. C}\ }\textbf {\bibinfo {volume} {70}},\ \bibinfo
  {pages} {015206} (\bibinfo {year} {2004})}\BibitemShut {NoStop}%
%%CITATION = NUCL-EX 0401030;%%
\bibitem [{\citenamefont {Qattan}\ \emph {et~al.}(2005)\citenamefont {Qattan}
  \emph {et~al.}}]{qattan05}%
  \BibitemOpen
  \bibfield  {author} {\bibinfo {author} {\bibfnamefont {I.~A.}\ \bibnamefont
  {Qattan}} \emph {et~al.},\ }\href@noop {} {\bibfield  {journal} {\bibinfo
  {journal} {Phys. Rev. Lett.}\ }\textbf {\bibinfo {volume} {94}},\ \bibinfo
  {pages} {142301} (\bibinfo {year} {2005})}\BibitemShut {NoStop}%
%%CITATION = NUCL-EX 0410010;%%
\bibitem [{\citenamefont {Punjabi}\ \emph {et~al.}(2005)\citenamefont {Punjabi}
  \emph {et~al.}}]{punjabi05}%
  \BibitemOpen
  \bibfield  {author} {\bibinfo {author} {\bibfnamefont {V.}~\bibnamefont
  {Punjabi}} \emph {et~al.},\ }\href@noop {} {\bibfield  {journal} {\bibinfo
  {journal} {Phys. Rev. C}\ }\textbf {\bibinfo {volume} {71}},\ \bibinfo
  {pages} {055202} (\bibinfo {year} {2005})}\BibitemShut {NoStop}%
\bibitem [{\citenamefont {Puckett}\ \emph {et~al.}(2010)\citenamefont {Puckett}
  \emph {et~al.}}]{puckett10}%
  \BibitemOpen
  \bibfield  {author} {\bibinfo {author} {\bibfnamefont {A.~J.~R.}\
  \bibnamefont {Puckett}} \emph {et~al.},\ }\href@noop {} {\bibfield  {journal}
  {\bibinfo  {journal} {Phys. Rev. Lett.}\ }\textbf {\bibinfo {volume} {104}},\
  \bibinfo {pages} {242301} (\bibinfo {year} {2010})}\BibitemShut {NoStop}%
%%CITATION = 1005.3419;%%
\bibitem [{\citenamefont {Puckett}\ \emph {et~al.}(2012)\citenamefont {Puckett}
  \emph {et~al.}}]{puckett11}%
  \BibitemOpen
  \bibfield  {author} {\bibinfo {author} {\bibfnamefont {A.~J.~R.}\
  \bibnamefont {Puckett}} \emph {et~al.},\ }\href {\doibase
  10.1103/PhysRevC.85.045203} {\bibfield  {journal} {\bibinfo  {journal} {Phys.
  Rev. C}\ }\textbf {\bibinfo {volume} {85}},\ \bibinfo {pages} {045203}
  (\bibinfo {year} {2012})}\BibitemShut {NoStop}%
\bibitem [{\citenamefont {Zhan}\ \emph {et~al.}(2011)\citenamefont {Zhan} \emph
  {et~al.}}]{zhan11}%
  \BibitemOpen
  \bibfield  {author} {\bibinfo {author} {\bibfnamefont {X.}~\bibnamefont
  {Zhan}} \emph {et~al.},\ }\href {\doibase 10.1016/j.physletb.2011.10.002}
  {\bibfield  {journal} {\bibinfo  {journal} {Phys. Lett.}\ }\textbf {\bibinfo
  {volume} {B705}},\ \bibinfo {pages} {59} (\bibinfo {year}
  {2011})}\BibitemShut {NoStop}%
\bibitem [{\citenamefont {Ron}\ \emph {et~al.}(2011)\citenamefont {Ron} \emph
  {et~al.}}]{Ron11}%
  \BibitemOpen
  \bibfield  {author} {\bibinfo {author} {\bibfnamefont {G.}~\bibnamefont
  {Ron}} \emph {et~al.},\ }\href@noop {} {\bibfield  {journal} {\bibinfo
  {journal} {Phys. Rev. C}\ }\textbf {\bibinfo {volume} {84}},\ \bibinfo
  {pages} {055204} (\bibinfo {year} {2011})}\BibitemShut {NoStop}%
\bibitem [{\citenamefont {Crawford}\ \emph {et~al.}(2007)\citenamefont
  {Crawford} \emph {et~al.}}]{crawford07}%
  \BibitemOpen
  \bibfield  {author} {\bibinfo {author} {\bibfnamefont {B.}~\bibnamefont
  {Crawford}} \emph {et~al.},\ }\href@noop {} {\bibfield  {journal} {\bibinfo
  {journal} {Phys. Rev. Lett.}\ }\textbf {\bibinfo {volume} {98}},\ \bibinfo
  {pages} {052301} (\bibinfo {year} {2007})}\BibitemShut {NoStop}%
\bibitem [{\citenamefont {Arrington}(2003)}]{arrington03a}%
  \BibitemOpen
  \bibfield  {author} {\bibinfo {author} {\bibfnamefont {J.}~\bibnamefont
  {Arrington}},\ }\href@noop {} {\bibfield  {journal} {\bibinfo  {journal}
  {Phys. Rev. C}\ }\textbf {\bibinfo {volume} {68}},\ \bibinfo {pages} {034325}
  (\bibinfo {year} {2003})}\BibitemShut {NoStop}%
\bibitem [{\citenamefont {Guichon}\ and\ \citenamefont
  {Vanderhaeghen}(2003)}]{guichon03}%
  \BibitemOpen
  \bibfield  {author} {\bibinfo {author} {\bibfnamefont {P.~A.~M.}\
  \bibnamefont {Guichon}}\ and\ \bibinfo {author} {\bibfnamefont
  {M.}~\bibnamefont {Vanderhaeghen}},\ }\href@noop {} {\bibfield  {journal}
  {\bibinfo  {journal} {Phys. Rev. Lett.}\ }\textbf {\bibinfo {volume} {91}},\
  \bibinfo {pages} {142303} (\bibinfo {year} {2003})}\BibitemShut {NoStop}%
%%CITATION = HEP-PH 0306007;%%
\bibitem [{\citenamefont {Blunden}\ \emph {et~al.}(2003)\citenamefont
  {Blunden}, \citenamefont {Melnitchouk},\ and\ \citenamefont
  {Tjon}}]{blunden03}%
  \BibitemOpen
  \bibfield  {author} {\bibinfo {author} {\bibfnamefont {P.~G.}\ \bibnamefont
  {Blunden}}, \bibinfo {author} {\bibfnamefont {W.}~\bibnamefont
  {Melnitchouk}}, \ and\ \bibinfo {author} {\bibfnamefont {J.~A.}\ \bibnamefont
  {Tjon}},\ }\href@noop {} {\bibfield  {journal} {\bibinfo  {journal} {Phys.
  Rev. Lett.}\ }\textbf {\bibinfo {volume} {91}},\ \bibinfo {pages} {142304}
  (\bibinfo {year} {2003})}\BibitemShut {NoStop}%
%%CITATION = NUCL-TH 0306076;%%
\bibitem [{\citenamefont {Chen}\ \emph {et~al.}(2004)\citenamefont {Chen},
  \citenamefont {Afanasev}, \citenamefont {Brodsky}, \citenamefont {Carlson},\
  and\ \citenamefont {Vanderhaeghen}}]{chen04}%
  \BibitemOpen
  \bibfield  {author} {\bibinfo {author} {\bibfnamefont {Y.~C.}\ \bibnamefont
  {Chen}}, \bibinfo {author} {\bibfnamefont {A.}~\bibnamefont {Afanasev}},
  \bibinfo {author} {\bibfnamefont {S.~J.}\ \bibnamefont {Brodsky}}, \bibinfo
  {author} {\bibfnamefont {C.~E.}\ \bibnamefont {Carlson}}, \ and\ \bibinfo
  {author} {\bibfnamefont {M.}~\bibnamefont {Vanderhaeghen}},\ }\href@noop {}
  {\bibfield  {journal} {\bibinfo  {journal} {Phys. Rev. Lett.}\ }\textbf
  {\bibinfo {volume} {93}},\ \bibinfo {pages} {122301} (\bibinfo {year}
  {2004})}\BibitemShut {NoStop}%
%%CITATION = HEP-PH 0403058;%%
\bibitem [{\citenamefont {Arrington}(2004{\natexlab{a}})}]{arrington04a}%
  \BibitemOpen
  \bibfield  {author} {\bibinfo {author} {\bibfnamefont {J.}~\bibnamefont
  {Arrington}},\ }\href@noop {} {\bibfield  {journal} {\bibinfo  {journal}
  {Phys. Rev.}\ }\textbf {\bibinfo {volume} {C69}},\ \bibinfo {pages} {022201}
  (\bibinfo {year} {2004}{\natexlab{a}})}\BibitemShut {NoStop}%
%%CITATION = NUCL-EX 0309011;%%
\bibitem [{\citenamefont {Arrington}\ \emph {et~al.}(2007)\citenamefont
  {Arrington}, \citenamefont {Melnitchouk},\ and\ \citenamefont
  {Tjon}}]{arrington07c}%
  \BibitemOpen
  \bibfield  {author} {\bibinfo {author} {\bibfnamefont {J.}~\bibnamefont
  {Arrington}}, \bibinfo {author} {\bibfnamefont {W.}~\bibnamefont
  {Melnitchouk}}, \ and\ \bibinfo {author} {\bibfnamefont {J.~A.}\ \bibnamefont
  {Tjon}},\ }\href@noop {} {\bibfield  {journal} {\bibinfo  {journal} {Phys.
  Rev. C}\ }\textbf {\bibinfo {volume} {76}},\ \bibinfo {pages} {035205}
  (\bibinfo {year} {2007})}\BibitemShut {NoStop}%
\bibitem [{\citenamefont {Carlson}\ and\ \citenamefont
  {Vanderhaeghen}(2007)}]{carlson07}%
  \BibitemOpen
  \bibfield  {author} {\bibinfo {author} {\bibfnamefont {C.~E.}\ \bibnamefont
  {Carlson}}\ and\ \bibinfo {author} {\bibfnamefont {M.}~\bibnamefont
  {Vanderhaeghen}},\ }\href {\doibase 10.1146/annurev.nucl.57.090506.123116}
  {\bibfield  {journal} {\bibinfo  {journal} {Ann. Rev. Nucl. Part. Sci.}\
  }\textbf {\bibinfo {volume} {57}},\ \bibinfo {pages} {171} (\bibinfo {year}
  {2007})}\BibitemShut {NoStop}%
%%CITATION = HEP-PH/0701272;%%
\bibitem [{\citenamefont {Arrington}\ \emph {et~al.}(2011)\citenamefont
  {Arrington}, \citenamefont {Blunden},\ and\ \citenamefont
  {Melnitchouk}}]{arrington11b}%
  \BibitemOpen
  \bibfield  {author} {\bibinfo {author} {\bibfnamefont {J.}~\bibnamefont
  {Arrington}}, \bibinfo {author} {\bibfnamefont {P.}~\bibnamefont {Blunden}},
  \ and\ \bibinfo {author} {\bibfnamefont {W.}~\bibnamefont {Melnitchouk}},\
  }\href {\doibase 10.1016/j.ppnp.2011.07.003} {\bibfield  {journal} {\bibinfo
  {journal} {Prog. Part. Nucl. Phys.}\ }\textbf {\bibinfo {volume} {66}},\
  \bibinfo {pages} {782} (\bibinfo {year} {2011})}\BibitemShut {NoStop}%
\bibitem [{\citenamefont {Blunden}\ \emph {et~al.}(2005)\citenamefont
  {Blunden}, \citenamefont {Melnitchouk},\ and\ \citenamefont
  {Tjon}}]{blunden05}%
  \BibitemOpen
  \bibfield  {author} {\bibinfo {author} {\bibfnamefont {P.~G.}\ \bibnamefont
  {Blunden}}, \bibinfo {author} {\bibfnamefont {W.}~\bibnamefont
  {Melnitchouk}}, \ and\ \bibinfo {author} {\bibfnamefont {J.~A.}\ \bibnamefont
  {Tjon}},\ }\href@noop {} {\bibfield  {journal} {\bibinfo  {journal} {Phys.
  Rev. C}\ }\textbf {\bibinfo {volume} {72}},\ \bibinfo {pages} {034612}
  (\bibinfo {year} {2005})}\BibitemShut {NoStop}%
%%CITATION = NUCL-TH 0506039;%%
\bibitem [{\citenamefont {Afanasev}\ and\ \citenamefont
  {Carlson}(2005)}]{afanasev05}%
  \BibitemOpen
  \bibfield  {author} {\bibinfo {author} {\bibfnamefont {A.~V.}\ \bibnamefont
  {Afanasev}}\ and\ \bibinfo {author} {\bibfnamefont {C.~E.}\ \bibnamefont
  {Carlson}},\ }\href@noop {} {\bibfield  {journal} {\bibinfo  {journal} {Phys.
  Rev. Lett.}\ }\textbf {\bibinfo {volume} {94}},\ \bibinfo {pages} {212301}
  (\bibinfo {year} {2005})}\BibitemShut {NoStop}%
\bibitem [{\citenamefont {Afanasev}\ \emph {et~al.}(2005)\citenamefont
  {Afanasev}, \citenamefont {Brodsky}, \citenamefont {Carlson}, \citenamefont
  {Chen},\ and\ \citenamefont {Vanderhaeghen}}]{afanasev05b}%
  \BibitemOpen
  \bibfield  {author} {\bibinfo {author} {\bibfnamefont {A.~V.}\ \bibnamefont
  {Afanasev}}, \bibinfo {author} {\bibfnamefont {S.~J.}\ \bibnamefont
  {Brodsky}}, \bibinfo {author} {\bibfnamefont {C.~E.}\ \bibnamefont
  {Carlson}}, \bibinfo {author} {\bibfnamefont {Y.-C.}\ \bibnamefont {Chen}}, \
  and\ \bibinfo {author} {\bibfnamefont {M.}~\bibnamefont {Vanderhaeghen}},\
  }\href@noop {} {\bibfield  {journal} {\bibinfo  {journal} {Phys. Rev. D}\
  }\textbf {\bibinfo {volume} {72}},\ \bibinfo {pages} {013008} (\bibinfo
  {year} {2005})}\BibitemShut {NoStop}%
\bibitem [{\citenamefont {Kondratyuk}\ and\ \citenamefont
  {Blunden}(2006)}]{kondratyuk06}%
  \BibitemOpen
  \bibfield  {author} {\bibinfo {author} {\bibfnamefont {S.}~\bibnamefont
  {Kondratyuk}}\ and\ \bibinfo {author} {\bibfnamefont {P.~G.}\ \bibnamefont
  {Blunden}},\ }\href@noop {} {\bibfield  {journal} {\bibinfo  {journal} {Nucl.
  Phys.}\ }\textbf {\bibinfo {volume} {A778}},\ \bibinfo {pages} {44} (\bibinfo
  {year} {2006})}\BibitemShut {NoStop}%
\bibitem [{\citenamefont {Kondratyuk}\ and\ \citenamefont
  {Blunden}(2007)}]{kondratyuk07}%
  \BibitemOpen
  \bibfield  {author} {\bibinfo {author} {\bibfnamefont {S.}~\bibnamefont
  {Kondratyuk}}\ and\ \bibinfo {author} {\bibfnamefont {P.~G.}\ \bibnamefont
  {Blunden}},\ }\href@noop {} {\bibfield  {journal} {\bibinfo  {journal} {Phys.
  Rev. C}\ }\textbf {\bibinfo {volume} {75}},\ \bibinfo {pages} {038201}
  (\bibinfo {year} {2007})}\BibitemShut {NoStop}%
\bibitem [{\citenamefont {Belushkin}\ \emph {et~al.}(2007)\citenamefont
  {Belushkin}, \citenamefont {Hammer},\ and\ \citenamefont
  {Meisner}}]{belushkin07}%
  \BibitemOpen
  \bibfield  {author} {\bibinfo {author} {\bibfnamefont {M.~A.}\ \bibnamefont
  {Belushkin}}, \bibinfo {author} {\bibfnamefont {H.-W.}\ \bibnamefont
  {Hammer}}, \ and\ \bibinfo {author} {\bibfnamefont {U.-G.}\ \bibnamefont
  {Meisner}},\ }\href@noop {} {\bibfield  {journal} {\bibinfo  {journal} {Phys.
  Rev. C}\ }\textbf {\bibinfo {volume} {75}},\ \bibinfo {pages} {035202}
  (\bibinfo {year} {2007})}\BibitemShut {NoStop}%
\bibitem [{\citenamefont {Borisyuk}\ and\ \citenamefont
  {Kobushkin}(2008)}]{borisyuk08}%
  \BibitemOpen
  \bibfield  {author} {\bibinfo {author} {\bibfnamefont {D.}~\bibnamefont
  {Borisyuk}}\ and\ \bibinfo {author} {\bibfnamefont {A.}~\bibnamefont
  {Kobushkin}},\ }\href {\doibase 10.1103/PhysRevC.78.025208} {\bibfield
  {journal} {\bibinfo  {journal} {Phys. Rev. C}\ }\textbf {\bibinfo {volume}
  {78}},\ \bibinfo {pages} {025208} (\bibinfo {year} {2008})}\BibitemShut
  {NoStop}%
\bibitem [{\citenamefont {Borisyuk}\ and\ \citenamefont
  {Kobushkin}(2012)}]{borisyuk12}%
  \BibitemOpen
  \bibfield  {author} {\bibinfo {author} {\bibfnamefont {D.}~\bibnamefont
  {Borisyuk}}\ and\ \bibinfo {author} {\bibfnamefont {A.}~\bibnamefont
  {Kobushkin}},\ }\href {\doibase 10.1103/PhysRevC.86.055204} {\bibfield
  {journal} {\bibinfo  {journal} {Phys.Rev.}\ }\textbf {\bibinfo {volume}
  {C86}},\ \bibinfo {pages} {055204} (\bibinfo {year} {2012})}\BibitemShut
  {NoStop}%
\bibitem [{\citenamefont {Borisyuk}\ and\ \citenamefont
  {Kobushkin}(2014)}]{borisyuk14}%
  \BibitemOpen
  \bibfield  {author} {\bibinfo {author} {\bibfnamefont {D.}~\bibnamefont
  {Borisyuk}}\ and\ \bibinfo {author} {\bibfnamefont {A.}~\bibnamefont
  {Kobushkin}},\ }\href@noop {} {\bibfield  {journal} {\bibinfo  {journal}
  {Phys. Rev. C}\ }\textbf {\bibinfo {volume} {89}},\ \bibinfo {pages} {025204}
  (\bibinfo {year} {2014})}\BibitemShut {NoStop}%
\bibitem [{\citenamefont {Tomalak}\ and\ \citenamefont
  {Vanderhaeghen}(2015)}]{tomalak14}%
  \BibitemOpen
  \bibfield  {author} {\bibinfo {author} {\bibfnamefont {O.}~\bibnamefont
  {Tomalak}}\ and\ \bibinfo {author} {\bibfnamefont {M.}~\bibnamefont
  {Vanderhaeghen}},\ }\href@noop {} {\bibfield  {journal} {\bibinfo  {journal}
  {Eur.~Phys.~J.}\ }\textbf {\bibinfo {volume} {A51}},\ \bibinfo {pages} {24}
  (\bibinfo {year} {2015})}\BibitemShut {NoStop}%
\bibitem [{\citenamefont {Zhou}\ and\ \citenamefont {Yang}(2015)}]{Zhou2014}%
  \BibitemOpen
  \bibfield  {author} {\bibinfo {author} {\bibfnamefont {H.-Q.}\ \bibnamefont
  {Zhou}}\ and\ \bibinfo {author} {\bibfnamefont {S.~N.}\ \bibnamefont
  {Yang}},\ }\href@noop {} {\bibfield  {journal} {\bibinfo  {journal} {Eur.
  Phys. J.}\ }\textbf {\bibinfo {volume} {A51}},\ \bibinfo {pages} {105}
  (\bibinfo {year} {2015})}\BibitemShut {NoStop}%
\bibitem [{\citenamefont {Arrington}(2004{\natexlab{b}})}]{arrington04b}%
  \BibitemOpen
  \bibfield  {author} {\bibinfo {author} {\bibfnamefont {J.}~\bibnamefont
  {Arrington}},\ }\href@noop {} {\bibfield  {journal} {\bibinfo  {journal}
  {Phys. Rev. C}\ }\textbf {\bibinfo {volume} {69}},\ \bibinfo {pages} {032201}
  (\bibinfo {year} {2004}{\natexlab{b}})}\BibitemShut {NoStop}%
%%CITATION = NUCL-EX 0311019;%%
\bibitem [{\citenamefont {Arrington}(2009)}]{arrington09b}%
  \BibitemOpen
  \bibfield  {author} {\bibinfo {author} {\bibfnamefont {J.}~\bibnamefont
  {Arrington}},\ }\href {\doibase 10.1063/1.3232022} {\bibfield  {journal}
  {\bibinfo  {journal} {AIP Conf. Proc.}\ }\textbf {\bibinfo {volume} {1160}},\
  \bibinfo {pages} {13} (\bibinfo {year} {2009})}\BibitemShut {NoStop}%
\bibitem [{\citenamefont {Mo}\ and\ \citenamefont {Tsai}(1969)}]{mo69}%
  \BibitemOpen
  \bibfield  {author} {\bibinfo {author} {\bibfnamefont {L.~W.}\ \bibnamefont
  {Mo}}\ and\ \bibinfo {author} {\bibfnamefont {Y.-S.}\ \bibnamefont {Tsai}},\
  }\href@noop {} {\bibfield  {journal} {\bibinfo  {journal} {Rev. Mod. Phys.}\
  }\textbf {\bibinfo {volume} {41}},\ \bibinfo {pages} {205} (\bibinfo {year}
  {1969})}\BibitemShut {NoStop}%
%%CITATION = RMPHA,41,205;%%
\bibitem [{\citenamefont {Maximon}\ and\ \citenamefont
  {Tjon}(2000)}]{maximon00}%
  \BibitemOpen
  \bibfield  {author} {\bibinfo {author} {\bibfnamefont {L.~C.}\ \bibnamefont
  {Maximon}}\ and\ \bibinfo {author} {\bibfnamefont {J.~A.}\ \bibnamefont
  {Tjon}},\ }\href@noop {} {\bibfield  {journal} {\bibinfo  {journal} {Phys.
  Rev.}\ }\textbf {\bibinfo {volume} {C62}},\ \bibinfo {pages} {054320}
  (\bibinfo {year} {2000})}\BibitemShut {NoStop}%
%%CITATION = NUCL-TH 0002058;%%
\bibitem [{\citenamefont {Ent}\ \emph {et~al.}(2001)\citenamefont {Ent} \emph
  {et~al.}}]{ent01}%
  \BibitemOpen
  \bibfield  {author} {\bibinfo {author} {\bibfnamefont {R.}~\bibnamefont
  {Ent}} \emph {et~al.},\ }\href@noop {} {\bibfield  {journal} {\bibinfo
  {journal} {Phys. Rev.}\ }\textbf {\bibinfo {volume} {C64}},\ \bibinfo {pages}
  {054610} (\bibinfo {year} {2001})}\BibitemShut {NoStop}%
%%CITATION = PHRVA,C64,054610;%%
\bibitem [{\citenamefont {Bernauer}\ \emph {et~al.}(2010)\citenamefont
  {Bernauer} \emph {et~al.}}]{bernauer10}%
  \BibitemOpen
  \bibfield  {author} {\bibinfo {author} {\bibfnamefont {J.}~\bibnamefont
  {Bernauer}} \emph {et~al.},\ }\href {\doibase 10.1103/PhysRevLett.105.242001}
  {\bibfield  {journal} {\bibinfo  {journal} {Phys. Rev. Lett.}\ }\textbf
  {\bibinfo {volume} {105}},\ \bibinfo {pages} {242001} (\bibinfo {year}
  {2010})}\BibitemShut {NoStop}%
\bibitem [{\citenamefont {Moteabbed}\ \emph {et~al.}(2013)\citenamefont
  {Moteabbed} \emph {et~al.}}]{Moteabbed}%
  \BibitemOpen
  \bibfield  {author} {\bibinfo {author} {\bibfnamefont {M.}~\bibnamefont
  {Moteabbed}} \emph {et~al.},\ }\href@noop {} {\bibfield  {journal} {\bibinfo
  {journal} {Phys. Rev. C}\ }\textbf {\bibinfo {volume} {88}},\ \bibinfo
  {pages} {025210} (\bibinfo {year} {2013})}\BibitemShut {NoStop}%
\bibitem [{\citenamefont {Tvaskis}\ \emph {et~al.}(2006)\citenamefont {Tvaskis}
  \emph {et~al.}}]{tvaskis06}%
  \BibitemOpen
  \bibfield  {author} {\bibinfo {author} {\bibfnamefont {V.}~\bibnamefont
  {Tvaskis}} \emph {et~al.},\ }\href@noop {} {\bibfield  {journal} {\bibinfo
  {journal} {Phys. Rev.}\ }\textbf {\bibinfo {volume} {C73}},\ \bibinfo {pages}
  {025206} (\bibinfo {year} {2006})}\BibitemShut {NoStop}%
%%CITATION = NUCL-EX 0511021;%%
\bibitem [{\citenamefont {Qattan}\ \emph {et~al.}(2011)\citenamefont {Qattan},
  \citenamefont {Alsaad},\ and\ \citenamefont {Arrington}}]{qattan11b}%
  \BibitemOpen
  \bibfield  {author} {\bibinfo {author} {\bibfnamefont {I.~A.}\ \bibnamefont
  {Qattan}}, \bibinfo {author} {\bibfnamefont {A.}~\bibnamefont {Alsaad}}, \
  and\ \bibinfo {author} {\bibfnamefont {J.}~\bibnamefont {Arrington}},\
  }\href@noop {} {\bibfield  {journal} {\bibinfo  {journal} {Phys. Rev. C}\
  }\textbf {\bibinfo {volume} {84}},\ \bibinfo {pages} {054317} (\bibinfo
  {year} {2011})}\BibitemShut {NoStop}%
%%CITATION = ARXIV:1109.1441;%%
\bibitem [{\citenamefont {Qattan}\ \emph {et~al.}(2015)\citenamefont {Qattan},
  \citenamefont {Arrington},\ and\ \citenamefont {Alsaad}}]{qattan15}%
  \BibitemOpen
  \bibfield  {author} {\bibinfo {author} {\bibfnamefont {I.~A.}\ \bibnamefont
  {Qattan}}, \bibinfo {author} {\bibfnamefont {J.}~\bibnamefont {Arrington}}, \
  and\ \bibinfo {author} {\bibfnamefont {A.}~\bibnamefont {Alsaad}},\
  }\href@noop {} {\bibfield  {journal} {\bibinfo  {journal} {Phys. Rev. C}\
  }\textbf {\bibinfo {volume} {91}},\ \bibinfo {pages} {065203} (\bibinfo
  {year} {2015})}\BibitemShut {NoStop}%
\bibitem [{\citenamefont {Yount}\ and\ \citenamefont {Pine}(1962)}]{yount62}%
  \BibitemOpen
  \bibfield  {author} {\bibinfo {author} {\bibfnamefont {D.}~\bibnamefont
  {Yount}}\ and\ \bibinfo {author} {\bibfnamefont {J.}~\bibnamefont {Pine}},\
  }\href@noop {} {\bibfield  {journal} {\bibinfo  {journal} {Phys. Rev.}\
  }\textbf {\bibinfo {volume} {128}},\ \bibinfo {pages} {1842} (\bibinfo {year}
  {1962})}\BibitemShut {NoStop}%
\bibitem [{\citenamefont {Browman}\ \emph {et~al.}(1965)\citenamefont
  {Browman}, \citenamefont {Liu},\ and\ \citenamefont {Schaerf}}]{browman65}%
  \BibitemOpen
  \bibfield  {author} {\bibinfo {author} {\bibfnamefont {A.}~\bibnamefont
  {Browman}}, \bibinfo {author} {\bibfnamefont {F.}~\bibnamefont {Liu}}, \ and\
  \bibinfo {author} {\bibfnamefont {C.}~\bibnamefont {Schaerf}},\ }\href@noop
  {} {\bibfield  {journal} {\bibinfo  {journal} {Phys. Rev.}\ }\textbf
  {\bibinfo {volume} {139}},\ \bibinfo {pages} {B1079} (\bibinfo {year}
  {1965})}\BibitemShut {NoStop}%
\bibitem [{\citenamefont {Anderson}\ \emph {et~al.}(1966)\citenamefont
  {Anderson}, \citenamefont {Borgia}, \citenamefont {Cassiday}, \citenamefont
  {DeWire}, \citenamefont {Ito},\ and\ \citenamefont {Loh}}]{anderson66}%
  \BibitemOpen
  \bibfield  {author} {\bibinfo {author} {\bibfnamefont {R.~L.}\ \bibnamefont
  {Anderson}}, \bibinfo {author} {\bibfnamefont {B.}~\bibnamefont {Borgia}},
  \bibinfo {author} {\bibfnamefont {G.~L.}\ \bibnamefont {Cassiday}}, \bibinfo
  {author} {\bibfnamefont {J.~W.}\ \bibnamefont {DeWire}}, \bibinfo {author}
  {\bibfnamefont {A.~S.}\ \bibnamefont {Ito}}, \ and\ \bibinfo {author}
  {\bibfnamefont {E.~C.}\ \bibnamefont {Loh}},\ }\href@noop {} {\bibfield
  {journal} {\bibinfo  {journal} {Phys. Rev. Lett.}\ }\textbf {\bibinfo
  {volume} {17}},\ \bibinfo {pages} {407} (\bibinfo {year} {1966})}\BibitemShut
  {NoStop}%
\bibitem [{\citenamefont {Bartel}\ \emph {et~al.}(1967)\citenamefont {Bartel},
  \citenamefont {Dudelzak}, \citenamefont {Krehbiel}, \citenamefont {McElroy},
  \citenamefont {Morrison}, \citenamefont {Schmidt}, \citenamefont {Walther},\
  and\ \citenamefont {Weber}}]{bartel67}%
  \BibitemOpen
  \bibfield  {author} {\bibinfo {author} {\bibfnamefont {W.}~\bibnamefont
  {Bartel}}, \bibinfo {author} {\bibfnamefont {B.}~\bibnamefont {Dudelzak}},
  \bibinfo {author} {\bibfnamefont {H.}~\bibnamefont {Krehbiel}}, \bibinfo
  {author} {\bibfnamefont {J.~M.}\ \bibnamefont {McElroy}}, \bibinfo {author}
  {\bibfnamefont {R.~J.}\ \bibnamefont {Morrison}}, \bibinfo {author}
  {\bibfnamefont {W.}~\bibnamefont {Schmidt}}, \bibinfo {author} {\bibfnamefont
  {V.}~\bibnamefont {Walther}}, \ and\ \bibinfo {author} {\bibfnamefont
  {G.}~\bibnamefont {Weber}},\ }\href@noop {} {\bibfield  {journal} {\bibinfo
  {journal} {Phys. Lett.}\ }\textbf {\bibinfo {volume} {B25}},\ \bibinfo
  {pages} {242} (\bibinfo {year} {1967})}\BibitemShut {NoStop}%
\bibitem [{\citenamefont {Cassiday}\ \emph {et~al.}(1967)\citenamefont
  {Cassiday}, \citenamefont {DeWire}, \citenamefont {Fischer}, \citenamefont
  {Ito}, \citenamefont {Loh},\ and\ \citenamefont {Rutherfoord}}]{cassiday67}%
  \BibitemOpen
  \bibfield  {author} {\bibinfo {author} {\bibfnamefont {G.~L.}\ \bibnamefont
  {Cassiday}}, \bibinfo {author} {\bibfnamefont {J.~W.}\ \bibnamefont
  {DeWire}}, \bibinfo {author} {\bibfnamefont {H.}~\bibnamefont {Fischer}},
  \bibinfo {author} {\bibfnamefont {A.}~\bibnamefont {Ito}}, \bibinfo {author}
  {\bibfnamefont {E.}~\bibnamefont {Loh}}, \ and\ \bibinfo {author}
  {\bibfnamefont {J.}~\bibnamefont {Rutherfoord}},\ }\href@noop {} {\bibfield
  {journal} {\bibinfo  {journal} {Phys. Rev. Lett.}\ }\textbf {\bibinfo
  {volume} {19}},\ \bibinfo {pages} {1191} (\bibinfo {year}
  {1967})}\BibitemShut {NoStop}%
\bibitem [{\citenamefont {Anderson}\ \emph {et~al.}(1968)\citenamefont
  {Anderson}, \citenamefont {Borgia}, \citenamefont {Cassiday}, \citenamefont
  {DeWire}, \citenamefont {Ito},\ and\ \citenamefont {Loh}}]{anderson68}%
  \BibitemOpen
  \bibfield  {author} {\bibinfo {author} {\bibfnamefont {R.~L.}\ \bibnamefont
  {Anderson}}, \bibinfo {author} {\bibfnamefont {B.}~\bibnamefont {Borgia}},
  \bibinfo {author} {\bibfnamefont {G.~L.}\ \bibnamefont {Cassiday}}, \bibinfo
  {author} {\bibfnamefont {J.~W.}\ \bibnamefont {DeWire}}, \bibinfo {author}
  {\bibfnamefont {A.~S.}\ \bibnamefont {Ito}}, \ and\ \bibinfo {author}
  {\bibfnamefont {E.~C.}\ \bibnamefont {Loh}},\ }\href@noop {} {\bibfield
  {journal} {\bibinfo  {journal} {Phys. Rev.}\ }\textbf {\bibinfo {volume}
  {166}},\ \bibinfo {pages} {1336} (\bibinfo {year} {1968})}\BibitemShut
  {NoStop}%
\bibitem [{\citenamefont {Bouquet}\ \emph {et~al.}(1968)\citenamefont
  {Bouquet}, \citenamefont {Benaksas}, \citenamefont {Grossetete},
  \citenamefont {Jean-Marie}, \citenamefont {Parrour}, \citenamefont {Poux},\
  and\ \citenamefont {Tchapoutian}}]{bouquet68}%
  \BibitemOpen
  \bibfield  {author} {\bibinfo {author} {\bibfnamefont {B.}~\bibnamefont
  {Bouquet}}, \bibinfo {author} {\bibfnamefont {D.}~\bibnamefont {Benaksas}},
  \bibinfo {author} {\bibfnamefont {B.}~\bibnamefont {Grossetete}}, \bibinfo
  {author} {\bibfnamefont {B.}~\bibnamefont {Jean-Marie}}, \bibinfo {author}
  {\bibfnamefont {G.}~\bibnamefont {Parrour}}, \bibinfo {author} {\bibfnamefont
  {J.~P.}\ \bibnamefont {Poux}}, \ and\ \bibinfo {author} {\bibfnamefont
  {R.}~\bibnamefont {Tchapoutian}},\ }\href@noop {} {\bibfield  {journal}
  {\bibinfo  {journal} {Phys. Lett.}\ }\textbf {\bibinfo {volume} {B26}},\
  \bibinfo {pages} {178} (\bibinfo {year} {1968})}\BibitemShut {NoStop}%
\bibitem [{\citenamefont {Mar}\ \emph {et~al.}(1968)\citenamefont {Mar} \emph
  {et~al.}}]{mar68}%
  \BibitemOpen
  \bibfield  {author} {\bibinfo {author} {\bibfnamefont {J.}~\bibnamefont
  {Mar}} \emph {et~al.},\ }\href@noop {} {\bibfield  {journal} {\bibinfo
  {journal} {Phys. Rev. Lett.}\ }\textbf {\bibinfo {volume} {21}},\ \bibinfo
  {pages} {482} (\bibinfo {year} {1968})}\BibitemShut {NoStop}%
%%CITATION = PRLTA,21,482;%%
\bibitem [{\citenamefont {Hartwig}\ \emph {et~al.}(1975)\citenamefont {Hartwig}
  \emph {et~al.}}]{hartwig75}%
  \BibitemOpen
  \bibfield  {author} {\bibinfo {author} {\bibfnamefont {S.}~\bibnamefont
  {Hartwig}} \emph {et~al.},\ }\href@noop {} {\bibfield  {journal} {\bibinfo
  {journal} {Lett. Nuovo Cim.}\ }\textbf {\bibinfo {volume} {12}},\ \bibinfo
  {pages} {30} (\bibinfo {year} {1975})}\BibitemShut {NoStop}%
%%CITATION = NCLTA,12,30;%%
\bibitem [{\citenamefont {Drell}\ and\ \citenamefont
  {Ruderman}(1957)}]{drell57}%
  \BibitemOpen
  \bibfield  {author} {\bibinfo {author} {\bibfnamefont {S.~D.}\ \bibnamefont
  {Drell}}\ and\ \bibinfo {author} {\bibfnamefont {M.}~\bibnamefont
  {Ruderman}},\ }\href@noop {} {\bibfield  {journal} {\bibinfo  {journal}
  {Phys. Rev.}\ }\textbf {\bibinfo {volume} {106}},\ \bibinfo {pages} {561}
  (\bibinfo {year} {1957})}\BibitemShut {NoStop}%
\bibitem [{\citenamefont {Drell}\ and\ \citenamefont {Fubini}(1959)}]{drell59}%
  \BibitemOpen
  \bibfield  {author} {\bibinfo {author} {\bibfnamefont {S.~D.}\ \bibnamefont
  {Drell}}\ and\ \bibinfo {author} {\bibfnamefont {S.}~\bibnamefont {Fubini}},\
  }\href@noop {} {\bibfield  {journal} {\bibinfo  {journal} {Phys. Rev.}\
  }\textbf {\bibinfo {volume} {113}},\ \bibinfo {pages} {741} (\bibinfo {year}
  {1959})}\BibitemShut {NoStop}%
\bibitem [{\citenamefont {Greenhut}(1969)}]{greenhut69}%
  \BibitemOpen
  \bibfield  {author} {\bibinfo {author} {\bibfnamefont {G.~K.}\ \bibnamefont
  {Greenhut}},\ }\href@noop {} {\bibfield  {journal} {\bibinfo  {journal}
  {Phys. Rev.}\ }\textbf {\bibinfo {volume} {184}},\ \bibinfo {pages} {1860}
  (\bibinfo {year} {1969})}\BibitemShut {NoStop}%
\bibitem [{\citenamefont {Arrington}\ \emph {et~al.}(2004)\citenamefont
  {Arrington} \emph {et~al.}}]{VEPP-3}%
  \BibitemOpen
  \bibfield  {author} {\bibinfo {author} {\bibfnamefont {J.}~\bibnamefont
  {Arrington}} \emph {et~al.},\ }\href@noop {} {\  (\bibinfo {year} {2004})},\
  \Eprint {http://arxiv.org/abs/0408020} {arXiv:0408020 [nucl-ex]} \BibitemShut
  {NoStop}%
%%CITATION = NUCL-EX/0408020;%%
\bibitem [{\citenamefont {Rachek}\ \emph {et~al.}(2015)\citenamefont {Rachek}
  \emph {et~al.}}]{Rachek2015}%
  \BibitemOpen
  \bibfield  {author} {\bibinfo {author} {\bibfnamefont {I.}~\bibnamefont
  {Rachek}} \emph {et~al.},\ }\href@noop {} {\bibfield  {journal} {\bibinfo
  {journal} {Phys. Rev. Lett.}\ }\textbf {\bibinfo {volume} {114}},\ \bibinfo
  {pages} {062005} (\bibinfo {year} {2015})}\BibitemShut {NoStop}%
\bibitem [{oly()}]{olympus}%
  \BibitemOpen
  \href@noop {} {}\bibinfo {note} {The Proposal and Technical Design Report for
  the OLYMPUS experiment can be found at
  http://web.mit.edu/OLYMPUS}\BibitemShut {NoStop}%
\bibitem [{\citenamefont {Henderson}\ \emph {et~al.}(2016)\citenamefont
  {Henderson} \emph {et~al.}}]{Henderson:2016dea}%
  \BibitemOpen
  \bibfield  {author} {\bibinfo {author} {\bibfnamefont {B.~S.}\ \bibnamefont
  {Henderson}} \emph {et~al.},\ }\href@noop {} {\bibfield  {journal} {\bibinfo
  {journal} {Phys. Rev. Lett.}\ }\textbf {\bibinfo {volume} {118}},\ \bibinfo
  {pages} {092501} (\bibinfo {year} {2017})}\BibitemShut {NoStop}%
%%CITATION = ARXIV:1611.04685;%%
\bibitem [{\citenamefont {Bernaur}\ \emph {et~al.}(2014)\citenamefont {Bernaur}
  \emph {et~al.}}]{OLtgt}%
  \BibitemOpen
  \bibfield  {author} {\bibinfo {author} {\bibfnamefont {J.}~\bibnamefont
  {Bernaur}} \emph {et~al.},\ }\href@noop {} {\bibfield  {journal} {\bibinfo
  {journal} {Nucl. Instr. Methods}\ }\textbf {\bibinfo {volume} {A 755}},\
  \bibinfo {pages} {20} (\bibinfo {year} {2014})}\BibitemShut {NoStop}%
\bibitem [{\citenamefont {Gilman}\ \emph {et~al.}(2013)\citenamefont {Gilman}
  \emph {et~al.}}]{museproposal}%
  \BibitemOpen
  \bibfield  {author} {\bibinfo {author} {\bibfnamefont {R.}~\bibnamefont
  {Gilman}} \emph {et~al.} (\bibinfo {collaboration} {MUSE Collaboration}),\
  }\href@noop {} {\  (\bibinfo {year} {2013})},\ \Eprint
  {http://arxiv.org/abs/1303.2160} {arXiv:1303.2160 [nucl-ex]} \BibitemShut
  {NoStop}%
\bibitem [{\citenamefont {Pohl}\ \emph {et~al.}(2010)\citenamefont {Pohl} \emph
  {et~al.}}]{pohl10}%
  \BibitemOpen
  \bibfield  {author} {\bibinfo {author} {\bibfnamefont {R.}~\bibnamefont
  {Pohl}} \emph {et~al.},\ }\href {\doibase 10.1038/nature09250} {\bibfield
  {journal} {\bibinfo  {journal} {Nature}\ }\textbf {\bibinfo {volume} {466}},\
  \bibinfo {pages} {213} (\bibinfo {year} {2010})}\BibitemShut {NoStop}%
%%CITATION = NATUA,466,213;%%
\bibitem [{\citenamefont {Pohl}\ \emph {et~al.}(2013)\citenamefont {Pohl},
  \citenamefont {Gilman}, \citenamefont {Miller},\ and\ \citenamefont
  {Pachucki}}]{pohl13}%
  \BibitemOpen
  \bibfield  {author} {\bibinfo {author} {\bibfnamefont {R.}~\bibnamefont
  {Pohl}}, \bibinfo {author} {\bibfnamefont {R.}~\bibnamefont {Gilman}},
  \bibinfo {author} {\bibfnamefont {G.~A.}\ \bibnamefont {Miller}}, \ and\
  \bibinfo {author} {\bibfnamefont {K.}~\bibnamefont {Pachucki}},\ }\href@noop
  {} {\bibfield  {journal} {\bibinfo  {journal} {Ann.~Rev.~Nucl.~Part.~Sci.}\
  }\textbf {\bibinfo {volume} {63}},\ \bibinfo {pages} {175} (\bibinfo {year}
  {2013})}\BibitemShut {NoStop}%
\bibitem [{\citenamefont {Mohr}\ \emph {et~al.}(2008)\citenamefont {Mohr},
  \citenamefont {Taylor},\ and\ \citenamefont {Newell}}]{mohr08}%
  \BibitemOpen
  \bibfield  {author} {\bibinfo {author} {\bibfnamefont {P.~J.}\ \bibnamefont
  {Mohr}}, \bibinfo {author} {\bibfnamefont {B.~N.}\ \bibnamefont {Taylor}}, \
  and\ \bibinfo {author} {\bibfnamefont {D.~B.}\ \bibnamefont {Newell}},\
  }\href {\doibase 10.1103//RevModPhys.80.633} {\bibfield  {journal} {\bibinfo
  {journal} {Rev. Mod. Phys.}\ }\textbf {\bibinfo {volume} {80}},\ \bibinfo
  {pages} {633} (\bibinfo {year} {2008})}\BibitemShut {NoStop}%
\bibitem [{\citenamefont {Rosenfelder}(2000)}]{rosenfelder00}%
  \BibitemOpen
  \bibfield  {author} {\bibinfo {author} {\bibfnamefont {R.}~\bibnamefont
  {Rosenfelder}},\ }\href@noop {} {\bibfield  {journal} {\bibinfo  {journal}
  {Phys. Lett.}\ }\textbf {\bibinfo {volume} {B479}},\ \bibinfo {pages} {381}
  (\bibinfo {year} {2000})}\BibitemShut {NoStop}%
%%CITATION = NUCL-TH 9912031;%%
\bibitem [{\citenamefont {Arrington}(2011)}]{arrington11c}%
  \BibitemOpen
  \bibfield  {author} {\bibinfo {author} {\bibfnamefont {J.}~\bibnamefont
  {Arrington}},\ }\href {\doibase 10.1103/PhysRevLett.107.119101} {\bibfield
  {journal} {\bibinfo  {journal} {Phys. Rev. Lett.}\ }\textbf {\bibinfo
  {volume} {107}},\ \bibinfo {pages} {119101} (\bibinfo {year}
  {2011})}\BibitemShut {NoStop}%
\bibitem [{\citenamefont {Bernauer}\ \emph {et~al.}(2011)\citenamefont
  {Bernauer} \emph {et~al.}}]{bernauer11}%
  \BibitemOpen
  \bibfield  {author} {\bibinfo {author} {\bibfnamefont {J.}~\bibnamefont
  {Bernauer}} \emph {et~al.},\ }\href {\doibase 10.1103/PhysRevLett.107.119102}
  {\bibfield  {journal} {\bibinfo  {journal} {Phys. Rev. Lett.}\ }\textbf
  {\bibinfo {volume} {107}},\ \bibinfo {pages} {119102} (\bibinfo {year}
  {2011})}\BibitemShut {NoStop}%
%%CITATION = PRLTA,107,119102;%%
\bibitem [{\citenamefont {Arrington}(2013)}]{arrington13}%
  \BibitemOpen
  \bibfield  {author} {\bibinfo {author} {\bibfnamefont {J.}~\bibnamefont
  {Arrington}},\ }\href@noop {} {\bibfield  {journal} {\bibinfo  {journal}
  {J.~Phys.~G}\ }\textbf {\bibinfo {volume} {40}},\ \bibinfo {pages} {115003}
  (\bibinfo {year} {2013})}\BibitemShut {NoStop}%
\bibitem [{\citenamefont {Arrington}\ and\ \citenamefont
  {Sick}(2015)}]{arrington-sick}%
  \BibitemOpen
  \bibfield  {author} {\bibinfo {author} {\bibfnamefont {J.}~\bibnamefont
  {Arrington}}\ and\ \bibinfo {author} {\bibfnamefont {I.}~\bibnamefont
  {Sick}},\ }\href@noop {} {\  (\bibinfo {year} {2015})},\ \Eprint
  {http://arxiv.org/abs/1505.02680} {arXiv:1505.02680 [nucl-ex]} \BibitemShut
  {NoStop}%
\bibitem [{\citenamefont {Higinbotham}\ \emph {et~al.}(2015)\citenamefont
  {Higinbotham}, \citenamefont {Kabir}, \citenamefont {Lin}, \citenamefont
  {Meekins}, \citenamefont {Norum},\ and\ \citenamefont
  {Sawatzky}}]{Higinbotham15}%
  \BibitemOpen
  \bibfield  {author} {\bibinfo {author} {\bibfnamefont {D.~W.}\ \bibnamefont
  {Higinbotham}}, \bibinfo {author} {\bibfnamefont {A.~A.}\ \bibnamefont
  {Kabir}}, \bibinfo {author} {\bibfnamefont {V.}~\bibnamefont {Lin}}, \bibinfo
  {author} {\bibfnamefont {D.}~\bibnamefont {Meekins}}, \bibinfo {author}
  {\bibfnamefont {B.}~\bibnamefont {Norum}}, \ and\ \bibinfo {author}
  {\bibfnamefont {B.}~\bibnamefont {Sawatzky}},\ }\href@noop {} {\  (\bibinfo
  {year} {2015})},\ \Eprint {http://arxiv.org/abs/1510.01293} {arXiv:1510.01293
  [nucl-ex]} \BibitemShut {NoStop}%
\bibitem [{\citenamefont {Griffioen}\ \emph {et~al.}(2015)\citenamefont
  {Griffioen}, \citenamefont {Carlson},\ and\ \citenamefont
  {Maddox}}]{Griffioen15}%
  \BibitemOpen
  \bibfield  {author} {\bibinfo {author} {\bibfnamefont {K.}~\bibnamefont
  {Griffioen}}, \bibinfo {author} {\bibfnamefont {C.}~\bibnamefont {Carlson}},
  \ and\ \bibinfo {author} {\bibfnamefont {S.}~\bibnamefont {Maddox}},\
  }\href@noop {} {\  (\bibinfo {year} {2015})},\ \Eprint
  {http://arxiv.org/abs/1510.01293} {arXiv:1510.01293 [nucl-ex]} \BibitemShut
  {NoStop}%
\bibitem [{\citenamefont {Adikaram}\ \emph {et~al.}(2015)\citenamefont
  {Adikaram} \emph {et~al.}}]{Adikaram15}%
  \BibitemOpen
  \bibfield  {author} {\bibinfo {author} {\bibfnamefont {D.}~\bibnamefont
  {Adikaram}} \emph {et~al.},\ }\href@noop {} {\bibfield  {journal} {\bibinfo
  {journal} {Phys. Rev. Lett.}\ }\textbf {\bibinfo {volume} {114}},\ \bibinfo
  {pages} {062003} (\bibinfo {year} {2015})}\BibitemShut {NoStop}%
\bibitem [{\citenamefont {Sober}\ \emph {et~al.}(2000)\citenamefont {Sober}
  \emph {et~al.}}]{clastagger}%
  \BibitemOpen
  \bibfield  {author} {\bibinfo {author} {\bibfnamefont {D.~I.}\ \bibnamefont
  {Sober}} \emph {et~al.},\ }\href@noop {} {\bibfield  {journal} {\bibinfo
  {journal} {Nucl. Instr. Methods}\ }\textbf {\bibinfo {volume} {A 440}},\
  \bibinfo {pages} {263} (\bibinfo {year} {2000})}\BibitemShut {NoStop}%
\bibitem [{\citenamefont {Mecking}\ \emph {et~al.}(2003)\citenamefont {Mecking}
  \emph {et~al.}}]{clasNIM}%
  \BibitemOpen
  \bibfield  {author} {\bibinfo {author} {\bibfnamefont {B.~A.}\ \bibnamefont
  {Mecking}} \emph {et~al.},\ }\href@noop {} {\bibfield  {journal} {\bibinfo
  {journal} {Nucl. Instr. Methods}\ }\textbf {\bibinfo {volume} {A 503}},\
  \bibinfo {pages} {513} (\bibinfo {year} {2003})}\BibitemShut {NoStop}%
\bibitem [{\citenamefont {Mestayer}\ \emph {et~al.}(2000)\citenamefont
  {Mestayer} \emph {et~al.}}]{clasDCNIM}%
  \BibitemOpen
  \bibfield  {author} {\bibinfo {author} {\bibfnamefont {M.~D.}\ \bibnamefont
  {Mestayer}} \emph {et~al.},\ }\href@noop {} {\bibfield  {journal} {\bibinfo
  {journal} {Nucl. Instr. Methods}\ }\textbf {\bibinfo {volume} {A 449}},\
  \bibinfo {pages} {81} (\bibinfo {year} {2000})}\BibitemShut {NoStop}%
\bibitem [{\citenamefont {Adams}\ \emph {et~al.}(2001)\citenamefont {Adams}
  \emph {et~al.}}]{clasCCNIM}%
  \BibitemOpen
  \bibfield  {author} {\bibinfo {author} {\bibfnamefont {G.}~\bibnamefont
  {Adams}} \emph {et~al.},\ }\href@noop {} {\bibfield  {journal} {\bibinfo
  {journal} {Nucl. Instr. Methods}\ }\textbf {\bibinfo {volume} {A 465}},\
  \bibinfo {pages} {414} (\bibinfo {year} {2001})}\BibitemShut {NoStop}%
\bibitem [{\citenamefont {Smith}\ \emph {et~al.}(1999)\citenamefont {Smith}
  \emph {et~al.}}]{clasSCNIM}%
  \BibitemOpen
  \bibfield  {author} {\bibinfo {author} {\bibfnamefont {E.~S.}\ \bibnamefont
  {Smith}} \emph {et~al.},\ }\href@noop {} {\bibfield  {journal} {\bibinfo
  {journal} {Nucl. Instr. Methods}\ }\textbf {\bibinfo {volume} {A 432}},\
  \bibinfo {pages} {265} (\bibinfo {year} {1999})}\BibitemShut {NoStop}%
\bibitem [{\citenamefont {Amarian}\ \emph {et~al.}(2001)\citenamefont {Amarian}
  \emph {et~al.}}]{clasECNIM}%
  \BibitemOpen
  \bibfield  {author} {\bibinfo {author} {\bibfnamefont {M.}~\bibnamefont
  {Amarian}} \emph {et~al.},\ }\href@noop {} {\bibfield  {journal} {\bibinfo
  {journal} {Nucl. Instr. Methods}\ }\textbf {\bibinfo {volume} {A 460}},\
  \bibinfo {pages} {239} (\bibinfo {year} {2001})}\BibitemShut {NoStop}%
\bibitem [{\citenamefont {Badier}\ \emph {et~al.}(1994)\citenamefont {Badier}
  \emph {et~al.}}]{Badier94}%
  \BibitemOpen
  \bibfield  {author} {\bibinfo {author} {\bibfnamefont {J.}~\bibnamefont
  {Badier}} \emph {et~al.},\ }\href@noop {} {\bibfield  {journal} {\bibinfo
  {journal} {Nucl. Instr. Methods}\ }\textbf {\bibinfo {volume} {A 348}},\
  \bibinfo {pages} {74} (\bibinfo {year} {1994})}\BibitemShut {NoStop}%
\bibitem [{\citenamefont {Olive}\ and\ \citenamefont {others
  (PDG)}(2014)}]{PDG}%
  \BibitemOpen
  \bibfield  {author} {\bibinfo {author} {\bibfnamefont {K.}~\bibnamefont
  {Olive}}\ and\ \bibinfo {author} {\bibnamefont {others (PDG)}},\ }\href@noop
  {} {\bibfield  {journal} {\bibinfo  {journal} {Chin. Phys.}\ }\textbf
  {\bibinfo {volume} {C 38}},\ \bibinfo {pages} {090001} (\bibinfo {year}
  {2014})}\BibitemShut {NoStop}%
\bibitem [{\citenamefont {Messel}\ and\ \citenamefont
  {Crawford}(1970)}]{Messel}%
  \BibitemOpen
  \bibfield  {author} {\bibinfo {author} {\bibfnamefont {H.}~\bibnamefont
  {Messel}}\ and\ \bibinfo {author} {\bibfnamefont {D.}~\bibnamefont
  {Crawford}},\ }\href@noop {} {\emph {\bibinfo {title} {Electron-Photon Shower
  Distribution Function Tables for Lead, Copper, and Air Absorbers}}},\
  \bibinfo {edition} {1st}\ ed.\ (\bibinfo  {publisher} {Pergamon Press},\
  \bibinfo {year} {1970})\BibitemShut {NoStop}%
\bibitem [{\citenamefont {Pasyuk}()}]{ELOSS}%
  \BibitemOpen
  \bibfield  {author} {\bibinfo {author} {\bibfnamefont {E.}~\bibnamefont
  {Pasyuk}},\ }\href@noop {} {\ }\bibinfo {note} {Available at
  https://misportal.jlab.org/ul/physics/hall-b/clas/index.cfm?note\_year=2007}\BibitemShut
  {NoStop}%
\bibitem [{\citenamefont {Weissbach}\ \emph {et~al.}(2006)\citenamefont
  {Weissbach}, \citenamefont {Hencken}, \citenamefont {Rohe}, \citenamefont
  {Sick},\ and\ \citenamefont {Trautmann}}]{weissbach06}%
  \BibitemOpen
  \bibfield  {author} {\bibinfo {author} {\bibfnamefont {F.}~\bibnamefont
  {Weissbach}}, \bibinfo {author} {\bibfnamefont {K.}~\bibnamefont {Hencken}},
  \bibinfo {author} {\bibfnamefont {D.}~\bibnamefont {Rohe}}, \bibinfo {author}
  {\bibfnamefont {I.}~\bibnamefont {Sick}}, \ and\ \bibinfo {author}
  {\bibfnamefont {D.}~\bibnamefont {Trautmann}},\ }\href@noop {} {\bibfield
  {journal} {\bibinfo  {journal} {Eur.Phys.J.}\ }\textbf {\bibinfo {volume}
  {A30}},\ \bibinfo {pages} {477} (\bibinfo {year} {2006})}\BibitemShut
  {NoStop}%
\bibitem [{\citenamefont {Afanasev}\ \emph {et~al.}(2017)\citenamefont
  {Afanasev}, \citenamefont {Blunden}, \citenamefont {Hasell},\ and\
  \citenamefont {Raue}}]{TPERev}%
  \BibitemOpen
  \bibfield  {author} {\bibinfo {author} {\bibfnamefont {A.}~\bibnamefont
  {Afanasev}}, \bibinfo {author} {\bibfnamefont {P.~G.}\ \bibnamefont
  {Blunden}}, \bibinfo {author} {\bibfnamefont {D.}~\bibnamefont {Hasell}}, \
  and\ \bibinfo {author} {\bibfnamefont {B.~A.}\ \bibnamefont {Raue}},\
  }\href@noop {} {\bibfield  {journal} {\bibinfo  {journal} {Prog. Part. Nucl.
  Phys.}\ ,\ \bibinfo {pages} {In press}} (\bibinfo {year} {2017})},\ \Eprint
  {http://arxiv.org/abs/1703.03874} {arXiv:1703.03874 [nucl-ex]} \BibitemShut
  {NoStop}%
\end{thebibliography}

\end{document}